\documentclass[fleqn,usenatbib]{mnras} % ,draft]

\usepackage{newtxtext,newtxmath}
\usepackage[T1]{fontenc}

\DeclareRobustCommand{\VAN}[3]{#2}
\let\VANthebibliography\thebibliography
\def\thebibliography{\DeclareRobustCommand{\VAN}[3]{##3}\VANthebibliography}

\usepackage{graphicx}	% Including figure files
\usepackage{amsmath}	% Advanced maths commands

\usepackage{amssymb}	% Extra maths symbols
\usepackage{subfigure}
\usepackage{float,lscape}

\title[G4Jy: Host-galaxy identification and redshifts]{The GLEAM 4-Jy (G4Jy) Sample: \newline III. Further host-galaxy identification, and redshift assessment}

\author[White et al. (2025c)]{Sarah V. White$^{1,2,3}$\thanks{sarahwhite.astro@gmail.com}, Precious K. Sejake$^{4}$, Kshitij Thorat$^{4}$, Heinz Andernach$^{5,6}$, Thomas M.O. Franzen$^{7}$,
\newauthor O. Ivy Wong$^{8,9}$, Anna D. Kapi{\'n}ska$^{10}$, Joseph R. Callingham$^{11,12}$, Christopher J. Riseley$^{13,14}$,  \newauthor Nick Seymour$^{3}$, Randall Wayth$^{15}$, Lister Staveley--Smith$^{9}$, Rajan Chhetri$^{3}$, Natasha Hurley-Walker$^{3}$,  \newauthor John Morgan$^{3}$, Paul Hancock$^{3}$, Francesco Massaro$^{16,17,18}$,
 Abigail Garc{\' i}a--P{\' e}rez$^{16,19,20}$, 
\newauthor Ana Jim{\' e}nez--Gallardo$^{20}$, and  Harold A. Pe{\~ n}a--Herazo$^{21}$  \\
$^{1}$South African Astronomical Observatory (SAAO), PO Box 9, Observatory, 7935, South Africa\\
$^{2}$Department of Physics and Electronics, Rhodes University, PO Box 94, Grahamstown, 6140, South Africa\\
$^{3}$International Centre for Radio Astronomy Research (ICRAR), Curtin University, Bentley, WA 6102, Australia\\
$^{4}$Department of Physics, University of Pretoria, Hatfield, Pretoria, 0028, South Africa\\
$^{5}$Th{\" u}ringer Landessternwarte, Sternwarte 5, D-07778 Tautenburg, Germany \\
$^{6}$Permanent address: Depto. de Astronom{\' i}a, DCNE, Univ. de Guanajuato, Callej{\' o}n de Jalisco s/n, C.P. 36023 Guanajuato, Mexico\\
$^{7}$Square Kilometre Array Observatory (SKAO), Jodrell Bank, Lower Withington, Macclesfield, SK11 9FT, UK\\
$^{8}$CSIRO Space \& Astronomy, PO Box 1130, Bentley, WA 6102, Australia\\
$^{9}$ICRAR, University of Western Australia (M468), 35 Stirling Highway, Crawley, WA 6009, Australia\\
$^{10}$National Radio Astronomy Observatory (NRAO), 1003 Lopezville Rd, Socorro NM 87801, USA\\
$^{11}$Netherlands Institute for Radio Astronomy (ASTRON), Oude Hoogeveensedijk 4, 7991 PD, Dwingeloo, The Netherlands\\
$^{12}$Anton Pannekoek Institute for Astronomy, University of Amsterdam, Science Park 904, 1098\,XH, Amsterdam, The Netherlands\\
$^{13}$Astronomisches Institut, Ruhr-Universität Bochum
Universit{\" a}tsstra$\beta$e 150, 44801 Bochum, Germany\\
$^{14}$Ruhr Astroparticle and Plasma Physics Center (RAPP Center), 44780 Bochum, Germany \\
$^{15}$SKAO, 26 Dick Perry Ave, Kensington, WA 6151, Australia \\
$^{16}$Dipartimento di Fisica, Universit{\` a} degli Studi di Torino, via Pietro Giuria 1, I-10125 Torino, Italy\\
$^{17}$Istituto Nazionale di Astrofisica (INAF) - Osservatorio Astrofisico di Torino, via Osservatorio 20, 10025 Pino Torinese, Italy\\
$^{18}$Istituto Nazionale di Fisica Nucleare (INFN) - Sezione di Torino, via Pietro Giuria 1, I-10125 Torino, Italy\\
$^{19}$Instituto Nacional de Astrof{\' i}sica, {\' O}ptica y Electr{\' o}nica, Luis Enrique Erro 1, Tonantzintla, Puebla 72840, M{\' e}xico\\
$^{20}$European Southern Observatory (ESO), Alonso de C{\' o}rdova 3107, Vitacura, Regi{\' o}n Metropolitana, Chile\\
$^{21}$East Asian Observatory (EAO), 660 N. A’oh{\= o}k{\= u} Place, Hilo, HI 96720, USA\\
}

\date{Accepted 2026 January 23. Received 2026 January 20; in original form 2025 July 28}

\pubyear{2015}

\begin{document}
\label{firstpage}
\pagerange{\pageref{firstpage}--\pageref{lastpage}}
\maketitle

% Abstract of the paper
\begin{abstract} % KEEP IT SIMPLE
In this paper we present 127 new host-galaxy identifications for G4Jy sources ($S_{\mathrm{151\,MHz}} > 4$\,Jy), based on radio images from MeerKAT, the Very Large Array Sky Survey (VLASS), and the Rapid ASKAP (Australian Square Kilometre Array Pathfinder) Continuum Survey (RACS). This includes identifications that result from visual inspection of radio contours on $K_{\mathrm{S}}$-band images, as opposed to the AllWISE-W1 images that were used for the original set of overlays when defining the G4Jy Sample (Papers I and II). Our aim is to achieve 100 per cent spectroscopic completeness for the sample, where all of the spectroscopy is available in digital form online. For now, we have gathered (i) digital optical spectroscopy for 34 per cent of the sample, (ii) photometric redshifts for an additional 21 per cent of the sample, and (iii) further redshifts found through the NASA/IPAC Extragalactic Database (but not recently verified). Our assessment of the redshifts includes visual inspection of all of the digital spectroscopy, and re-fitting redshift templates where necessary. The resulting redshift range is (currently) $0.0 < z < 3.6$. We also present 151-MHz luminosities {\color{black}and} linear sizes for the G4Jy Sample, based on initial analysis. 
\end{abstract}

% More-in-depth multiwavelength analysis is presented in the accompanying paper, Paper IV, along with the `multiwavelength' version of the G4Jy catalogue.

% 16 WORDS: and that the optical morphology of candidate quasars is in accordance with the well-studied $K$--$z$ relation. 

% SkyMapper DR4 would give us u,v photometry
% $YJHK$ photometry courtesy of the VISTA (Visible and Infrared Survey Telescope for Astronomy) Hemisphere Survey, where available

% Select between one and six entries from the list of approved keywords.
% Don't make up new ones.

%%%%%%%%%%%%%%%%%%%%%%%%%%%%%%%%%%%%%%%%%%%%%%%%%%

%%%%%%%%%%%%%%%%% BODY OF PAPER %%%%%%%%%%%%%%%%%%

\section{Introduction}

Active galactic nuclei (AGN) play a crucial role in galaxy evolution. They are so-called because they are at the centre of galaxies where there is gas accreting onto the supermassive black-hole \citep[e.g.][]{EHT2019}, which produces emission (across the electromagnetic spectrum) that cannot be explained by the emission from the host galaxy alone. The interaction between the AGN and the host is demonstrated through {\it suppressed} star-formation \citep[e.g.][]{Davies2020, Lammers2023} and {\it enhanced} star-formation \citep[e.g.][]{Ishibashi2012, Silk2013}, with studies yet to reveal the overall impact of the AGN on its surroundings. We are also yet to determine whether the similar histories of black-hole accretion and star-formation activity across cosmic time \citep[e.g.][]{Madau2014} are a result of interaction, or due to there being a common reservoir of gas that fuels both of these processes.      

So how do we create {\color{black}a} sample that will give us a complete picture over a wide range in redshift? An obvious step is to turn to radio-selection, since this waveband enables both star-formation and AGN activity to be detected \citep[e.g.][]{McAlpine2015,White2023} independent of dust obscuration \citep[e.g.][]{DeBreuck2000,Collier2014, Singh2014}, and also independent of high column densities that can affect X-ray observations \citep[e.g.][]{Prieto1996,Massaro2023b}. The difficulty is disentangling the contribution from the AGN from the star-formation-related contribution -- but the same can be said for any sample selected in a given waveband. For example, the AGN can still contribute to the far-infrared (FIR) emission that is typically assumed to be entirely due to re-radiation from star-formation activity \citep[e.g.][]{Symeonidis2022,White2025b}. 

We note that AGN samples selected at mid/high radio frequencies ($\gtrsim1$\,GHz) can encounter a bias with respect to orientation of the radio-jet axis \citep{Lister2003}, but this can be overcome through radio-selection that is at low frequencies instead \citep[e.g.][]{Barthel1989}. This is because the AGN's radio lobes dominate the emission at low radio-frequencies. Not being subject to relativistic beaming \citep{Rees1966}, they therefore provide an isotropic view of AGN \citep[e.g.][]{Hardcastle2020}, in addition to revealing past AGN activity \citep[such as `remnant' radio-galaxies, e.g.][]{Mahatma2018}.

Most notably, the revised Third Cambridge Catalogue of Radio Sources \citep[3CRR;][]{Laing1983} is selected at low frequencies ($S_{\mathrm{178\,MHz}} > 10.9$\,Jy), and complete optical spectroscopy for the sample has enabled pioneering research into the properties/behaviour of powerful radio-galaxies \citep[e.g.][]{Heckman1986}. Specifically, these authors \citep{Laing1983} showed that the sources show strong space-density evolution, and that there is little difference in the evolutionary properties of (normal) galaxies and quasars when matched in luminosity or redshift. This is {\color{black}consistent with} the Unification Model of radio sources \citep[e.g.][]{Barthel1989,Urry1995}.  

Given that the Square Kilometre Array (SKA) and its precursor telescopes -- such as the Murchison Widefield Array \citep[MWA;][]{Tingay2013} --  are located in the southern hemisphere, it is unfortunate that the 173 sources making up the 3CRR sample are restricted to the northern hemisphere (Dec. $\gtrsim$\,+10\,deg). Therefore, in preparation for the SKA, \citet{White2018,White2020a,White2020b} created the G4Jy Sample, which is over 10 times larger than 3CRR (comprising 1,863 sources) and covers the southern sky (Dec. $<$\,+30\,deg). Its low-radio-frequency selection ($S_{\mathrm{151\,MHz}} > 4$\,Jy) also makes it {\color{black}{\it unbiased with respect to the viewing angle of the radio-jet axis}}, like 3CRR. However, as discussed by \citet{White2025a} and Sejake et al. (submitted), analysis of the sample has been slowed by the paucity of all-sky optical spectroscopy in the southern hemisphere. There is the 6-degree Field Galaxy Survey \citep[6dFGS;][]{Jones2004} -- which enabled \citet{Mauch2007} to (i) determine the local ($z\lesssim0.1$) radio luminosity function for `radio-loud' AGN and star-forming galaxies, (ii) study the FIR--radio correlation, and (iii) investigate the bivariate radio--$K$-band luminosity function -- but 6dFGS is only complete to an optical magnitude of $r_{\mathrm{F}}<15.6$ \citep{Jones2009}. 
% (following visual inspection of optical spectroscopy)

% https://classic.sdss.org/dr7/coverage/
%https://skyserver.sdss.org/dr1/en/help/browser/constants.asp?n=SDSSConstants
%https://classic.sdss.org/dr7/products/general/target_quality.php
Contrast this with the Sloan Digital Sky Survey \citep[SDSS;][]{York2000}, which covers $\sim$one third of the northern sky and has a limiting (Petrosian) magnitude of $r'<17.8$ \citep{Strauss2002}. For their seminal work, \citet{Best2012} combined the 7th data release \citep[DR7;][]{Abazajian2009} with radio imaging from the NRAO (National Radio Astronomy Observatory) VLA (Very Large Array) Sky Survey \citep[NVSS;][]{Condon1998} and the Faint Images of the Radio Sky at Twenty centimeters (FIRST) survey \citep{Becker1995}. They were then able to classify sources into high-excitation radio-galaxies (HERGs) and low-excitation radio-galaxies (LERGs), similar to the `strong-line' and `weak-line' criteria applied by \citet{Hine1979} to 3CR sources \citep{Bennett1962}. \citet{Hine1979} show that `strong-line' HERGs tend to have larger radio luminosities than `weak-line' LERGs (albeit affected by Malmquist bias), whilst {\color{black}\citet{Hardcastle2007}} take the classification a step further by suggesting that the radio-galaxy sub-populations are borne out of different black-hole accretion modes being in operation. {\color{black}Furthermore, \citet{Best2012} find that HERGs and LERGs have} distinct host-galaxy properties, with LERGs tending to have more-massive black holes and stellar masses, as well as older stellar populations. 
%(as indicated via the 4000\,\AA\ break).   

We note that SDSS DR7 \citep{Abazajian2009} has also been used by \citet{Sabater2019}, this time in combination with low-frequency radio data from the first data release of the LOFAR Two-metre Sky Survey \citep[LoTSS
DR1;][]{Shimwell2019,Williams2019}. They find that, in their local sample ($z<0.3$), the most-massive galaxies have AGN that are always switched `on' in the radio. The same may be true for higher-redshift massive galaxies, noting that there is a tendency for `radio-loud' AGN to reside in dense environments \citep{Hine1979,Venemans2007,Wylezalek2013, Hlavacek-Larrondo2015}.

How radio galaxies interact with their environment \citep[e.g.][]{Hardcastle2013,Hardcastle2014} requires further investigation, both in terms of the density and turbulence of the surrounding medium, and in terms of how much gas is available for black-hole accretion in the first place \citep[e.g.][]{Ruffa2019,Raouf2023}. This brings us to noting that the Atacama Large Millimeter/submillimeter Array (ALMA), like the SKA, is located in the southern hemisphere, making G4Jy sources the perfect candidates for follow-up. Of course, to simplify observation studies, it helps to know the redshifts of these galaxies ahead of time, hence our considerable effort in this work (Paper III) in collating and assessing redshifts for the G4Jy Sample. Combined with additional multiwavelength analysis (see Paper IV), the G4Jy Sample will enhance models of powerful AGN and allow for more-complete testing than ever before \citep[e.g.][]{Turner2015}.

\subsection{Paper outline} 

This paper, which is the third in the G4Jy paper series, is outlined as follows: Section~\ref{sec:data} describes the multiwavelength datasets that we consider for this work, which includes new host-galaxy identification (a continuation of G4Jy Paper II) in Section~\ref{sec:IDs}. This is followed by contributions towards a multiwavelength catalogue and a summary of the redshift collation {\color{black}(Section~\ref{sec:catalogue})}. We then present initial results of studying the radio properties of G4Jy sources in Section~\ref{sec:results}. We describe our conclusions in Section~\ref{sec:conclusions}, and provide additional information in Appendices~\ref{app:FornaxA}--\ref{app:desi_spectra}. More-in-depth multiwavelength analysis is presented in the accompanying paper, Paper IV \citep{White2025d}, along with the `multiwavelength' version of the G4Jy catalogue.

J2000 co-ordinates and AB magnitudes are used throughout this work, except where {\it WISE} magnitudes are considered (e.g. Section~\ref{sec:MIR_data}) -- these are in the Vega system. We apply a $\Lambda$CDM cosmology, with $H_{0} = 70$\,km\,s$^{-1}$\,Mpc$^{-1}$, $\Omega_{m}=0.3$, $\Omega_{\Lambda}=0.7$.

%Simple mathematics can be inserted into the flow of the text e.g. $2\times3=6$
%or $v=220$\,km\,s$^{-1}$, but more complicated expressions should be entered as a numbered equation:
%\begin{equation}
%    x=\frac{-b$\pm$\sqrt{b^2-4ac}}{2a}.
%	\label{eq:quadratic}
%\end{equation}
%Refer back to them as e.g. equation~(\ref{eq:quadratic}).

%Figures are referred to as e.g. Fig.~\ref{fig:example_figure}, and tables as
%e.g. Table~\ref{tab:example_table}.

%\section{Data}

%In this section we describe the deep, multiwavelength datasets that are used for the selection and analysis of sources over two fields:
%\begin{enumerate}
 %   \item The Cosmic Evolution Survey region (COSMOS; \citealt{Scoville2007}), centred at 10:00:28.6, +02:12:21.0 (J2000) and covering 1.6 sq deg. 
%    \item The XMM-{\it Newton} Large Scale Structure (XMM-LSS) field \citep{Pierre2004}. This is centred at 02:18:00.0, $-$7:00:00.0 (J2000), and we consider the area over which deep near-infrared data is available, which covers 3.5 sq deg. 
%\end{enumerate}
% Where the positions came from:
% https://cosmos.astro.caltech.edu/page/astronomers
% https://vela.astro.uliege.be/themes/spatial/xmm/LSS/srv_char_e.html

\section{Data}
\label{sec:data}

In this section we describe the multiwavelength datasets that are used for this work. For the radio data (Section~\ref{sec:radio_data}), we provide an overview and direct the reader to further details that can be found in \citet{White2020a,White2020b} and \citet{Sejake2023}. Also important for host-galaxy identification and sample analysis are mid-infrared, near-infrared, and optical photometry, as well as the optical spectroscopy (Section~\ref{sec:optical_spectroscopy}) that is crucial for determining reliable redshifts for the sample.

\subsection{Radio data}
\label{sec:radio_data}

The G4Jy Sample \citep{White2020a, White2020b} is a complete selection of all radio-sources in the Southern sky (Dec. $<+30$\,deg, $|b| > 10$\,deg) that are brighter than 4\,Jy at 151\,MHz, as measured for the GLEAM Extragalactic Catalogue \citep{Wayth2015,HurleyWalker2017}. However, due to the difficulty in calibration and imaging, some of the very brightest sources (such as Fornax~A) were excised from the catalogue (see Appendix~\ref{app:FornaxA}). To improve completeness of the sample, \citet{White2020a, White2020b} carried out thorough searches for radio-sources that have their $S_{\mathrm{151\,MHz}}>4$-Jy radio emission spread across multiple GLEAM components, at varying angular separation (noting that the MWA Phase-I beam offers a spatial resolution of $\sim$2\,arcmin). Furthermore, to better understand the radio morphology of these sources (which include a cluster relic and the Flame Nebula), and to aid host-galaxy identification, the following higher-resolution images were employed for visual inspection\footnote{The full set of overlays (v8 and v9) can be downloaded from the G4Jy repository on Zenodo: \url{https://zenodo.org/communities/g4jy/records}.}:
\begin{enumerate}
    \item The TIFR GMRT Sky Survey (TGSS) first alternative data release \citep[ADR1;][]{Intema2017}, at a spatial resolution of $\sim$25\,arcsec. This provides coverage for sources above Dec.\ $=-53$\,deg, to a 7-$\sigma$ depth of $\sim$12\,mJy beam$^{-1}$ (for the corresponding catalogue).
    \item The Sydney University Molonglo Sky Survey \citep[SUMSS;][]{Mauch2003, Murphy2007}, at $\sim$45-arcsec resolution. This has a sky coverage of Dec. $<-30$\,deg, $|b| > 10$\,deg, and we use the imaging (and catalogue positions) for G4Jy sources at Dec.\ $<-39.5$\,deg.
    \item the NRAO (National Radio Astronomy Observatory) VLA (Very Large Array) Sky Survey \citep[NVSS;][]{Condon1998}, at $\sim$45-arcsec resolution. This covers the entire sky north of Dec.\ $=-40$\,deg, and we use the imaging (and catalogue positions) for G4Jy sources at Dec.\ $>-39.5$\,deg.
\end{enumerate}

 {\color{black}In addition, the} G4Jy catalogue \citep{White2020a, White2020b} presents summed flux-densities at each of the 20 sub-band frequencies of the GLEAM Survey \citep{Wayth2015}, spanning 72 to 231\,MHz, which provides excellent constraints of the low-frequency emission from these powerful sources. The current work includes host-galaxy identification for the sample (Section~\ref{sec:IDs}) that has since been aided by the following supplementary datasets:

\begin{enumerate}
\setcounter{enumi}{3}
    \item MeerKAT \citep{Jonas2016} follow-up of sources that have enigmatic radio-morphologies \citep{Sejake2023}. These 5-minute snapshot images are at a spatial resolution of $\sim$7\,arcsec, with core flux-densities of the G4Jy sources ranging from 37\,$\upmu$Jy\,beam$^{-1}$ to 2.5\,Jy\,beam$^{-1}$ at 1.3\,GHz.
    % Just picking the lowest and highest core-flux densities in Table A1 of Sejake et al. (2023)
    \item `Quick Look' \citep[e.g.][]{Gordon2020} and `Median Stack' (Dong et al., in prep.) images from the Karl G. Jansky VLA Sky Survey (VLASS; \citealt{Lacy2020}), providing $\sim$2.5-arcsec spatial-resolution imaging at 2--4\,GHz. The survey is ongoing, with an aim of reaching a 1-$\sigma$ depth of 70\,$\upmu$Jy\,beam$^{-1}$ in the final, co-added data. For comparison, the typical noise-level in the Median-Stack images is 80--90\,$\upmu$Jy\,beam$^{-1}$, dependent on the level of radio-frequency interference (M.~Lacy, private communication). 
    \item Images and catalogue positions from the Rapid ASKAP (Australian Square Kilometre Array Pathfinder) Continuum Survey \citep{McConnell2020} at 887.5\,MHz \citep[RACS-low1;][]{Hale2021}. Its sky coverage is $-80<$\,Dec./deg\,$<+30$ ($|b|>5$\,deg), with images at 25\,arcsec in spatial resolution and having 95-per-cent point-source completeness at an integrated flux-density of $\sim$3\,mJy.  % Astrometric offsets are < 0.8 arcsec in RA and Dec
\end{enumerate}

We note that updating the angular sizes in the G4Jy catalogue is beyond the scope of the current paper. For the moment, the sizes continue to be based upon the 45-arcsec imaging provided by the combination of SUMSS and NVSS (for the aforementioned Declination ranges).

\subsection{Mid-infrared data}
\label{sec:MIR_data}

We also continue to use W1 mid-infrared images from AllWISE \citep{Cutri2013}, to avoid our host-galaxy identifications being biased against the most dust-obscured sources \citep{White2020a,White2020b}. These data are from the all-sky survey conducted in four bands -- W1 (3.4\,$\upmu$m), W2 (4.6\,$\upmu$m), W3 (12\,$\upmu$m), and W4 (22\,$\upmu$m) -- by the {\it Wide-field Infrared Survey Explorer} ({\it WISE}; \citealt{Wright2010}). The corresponding resolutions are 6.1, 6.4, 6.5, and 12.0\,arcsec, and the 5-$\sigma$ sensitivities are 0.054, 0.071, 0.73, and 5.0\,mJy, respectively. In Vega magnitudes, the $>$95-per-cent completeness levels are 17.1, 15.7, 11.5, and 7.7\,mag \citep{Cutri2013}\footnote{\url{https://wise2.ipac.caltech.edu/docs/release/allwise/expsup/sec2_1.html}}.

In cases where we could not identify a mid-infrared counterpart in the AllWISE catalogue, we instead consulted the CatWISE2020 catalogue by \citet{Marocco2021}. This is based on combining W1 and W2 data from {\it WISE} with those from {\it NEOWISE(R)} \citep{Mainzer2014}, which is the reactivation of the original spacecraft after it had been put into hibernation mode (its subsequent, main mission being to detect, track and characterise near-Earth objects; `NEO's). The greater sensitivity in the combined dataset is reflected with 90-per-cent completeness levels of 17.7\,mag in the W1 band and 17.5\,mag in the W2 band. The greater number of detected sources is also owed to the source-finder, {\sc crowdsource} \citep{Schlafly2018}, which is better-able to model crowded fields than that used for AllWISE source-finding.

\subsection{Near-infrared data}
\label{sec:NIR_data} 

We use imaging from the VISTA (Visible and Infrared Survey Telescope for Astronomy) Hemisphere Survey \citep[VHS;][]{McMahon2013, McMahon2021} to help with host-galaxy identification (e.g. Section~\ref{sec:new_IDs}), and note that the 5-$\sigma$ point-source limiting magnitudes for VHS are: $Y = 21.1, J=20.8, H=20.6$, and $K_{\mathrm{s}}=20.0$\,mag (in the AB-magnitude system). {\color{black}The survey covers the whole southern hemisphere ($\sim$20,000\,deg$^2$), and so overlaps with 70 per cent of the G4Jy Sample.}  Conducted on the VISTA telescope \citep{Sutherland2015}, the resolution of the VHS data is limited by the seeing conditions at the Cerro Paranal Observatory. During the course of the survey the median seeing was $\sim$1\,arcsec (R. McMahon, private communication), which informs the positional cross-matching radius that we use for acquiring redshifts.

\subsection{Optical spectroscopy}
\label{sec:optical_spectroscopy}

The multiwavelength G4Jy catalogue (completed as part of Paper IV) contains spectroscopic redshifts from multiple surveys, these being:
\begin{itemize}
    \item the 6-degree Field Galaxy Survey \citep[6dFGS;][]{Jones2004,Jones2009} on the UK Schmidt Telescope \citep{Tritton1978}, {\color{black}which covers $\sim$17,000\,deg$^2$ of the southern sky at Dec.$<0.0$\,deg, $|b|>10$\,deg,}
    \item the Sloan Digital Sky Survey \citep[SDSS, Data Releases 12 and 16;][]{Gunn1993,Alam2015,Ahumada2020} on the Sloan Foundation 2.5-m Telescope \citep{Gunn2006}, the Ir{\' e}n{\' e}e du Pont Telescope \citep{Bowen1973}, and the NMSU 1-Meter Telescope \citep{Holtzman2010},
    \item the first data release \citep{AbdulKarim2025} of the Dark Energy Spectroscopic Instrument \citep[DESI;][]{Levi2013} survey on the Nicholas U. Mayall 4-meter Telescope, {\color{black}over a 14,000-deg$^2$ footprint,} and
    \item follow-up of the G4Jy Sample (2020-1-MLT-008, PI: White) on the Southern African Large Telescope \citep[SALT;][]{Buckley2006}, {\color{black}for sources at $-76<$\,Dec./deg\,$<+11$}.
\end{itemize}
  The latter is using the Robert Stobie Spectrograph \citep[RSS;][]{Burgh2003,Kobulnicky2003}, which covers the wavelength range $\sim$4500--7500\,\AA\ at a spectral resolution of R = $\sim$660--930. Data Releases 1 and 2 (\citealt{White2025a}; Sejake et al. submitted) have been made available to the community through the Zenodo repository for the G4Jy Sample\footnote{\url{https://zenodo.org/communities/g4jy}}, with processing of SALT data also completed for a third data release (Sejake et al., in prep.).

In addition, spectroscopy is being collated for the ``3CR equivalent'' subset of the G4Jy Sample: G4Jy-3CRE \citep{Massaro2023,GarciaPerez2024}. These are the radio-brightest sources in the sample ($S_{\mathrm{181\,MHz}}\gtrsim$\,9.0\,Jy) at Dec. $<-5$\,deg, with the following instruments being employed: 
\begin{itemize}
    \item the Cerro Tololo Ohio State Multi-Object Spectrograph (COSMOS; \citealt{Martini2014}) on the V{\' i}ctor Blanco Telescope \citep{Blanco1975},
    \item the EFOSC2 spectrograph \citep{Melnick1995} on the New Technology Telescope \citep[NTT;][]{Tarenghi1989},
    \item the Goodman High Throughput Spectrograph \citep{Clemens2004} on the Southern Astrophysical Research Telescope (SOAR),
    \item the Boller \& Chivens spectrograph \citep{HeydariMalayeri1989} on the 2.1-m telescope of the Observatorio Astron{\' o}mico Nacional at San Pedro M{\' a}rtir (OAN-SPM), and
    \item the Multi Unit Spectroscopic Explorer \citep[MUSE;][]{Bacon2010} on the Very Large Telescope \citep[VLT;][]{VLT1998}.
\end{itemize}

%A summary of the properties of the different spectrographs can be found in Table~\ref{tab:spectrographs}. If room, and time.

\section{Host-galaxy identification}
\label{sec:IDs}

As part of defining the G4Jy Sample, \citet{White2020a,White2020b} erred on the side of caution when identifying the host galaxies of the radio emission for 1,863 G4Jy sources. Noting ambiguity that could only be resolved with better imaging than that available at the time, they provided 1,606 identifications as part of the original G4Jy catalogue. For this work, we start by adding the 86 identifications enabled by MeerKAT follow-up of sources with particularly unusual radio morphology \citep{Sejake2023}\footnote{12 of the 98 identifications provided by \citet{Sejake2023} were confirmations of existing identifications in the G4Jy catalogue.}. In this section we: (i) continue to add new identifications, (ii) describe a `tentative' identification, (iii) draw attention to sources that are largely limited by the depth of the optical/near-infrared imaging, and (iv) note revisions to existing host-galaxy identifications (based on recent literature).

Like \citet{White2020a,White2020b}, we rely on careful visual inspection of radio contours, with emphasis placed on the higher resolution afforded by MeerKAT follow-up and VLASS\footnote{As a caveat, we acknowledge that the `quick-look' VLASS images are not deemed `science ready' but they provide valuable impressions of the radio morphology at $\sim2.5$-arcsec resolution, through 3-GHz observations}. For the underlying greyscale image, we first consider mid-infrared (MIR) images from AllWISE to avoid biases with respect to dust obscuration, and to maintain consistency (where possible) with the original set of G4Jy overlays. Where AllWISE-catalogue positions do not yield a result, we then consider $K$-band images from various near-infrared (NIR) surveys, followed by $H$-band NIR images and optical images from PanSTARRS. Whilst the latter {\it are} significantly biased by dust obscuration, they afford a higher resolution that can help to distinguish between candidate host-galaxies in particularly-dense fields. 

For sources identified in AllWISE \citep{Cutri2013}, we follow the naming convention where the (J2000) sexagesimal name is preceded with `WISEA'. Those identified in 2MASS \citep{Cutri2003}, VHS DR5 \citep{McMahon2013,McMahon2021}, and PanSTARRS \citep{Chambers2016} have a prefix of `2MASS', `VHS', and `PSO', respectively. We also include CatWISE2020 positions \citep{Marocco2021}, for sources identified in VHS and PanSTARRS, through 1-arcsec positional crossmatching via TOPCAT \citep{Taylor2005}. For these, we use the prefix `CWISE' for the `host name' column in the G4Jy catalogue, and find that there are CatWISE2020 counterparts for 31 of the 47 positions that are considered. For the remainder, we assign a sexagesimal name prefixed by `VHS' or `PSO', as appropriate.

\subsection{New identifications}
\label{sec:new_IDs}

%Thanks to Katlego's overlays. And some VLASS overlays that Natasha and I did.

\begin{figure}
\centering
\includegraphics[scale=0.23]{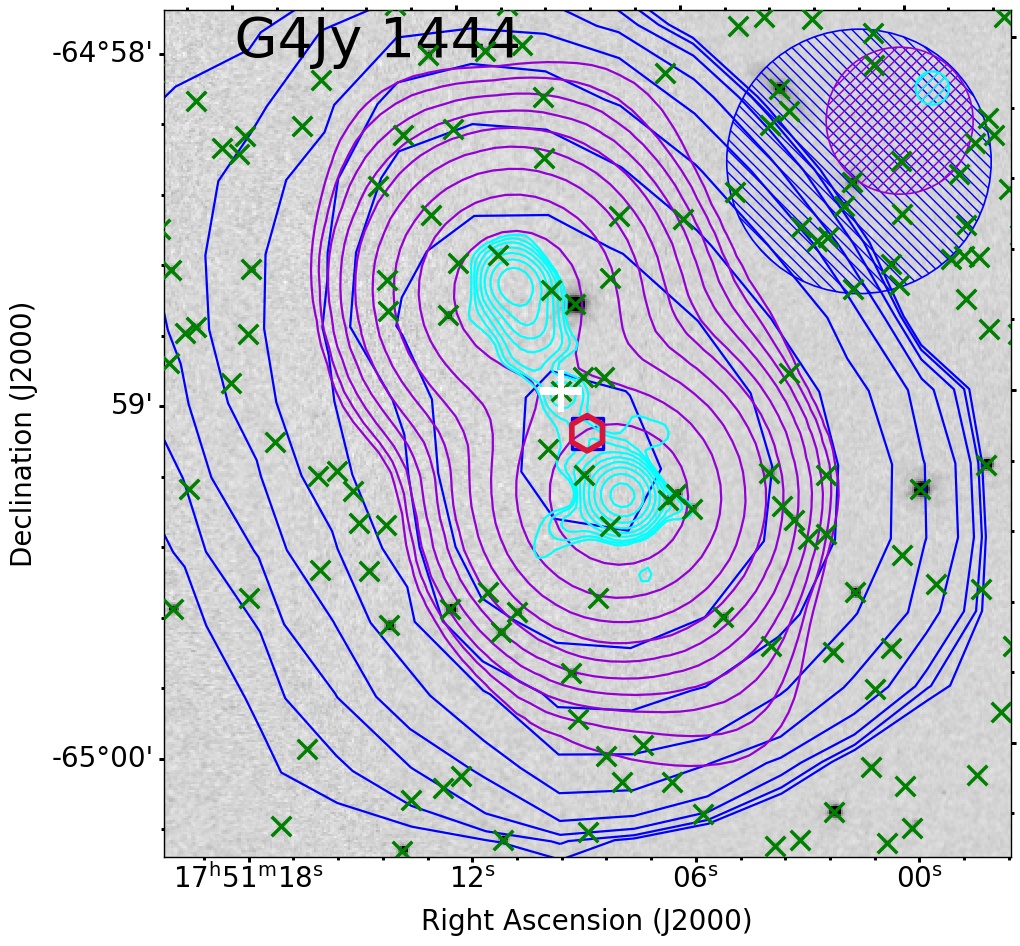}
\caption{ An example overlay showing how different sets of radio contours (SUMSS [843\,MHz] in blue, RACS-low1 [888\,MHz] in purple, and MeerKAT [1300\,MHz] in cyan) were used to assess the morphology of G4Jy~1444, with the respective beam-sizes of the different radio surveys shown in the upper right-hand corner. The underlying, inverted-greyscale image is a $K$-band image from VHS, with green plusses (`+') marking VHS catalogue positions within 3\,arcmin of the radio-centroid position (red hexagon). This enabled us to identify the appropriate host galaxy of the radio emission (white `+'), which was followed by thorough checks against the literature (Section~\ref{sec:new_IDs}) before being included in the updated G4Jy catalogue.}
\label{fig:contours}
\end{figure}

Figure~\ref{fig:contours} shows an example of how MeerKAT imaging \citep{Sejake2023} allows the radio-core of {\bf G4Jy 1444} to be detected, which is especially helpful in pinpointing the host galaxy of the radio emission (CWISE J175109.26$-$645858.3). Nearby are other candidates that are also located between the radio lobes of the source, but their non-coincidence with the core means that they can be safely ruled out. For particularly southern sources (like this one), beyond the reach of TGSS, RACS-low1 provides important intermediate-resolution imaging, showing that this is a single extended radio-galaxy rather than two unresolved radio-sources located very close to one another. Further examples are shown in Figure~\ref{fig:maintext_Kband_IDs}, with the remainder in Appendix~\ref{app:overlays}.

% \footnote{\url{https://cds.u-strasbg.fr/cgi-bin/Dic-Simbad?/5971713}}

\setkeys{Gin}{draft=false}
\setkeys{Gin}{draft}

{\bf G4Jy 41} shows pointlike morphology in each of the radio images that we have inspected, including VLASS and the `low'- and `mid'-frequency images from the Rapid ASKAP Continuum Survey (RACS; \citealt{McConnell2020,Hale2021,Duchesne2023}). However, as noted by \citet{White2020b}, a bright star southeast of the host means that it is not possible to provide an AllWISE/CatWISE2020 identification. Instead, we turn to the Dark Energy Survey, which allows us to identify the host as DES J002125.38$-$005540.3 \citep{Abbott2021}. As yet, there are only photometric redshifts for this source: $z_{\mathrm{ph}}=0.95$ by \citet{Zhou2021} and $z_{\mathrm{ph}}=1.062$ by \citet{Duncan2022}.
 
% Heinz: "in VLASS, so LAS < 2", hosted by ,  r=23.24 mag, with z\_phot\_DESI\_DR9,2022MNRAS.512.3662D=0.95,1.06. As you say, the WISE images are affected by a bright star ~1.15' SE of the host."

In Figure~\ref{fig:maintext_Kband_IDs}a, we see how imaging from VLASS also allows us to detect the core of the radio galaxy ({\bf G4Jy 562}), again differentiating the correct host of the radio emission (VHS J053558.90$-$045537.7). This time, the inverted-greyscale image is from the $K_{S}$ band of the VHS, and the source is at high-enough Declination to have coverage from TGSS (yellow contours). The additional emission seen towards the southeast of the source is believed to be an example of imaging artefacts in VLASS, and so should be treated with caution.

{\bf G4Jy 776} (Figure~\ref{fig:maintext_Kband_IDs}b) is a reminder that the radio emission from radio-bright sources can be very asymmetric, with bright compact emission (believed to be the core) coinciding with the position of a particularly MIR-bright source (CWISE J093518.20+020415.5). The adjacent compact emission (also marked by an orange star in the figure) is likely to be an inner jet, whose energy dissipation may be affected by a heterogeneous surrounding medium. Alternatively, {\color{black} the difference in brightness of the outer lobes may be due to a hotspot in the southwestern lobe being beamed at 3\,GHz \citep[e.g.][]{Georganopoulos2003}.} 

The VLASS contours for {\bf G4Jy 983} (PKS 1215+039) show that this is a bent-tail radio-galaxy, with extended emission that gets `pushed back' from the direction of motion, resulting in a narrow-angle-tail (NAT) morphology (Figure~\ref{fig:maintext_Kband_IDs}c). The `pinching' of the inner parts of the radio jets allows us to identify the host galaxy (CWISE J121731.43+033656.8), distinguishing it from another candidate that coincides with the collimated part of the southern jet. Like G4Jy 776, this source is recorded in the catalogues of \citet{Taylor2009} and \citet{Farnes2014} for its polarisation properties. 
% Removed: This is in agreement with the optical identification provided by \citet{Schilizzi1975}. -- because they actually provide multiple optical positions for 1215+03

{\bf G4Jy 1325} is another example of radio-core detection leading to unambiguous identification of the host galaxy, VHS J162100.60$-$234203.6 (Figure~\ref{fig:maintext_Kband_IDs}d). We note that it is significantly offset from the brightness-weighted centroid (which is calculated from NVSS/SUMSS components) but still lies in line with the hotspots of both lobes. Also known as PKS~1617$-$235, \citet{Bolton1975} refer to the host as being a ``highly obscured'' elliptical, but do not provide co-ordinates for their identification in the Palomar Sky Survey. 

Very often, the radio images from higher-frequency surveys are able to resolve apparently `pointlike' radio-sources into multiple components. This is not the case for {\bf G4Jy 1396} (PKS 1712$-$033), where each set of radio contours indicates that this is an unresolved source with `single' radio morphology (Figure~\ref{fig:maintext_Kband_IDs}e). Thankfully the depth of the VHS $K_{S}$-band is such that the host is easily identifiable amongst the candidate positions. 

For {\bf G4Jy 1835} (PKS 2326$-$196), the 3-GHz VLASS contours much-resemble the 5-GHz contours mapped by \citet{Reid1999}. The host galaxy, CWISE J232933.07$-$192252.1, is clearly where the inner parts of the radio jets `pinch' together (Figure~\ref{fig:maintext_Kband_IDs}f), and we note that this is another polarised source studied by \citet{Taylor2009} and \citet{Farnes2014}.

\setkeys{Gin}{draft=false}

\begin{figure*}
%\vspace{-0.7cm}
\centering

\subfigure[G4Jy~562]{
	\includegraphics[width=0.4\linewidth]{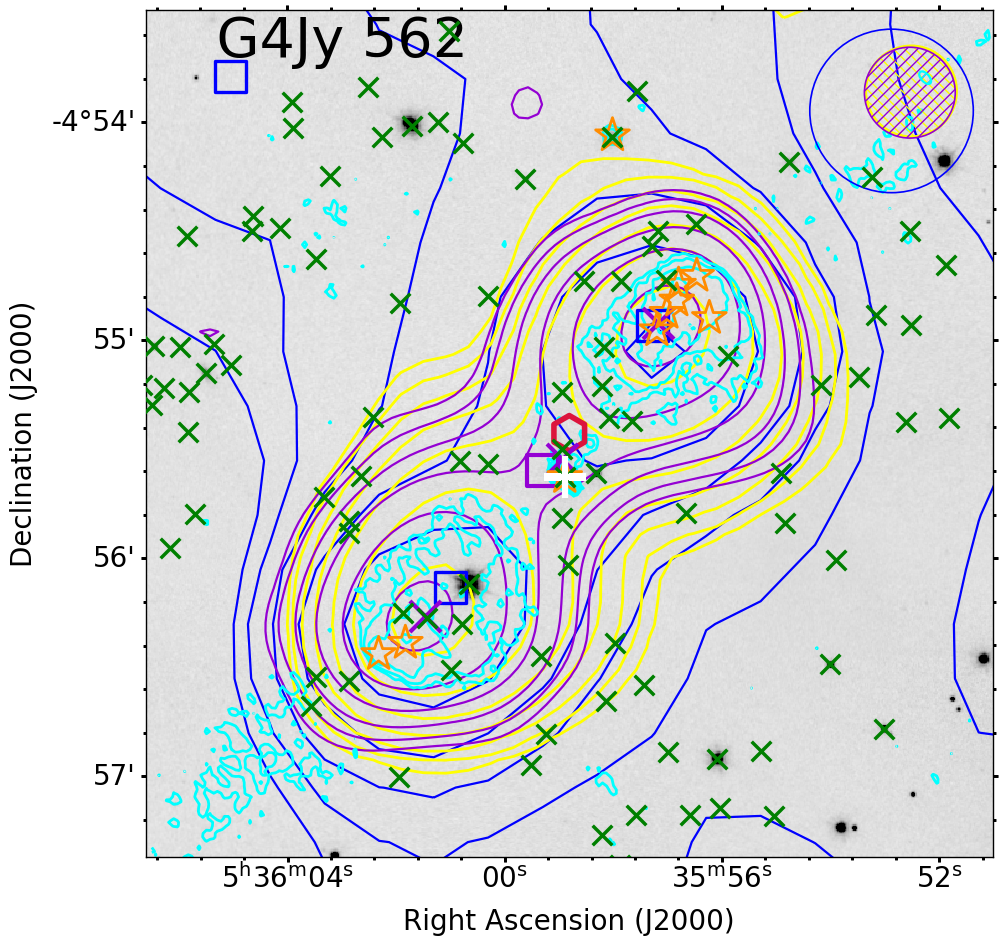}
	}
\subfigure[G4Jy~776 (MIR image from AllWISE)]{
	\includegraphics[width=0.4\linewidth]{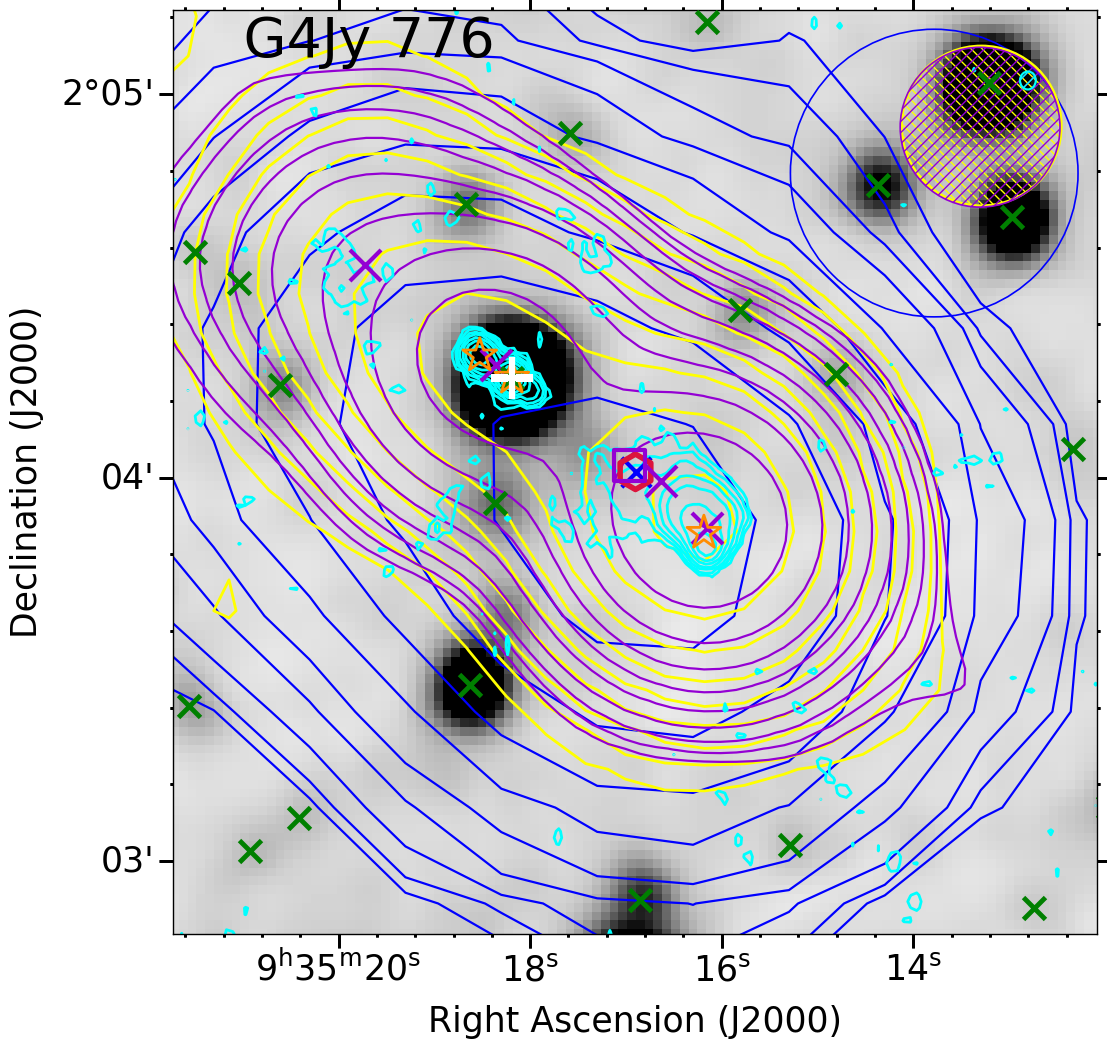} 
	}  \\[-0.05cm]
\subfigure[G4Jy~983 (MIR image from AllWISE)]{
	\includegraphics[width=0.4\linewidth]{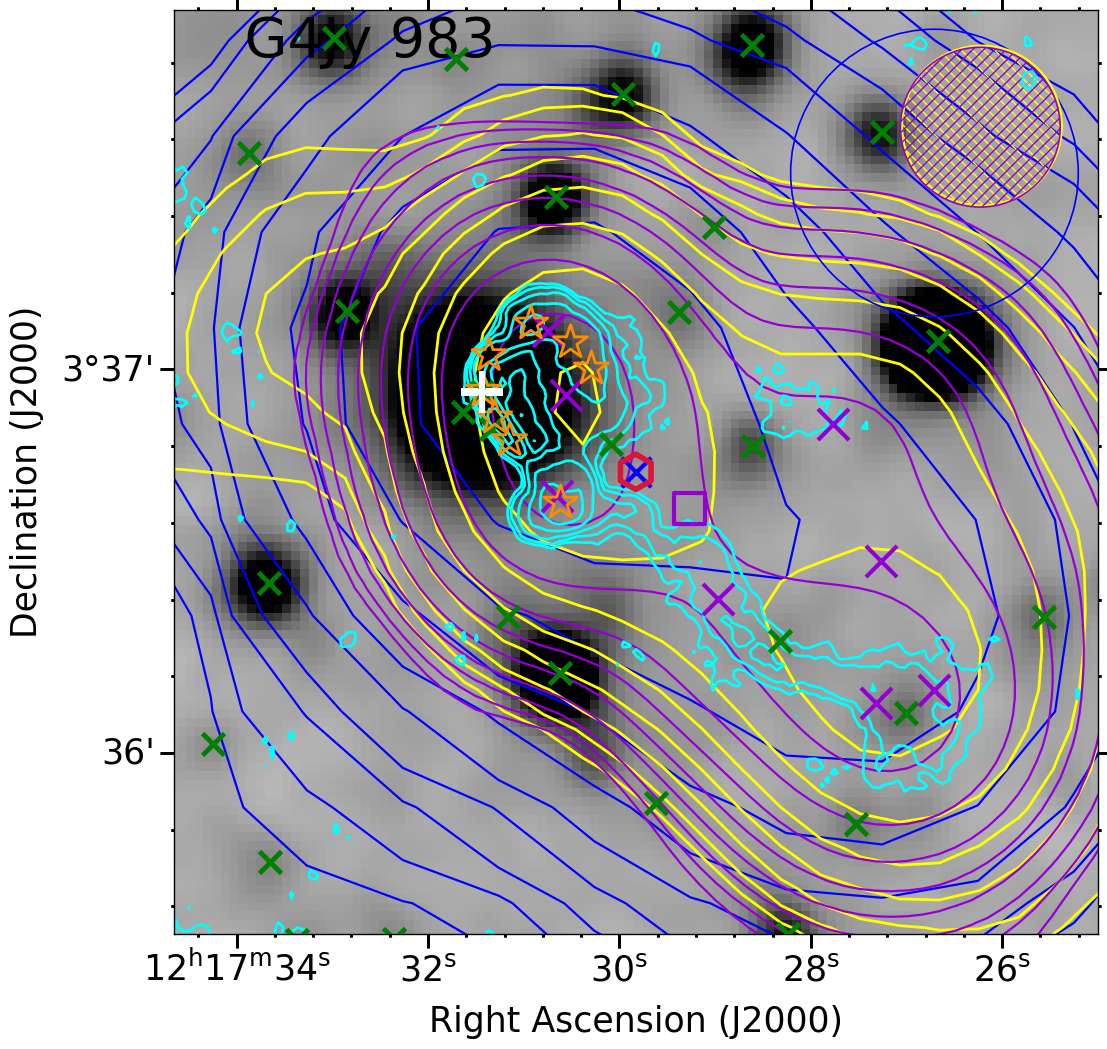} 
	} 
\subfigure[G4Jy~1325 ($H$-band image from UKIDSS)]{
	\includegraphics[width=0.425\linewidth]{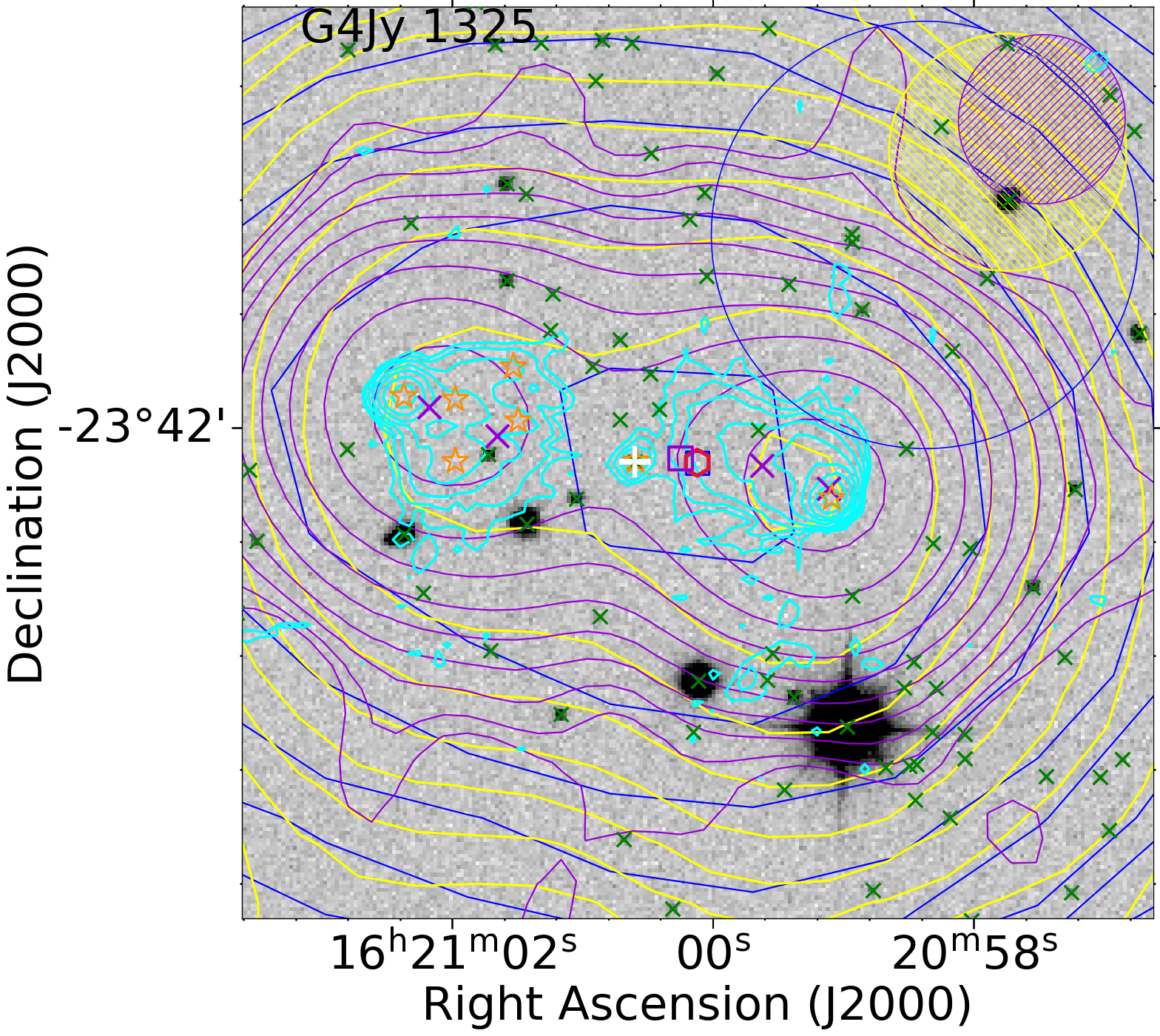} 
	}  \\[-0.05cm]
 \subfigure[G4Jy~1396]{
	\includegraphics[width=0.4\linewidth]{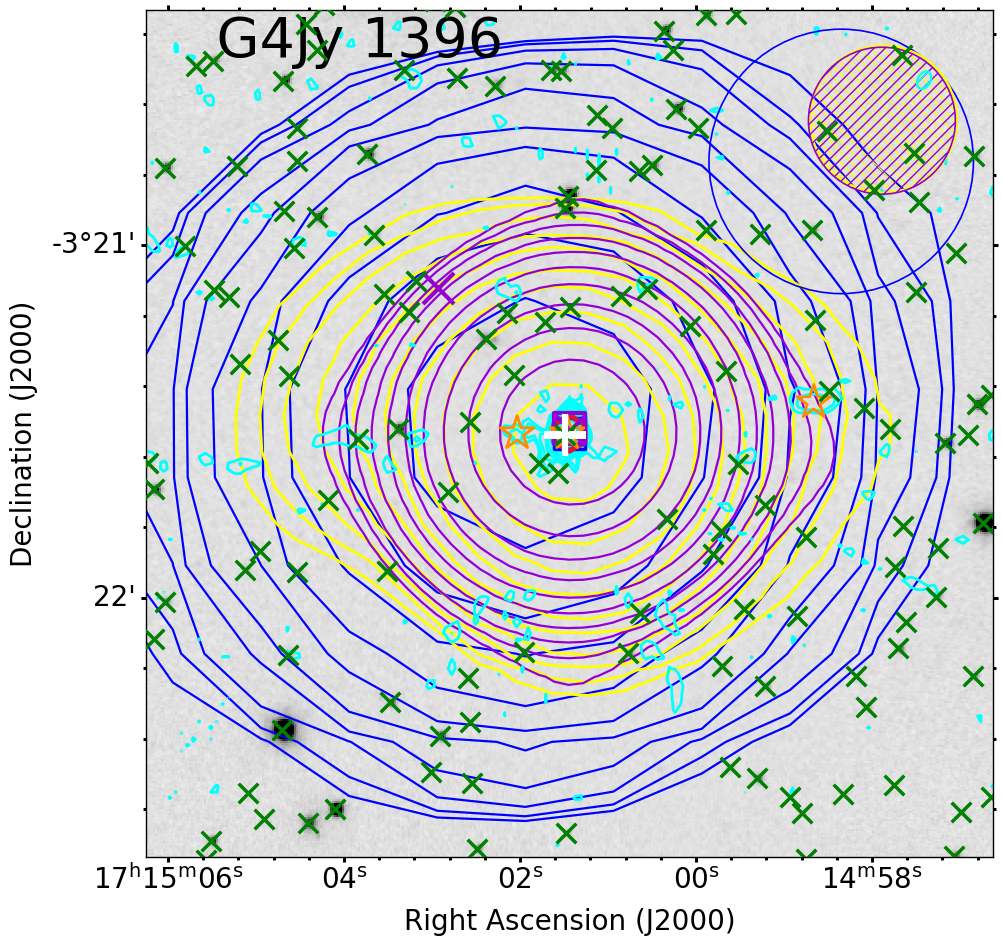} 
	} 
     \subfigure[G4Jy~1835 (MIR image from AllWISE)]{
	\includegraphics[width=0.41\linewidth]{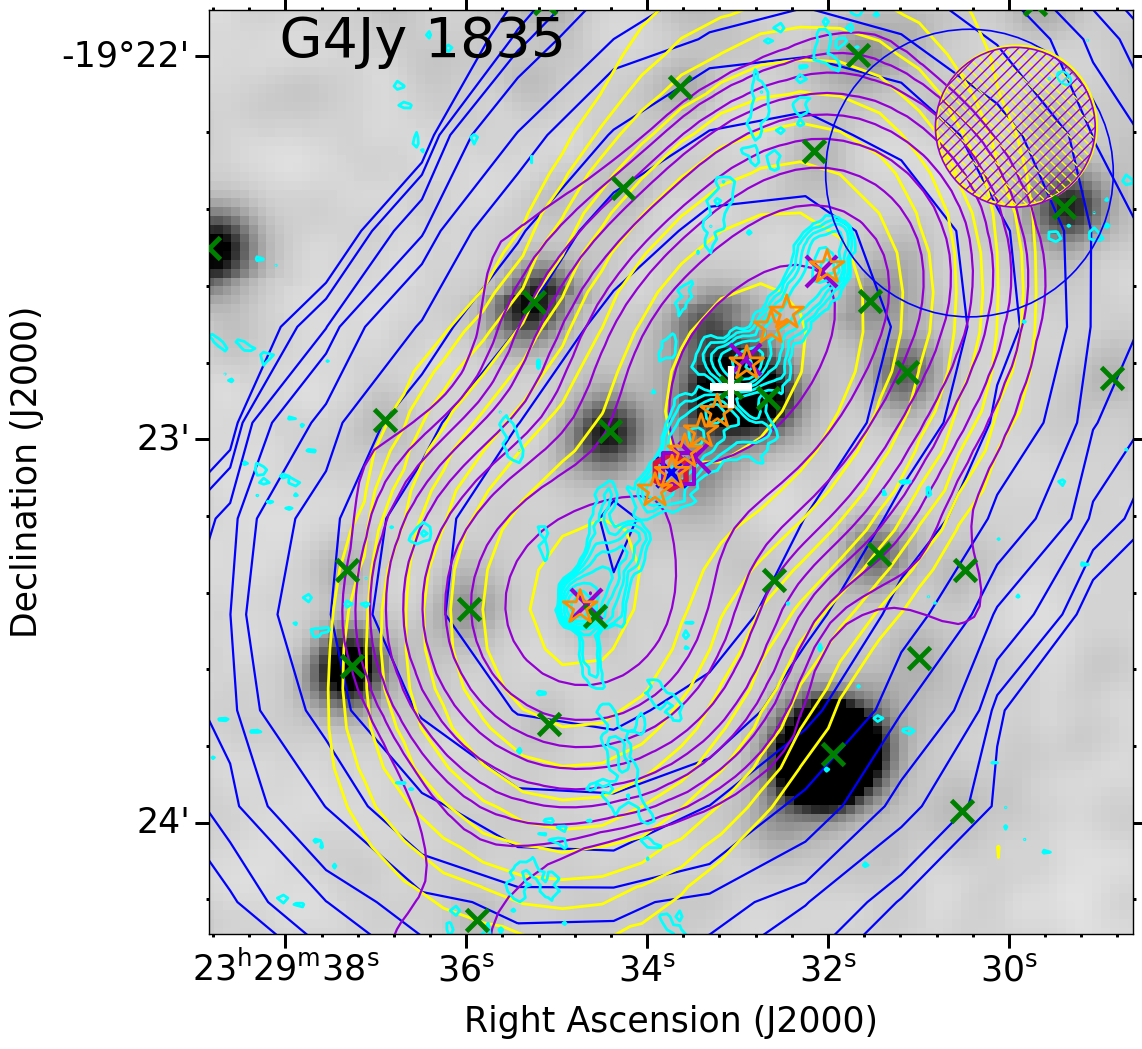} 
	} 

\caption{ Overlays of radio contours [and catalogue positions] from SUMSS/NVSS (blue), RACS-low1 (purple), and TGSS (yellow) on $K$-band images from VHS (unless otherwise stated). A red hexagon indicates the radio-brightness-weighted centroid calculated by \citet{White2020b}. Also plotted are cyan contours based on radio images from VLASS, with their peak-brightness positions indicated by orange stars. The relative synthesised-beam sizes are shown in the upper-right corner of each panel. Green crosses (`$\times$') indicate positions from the catalogue corresponding to the inverted grey-scale image, and a white `+' marks the position of the galaxy hosting the radio emission. \label{fig:maintext_Kband_IDs}}
\end{figure*}

\setkeys{Gin}{draft}

\subsection{A tentative identification}
\label{sec:tentative}

Note that the host-galaxy identification in this section is described in order to aid note-keeping, and will not appear in the G4Jy catalogue until we can confirm detection of the radio core. 

\begin{figure*}
\centering
\includegraphics[scale=0.4]{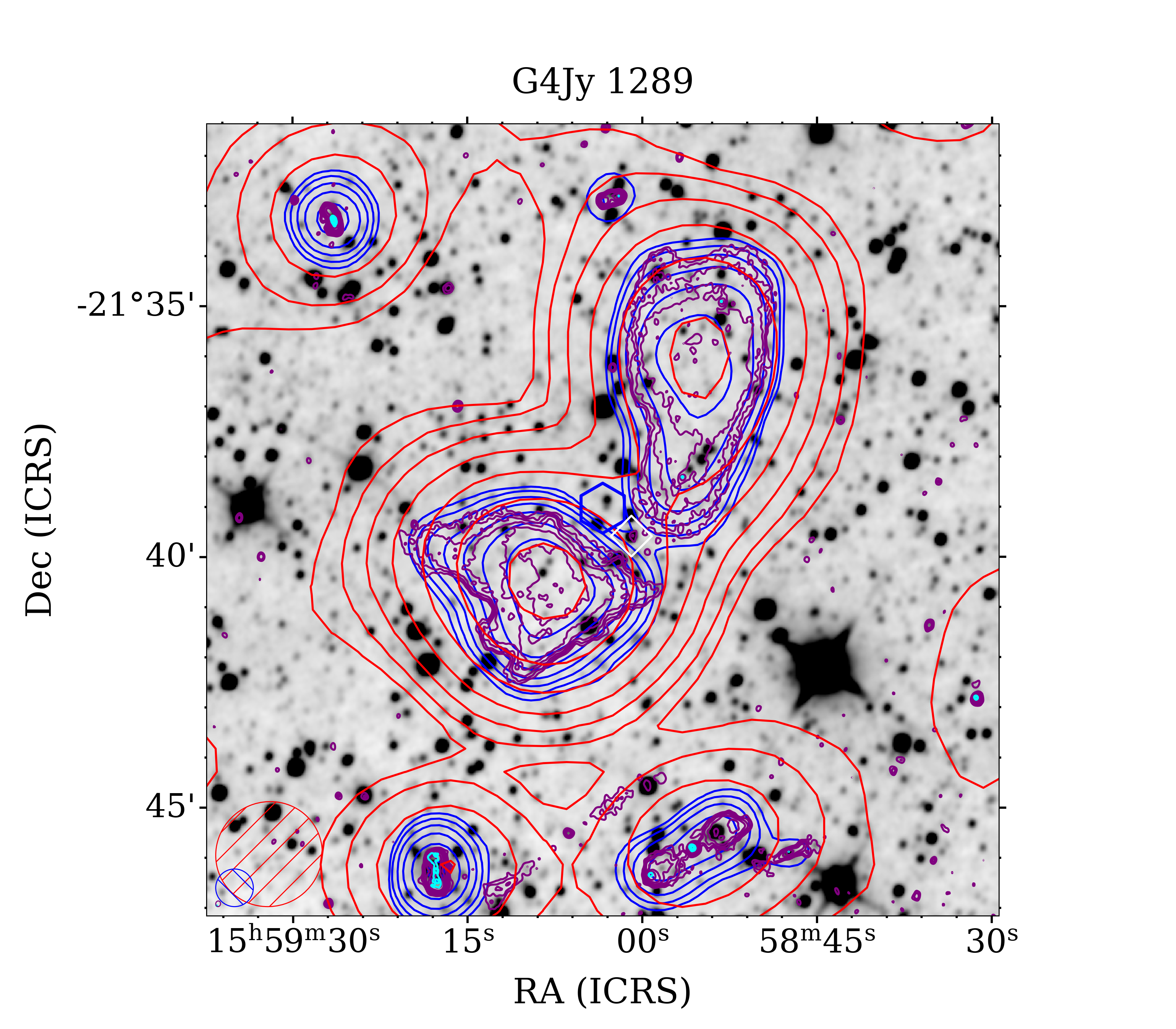}
\caption{ An overlay for G4Jy 1289, showing different sets of radio contours: GLEAM (200\,MHz) in red, NVSS (1400\,MHz) in blue, MeerKAT (1300\,MHz) in purple, and VLASS (3000\,MHz) in cyan. The respective beam-sizes of the different radio surveys are shown in the lower left-hand corner, where the scale of the image is such that VLASS contours of compact sources appear as filled cyan circles. The underlying, inverted-greyscale image is a W1-band image from AllWISE, with a blue hexagon marking the radio-centroid position (see \citealt{White2020b} for further details). VLASS detects two point sources in the western lobe, and we indicate our (tentative) host-galaxy identification (Section~\ref{sec:tentative}) with a white diamond (to avoid the marker covering the small contours).}
\label{fig:g4jy1289}
\end{figure*}

This is particularly necessary for {\bf G4Jy 1289}, where a faint 3-$\sigma$ detection appears between the extended lobes in the MeerKAT image (Figure~\ref{fig:g4jy1289}). The complication is that the only point-like source in the VLASS image (near the radio centroid) does not coincide with this `core' (marked by WISEA J155900.91$-$213935.5; white diamond in Figure~\ref{fig:g4jy1289}) but with compact radio-emission (UGCS J155856.54$-$213824.3) in the southern part of the western lobe (G4Jy 1289A; GLEAM J155855$-$213608). While \citet{Andernach2025} proposed UGCS J155856.54$-$213824.3 as the host galaxy, we believe that the latter is likely to be coincidence with an inner hotspot, as similar compact emission is seen (in both the MeerKAT and VLASS images) in the outer region of the western lobe [at R.A. (J2000) = 15:58:53.18, Dec. (J2000) = $-$21:34:55.2] -- i.e. likely an outer hotspot. (There is also the possibility that either or both `hotspots' are actually unrelated, compact radio-sources.) If we are correct in our AllWISE identification, then G4Jy~1289 remains with unknown linear-size because we cannot find a redshift for this source in published work. Meanwhile, \citet{Andernach2025} mention the possibility of the brighter galaxy 2MASX J15590200$-$2140032 being the host of G4Jy 1289, which they infer from the method described by \citet{Quici2025} who do not mention the host name explictly. If this identification is correct then the radio galaxy is 862\,kpc in linear size, as calculated from an angular size of 9.6\,arcmin (based on the MeerKAT image by \citealt{Sejake2023}) and a redshift of $z_{\mathrm{ph}} = 0.0792$ \citep{Bilicki2016}. Here, we have taken the median of the available photometric redshifts: $z_{\mathrm{ph}} = 0.0770$ \citep{Bilicki2014}, $z_{\mathrm{ph}} = 0.0792$ \citep{Bilicki2016}, and $z_{\mathrm{ph}} = 0.087$ \citep{Beck2021}. 
% "This [redshift] gives a scale of 1.496 kpc/"." 1.496 * 9.6 * 60 = 862 kpc

The reason for showing a large overlay in Figure~\ref{fig:g4jy1289} is also to draw attention to the extended radio-galaxies that are south of G4Jy 1289. The consistent orientation of the radio-jet axis for three of these adjacent sources is an example of what is referred to as `cosmic alignment' \citep{Taylor2016}\footnote{However, alignment in the ELAIS-N1 field has been refuted by \citet{Simonte2023} through 2D and 3D analyses of the radio-jet position-angles.}. This is thought to originate from the cosmic-web structure having an impact on the orientation of {\color{black}the black-hole spin axis, through angular momentum, and therefore the radio emission that is ejected parallel to this axis}. We hope that this prompts further investigation.

\setkeys{Gin}{draft=false}

\setkeys{Gin}{draft}

%[Other sources could be 'very good' identifications for sources that we want to confirm with MeerKAT. Heinz will likely contribute here.]

\subsection{G4Jy sources that remain unidentified}
\label{sec:unidentified}

\begin{figure*}
%\vspace{-0.7cm}
\centering

\subfigure[G4Jy~527 (VLASS contours on a VHS $K$-band image)]{
	\includegraphics[width=0.42\linewidth]{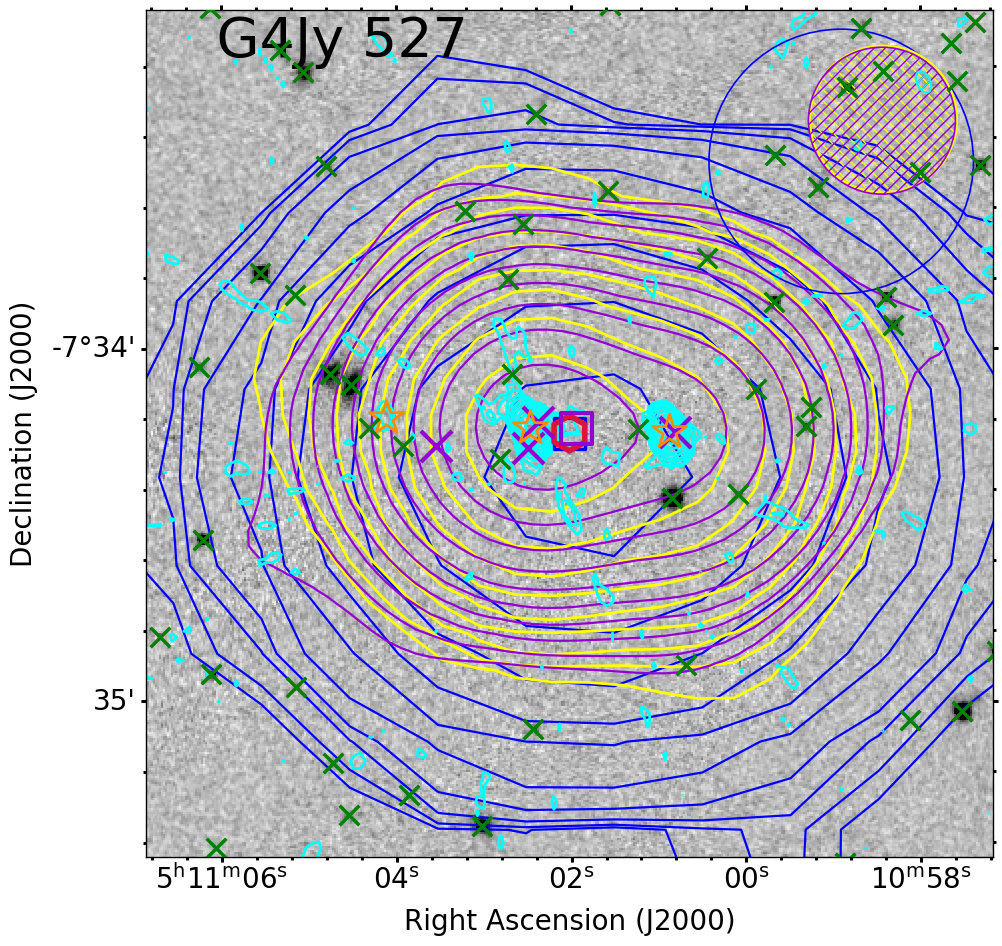}
	}
\subfigure[G4Jy~527 (MeerKAT contours on a PanSTARRS $r$-band image)]{
	\includegraphics[width=0.44\linewidth]{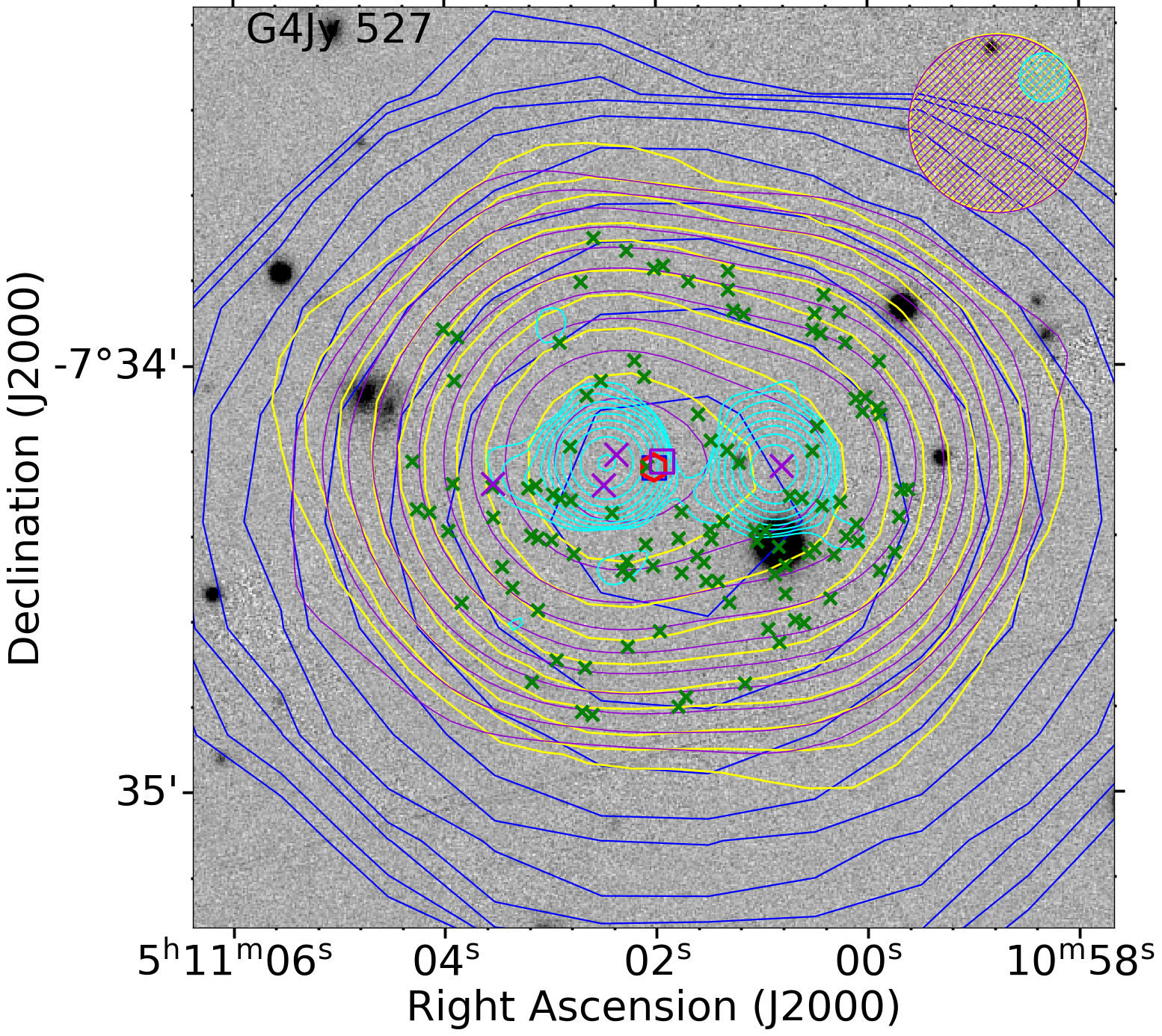} 
	} 

\caption{ Overlays of radio contours [and catalogue positions] from NVSS (blue), RACS-low1 (purple), and TGSS (yellow) on inverted greyscale images from (a) VHS and (b) PanSTARRS. A red hexagon indicates the radio-brightness-weighted centroid calculated by \citet{White2020b}. Also plotted, are cyan contours based on radio images from (a) VLASS and (b) MeerKAT-2019 follow-up \citep{Sejake2023}, with VLASS peak-brightness positions indicated by orange stars. The relative synthesised-beam sizes are shown in the upper-right corner of each panel, and green crosses (`$\times$') indicate positions from the catalogue corresponding to the inverted greyscale image. G4Jy 527 is an example of a source for which we are yet unable to provide a host-galaxy identification (Section~\ref{sec:unidentified}). \label{fig:compare_cyan_contours}}
\end{figure*}

In this sub-section we consider sources that remain unidentified. In these cases, the host-galaxy identification is not so much limited by the resolution or sensitivity of the radio data, but by the depth of the optical/NIR/MIR data. In Figure~\ref{fig:compare_cyan_contours} we show two overlays of G4Jy 527 (PKS 0508$-$076) -- one with VLASS contours (2.5-arcsec resolution) on a $K$-band image from VHS, and the other with MeerKAT contours (7-arcsec resolution) on an $r$-band image from PanSTARRS. The radio galaxy clearly has `double' morphology \citep{White2020b}, but the greater number of candidates in the optical image leads to ambiguity. Moreover, the MeerKAT contours appear to constrain where we would expect to see the host galaxy, but no optical (nor NIR) catalogue position coincides with that location.

\setkeys{Gin}{draft=false}

{\color{black}As further examples,} extended `double's (G4Jy 566, G4Jy 733, G4Jy 1362) and a compact `double' (G4Jy 1243) show no appropriate candidates between the two lobes, whilst {\color{black}two} pointlike radio-galaxies (G4Jy 676, G4Jy 1491) also have no candidate host-galaxy within the vicinity of the peak emission. In the case of G4Jy 1491, the RACS-low1 and TGSS contours lead us to question whether we are just seeing one of the radio lobes belonging to an extended radio-galaxy. However, inspection of the full field\footnote{\url{https://archive-gw-1.kat.ac.za/public/repository/10.48479/wyab-t838/data/Overlays/GLEAM_J183356-394023_SUMSS_G4Jy1491_10-arcmin_12_sigma_overlay.png}} shows that the extension of contours towards the south is the result of an unrelated radio galaxy (already noted by \citealt{Sejake2023}).

%[Saturation for G4Jy 1439? Are the green crosses actually all from AllWISE??]
% G4Jy 1439 is 3C 362, so should have an ID somewhere. Is that ID wrong then? No, can see the ID in an unWISE image. See notes on 29/3/2025

\setkeys{Gin}{draft}

As a reminder, the G4Jy catalogue \citep{White2020a,White2020b} contains a `host flag' column that indicates the following:
\begin{itemize}
    \item ‘i’ – a host galaxy has been identified in the AllWISE catalogue, with the position and mid-infrared magnitudes (W1, W2, W3, W4) recorded as part of the G4Jy catalogue (Section~\ref{sec:MIR_data}),
\item ‘u’ – it is unclear which AllWISE source is the most-likely host galaxy, due to the complexity of the radio morphology and/or the spatial distribution of mid-infrared sources (leading to ambiguity),
    \item ‘m’ – identification of the host galaxy is limited by the mid-infrared data, with the relevant source either being too faint to be detected in AllWISE, or affected by bright mid-infrared emission nearby,
\item ‘n’ – no AllWISE source should be specified, given the type of radio emission involved (e.g. cluster-relic radio emission).
\end{itemize}
For this work, we relax the meanings to refer to the host galaxy, irrespective of whether we are considering optical, NIR, or MIR (i.e. AllWISE) images. As such, the sources presented in Section~\ref{sec:new_IDs} are newly-assigned `host flag' = `i', and the sources {\color{black}mentioned} in this sub-section are labelled with `m'.

\subsection{Revised identifications}
\label{sec:revised_IDs}

Next, we revisit the radio-brightest galaxies in the G4Jy Sample, to ensure consistency with further follow-up being done for the G4Jy-3CRE (3CR-equivalent) subset \citep{Massaro2023,Massaro2023b,GarciaPerez2024} and by other groups \citep[e.g.][]{Maselli2022,Maselli2024}. We also note revised identifications for host galaxies where better radio-imaging shows that \citet{White2020a,White2020b} were previously mistaken.

\subsubsection{Cross-check with \citet{Massaro2023}}
\label{sec:massaro}

The G4Jy-3CRE subset \citep{Massaro2023} consists of 264 G4Jy sources that are equivalent in flux-density ($S_{\mathrm{178\,MHz}}> 9.0$\,Jy) to the first revised version of the Third Cambridge catalogue of radio sources  \citep[3CR survey;][]{Bennett1962, Spinrad1985} of the northern sky. To avoid overlap in sky coverage, a Declination cut of $< -$5$^{\circ}$ has been applied to the G4Jy catalogue, thus creating a sample of the radio-brightest sources in the southern hemisphere. To facilitate future follow-up, e.g. in the X-ray \citep[see][]{Massaro2023b}, it is imperative that the host galaxy is correctly identified.

Firstly, MeerKAT imaging \citep{Sejake2023} allowed us to specify the AllWISE positions for: G4Jy~113, G4Jy~530, G4Jy~611, G4Jy~672, G4Jy~680, G4Jy 1262, G4Jy 1365, G4Jy~1518, and G4Jy~1740, and we ensure that these positions are in agreement with the optical positions specified in the G4Jy-3CRE catalogue (version 9.1). We incorporate additional MIR and optical positions from this catalogue \citep{Massaro2023}, but disagree for sources listed in Table~\ref{tab:3cre_sources}, as discussed below.

%\begin{landscape}
\begin{table*}%[]
    \centering
    \begin{tabular}{c|c|c|c|c|c}
    \hline
    Source name & AllWISE host-galaxy name & Optical identification, J2000 & AllWISE host-galaxy name  & G4Jy & Brief comment
      \\ 
    & \citep{Massaro2023}  & \citep{Massaro2023} & (this work) & host-flag &  on columns 2 and 3
      \\ 
       \hline
 G4Jy 120 & --  &   -- & J010522.21$-$450517.2 & i & IDs are possible \\
  G4Jy 129 & --  &   -- & J011141.93$-$685937.5 & i &IDs are possible\\

G4Jy 162 & (artefact)  & 01:30:27.55$-$26:09:50.6  & -- & u  & Uncertainty over optical ID  \\
        G4Jy 350 & J032314.09$-$881605.2  & 03:23:10.63$-$88:16:06.3 & J032259.32$-$881600.4 & i & Both IDs are incorrect \\
        G4Jy 524 & J051032.83$-$183839.0  & 05:10:31.95$-$18:38:44.0 & J051031.94$-$183844.0 & i & Agree with the optical ID \\
          G4Jy 730 & (artefact)  &   08:53:00.28$-$20:47:26.7 & J085300.25$-$204730.7 & i & Agree with the optical ID \\

        G4Jy 837 &  J102003.93$-$425130.0  & 10:20:03.94$-$42:51:31.5 & J102003.93$-$425130.0 & i  & Agree with both IDs  \\
        G4Jy 939 & J114134.22$-$285048.0  & 11:41:34.25$-$28:50:48.0 & -- & m & IDs are not yet possible  \\
        G4Jy 1197 &J145235.55$-$131114.0  & 14:52:35.30$-$13:11:20.7 & J145235.30$-$131120.4 & i & Agree with the optical ID  \\

        G4Jy 1302 &  J160512.68$-$285912.1  & 16:05:13.06$-$28:59:14.6 & J160512.68$-$285912.1 & i  & Incorrect optical position \\
        G4Jy 1401 & J172011.01$-$070132.2  & 17:20:11.00$-$07:01:32.3 & -- & u & Uncertainty over both IDs   \\
  G4Jy 1455 & --  &   -- & J180117.98$-$662301.8 & i & IDs are possible \\

        G4Jy 1498 & (undetected -- faint)  & 18:37:41.47$-$43:35:41.4 & -- & u & Uncertainty over optical ID \\

                G4Jy 1590 &JJ195816.66$-$550934.9  & 19:58:16.68$-$55:09:35.5 & --  & u & Uncertainty over both IDs   \\

G4Jy 1843 & --  &   -- & J233510.30$-$663655.7 & i &IDs are possible \\
        G4Jy 1854 &J235050.37$-$245703.9  & 23:50:49.81$-$24:57:03.6 & J235049.80$-$245703.5 & i & Agree with the optical ID  \\
    \hline
    \end{tabular}
    \caption{This table summarises the difficulty of obtaining host-galaxy identifications (Section~\ref{sec:revised_IDs}), documenting the efforts by \citet{Massaro2023} and White et al. (2020a,b; this work) for the G4Jy-3CRE subset. The G4Jy `host flag' column refers to the latest version of the G4Jy catalogue (this work).} 
    \label{tab:3cre_sources}
\end{table*}
%\end{landscape}

For {\bf G4Jy 120}, we believe that PKS B0103$-$453 is a true association/cross-identification \citep[cf.][]{Massaro2023}. Whilst the radio core is not detected in the X-ray, at 0.5--10\,keV \citep{Massaro2023b}, a MeerKAT image \citep{Sejake2023}\footnote{\url{https://archive-gw-1.kat.ac.za/public/repository/10.48479/wyab-t838/data/Overlays/GLEAM_J010521-450527_SUMSS_G4Jy120_10-arcmin_12_sigma_overlay.png}} suggests that the host galaxy can be identified in AllWISE. A radio image of resolution better than $\sim$7\,arcsec would be needed to confirm or reject this identification.
% B&H Paper II -- MRC B0103−453: Because the radio structure is extended and does not show a core (Jones & McAdam 1992), a higher-resolution radio image will be necessary for a secure ID. Our candidate galaxy lies 17′′ from the MOST centroid.

Similarly, \citet{Massaro2023} provide neither an AllWISE identification nor an optical identification for {\bf G4Jy 129}. However, \citet{Sejake2023} again demonstrate that MeerKAT imaging helps with an AllWISE identification\footnote{\url{https://archive-gw-1.kat.ac.za/public/repository/10.48479/wyab-t838/data/Overlays/GLEAM_J011142-685956_SUMSS_G4Jy129_10-arcmin_24_sigma_overlay.png}}. This source is also known as PKS B0110$-$692 and has been notably difficult to cross-identify in the optical \citep{Jones1992,Burgess2006}.

For {\bf G4Jy 162}, we agree that an artefact in the mid-infrared image means that it is not possible to identify an AllWISE source as the host galaxy of the radio emission. Instead, \citet{Massaro2023} show that an appropriate candidate exists in an $r$-band image from PanSTARRS. They also note that, at $z=2.34665 \pm 0.00027$ \citep{Nesvadba2017}, this is ``the most distant source listed in the G4Jy-3CRE sample to date''. Also named PKS B0128$-$264, the VLASS contours are consistent with the double-lobe structure mapped by \citet{Kapahi1998} at 5\,GHz, and by \citet{Best1999} at 1.4\,GHz, with none detecting the radio core. Therefore, we assign a host flag of `u' to encourage re-observation. 

There are four AllWISE sources that lie along the radio-jet axis of {\bf G4Jy 350}, and so are viable candidates for hosting the radio emission. The uncertainty over the correct host is resolved by MeerKAT \citep{Sejake2023}, with the jets appearing to originate from WISEA J032259.32$-$881600.4\footnote{\url{https://archive-gw-1.kat.ac.za/public/repository/10.48479/wyab-t838/data/Overlays/GLEAM_J032318-881613_SUMSS_G4Jy350_10-arcmin_24_sigma_overlay.png}}. Therefore, both the AllWISE and optical positions in the G4Jy-3CRE catalogue \citep{Massaro2023} require updating.

An overlay of VLASS contours allowed \citet{Massaro2023} to correctly locate the optical identification for {\bf G4Jy 524} (PKS B0508$-$187), this being coincident with the radio-core position. However, the G4Jy-3CRE catalogue still reports the original AllWISE identification made by \citet{White2020a,White2020b}. This should be updated to J051031.94$-$183844.0, based upon the higher-resolution radio data from VLASS. 
% Apparently this source is in the 6dFGS catalogue: 6dF J0510323-183847, but no redshift is quoted in NED. Oh, looks too offset
% "G4Jy 524: is a radio galaxy (a.k.a. MRC 0508-187 and TXS 0508-187) with a lobe dominated radio structure. The optical image shows the presence of several nearby companion galaxies. G4Jy 524 is also listed in a sample of ultra steep spectrum radio sources (De Breuck et al. 2000)."

The MeerKAT follow-up by \citet{Sejake2023} also helps with identifying the host galaxy for {\bf G4Jy 730}. This AllWISE source (J085300.25$-$204730.7; see Table~\ref{tab:3cre_sources}) is detected despite the presence of artefacts connected to a bright nearby star. However, we also note the optical position quoted by \citet{Massaro2023}, which is very slightly further north than the AllWISE position and coincides with the brightness-weighted centroid calculated by \citet{White2020a,White2020b}. Also known as PKS B0850$-$206, G4Jy 730 was found by \citet{Best1999} to have a redshift of $z = 1.337\pm0.002$ and was classified as a {\it broad-line radio galaxy} (as opposed to a quasar).

\begin{figure*}
%\vspace{-0.7cm}
\centering
\subfigure[G4Jy~837 (original G4Jy overlay)]{
	\includegraphics[width=0.465\linewidth]{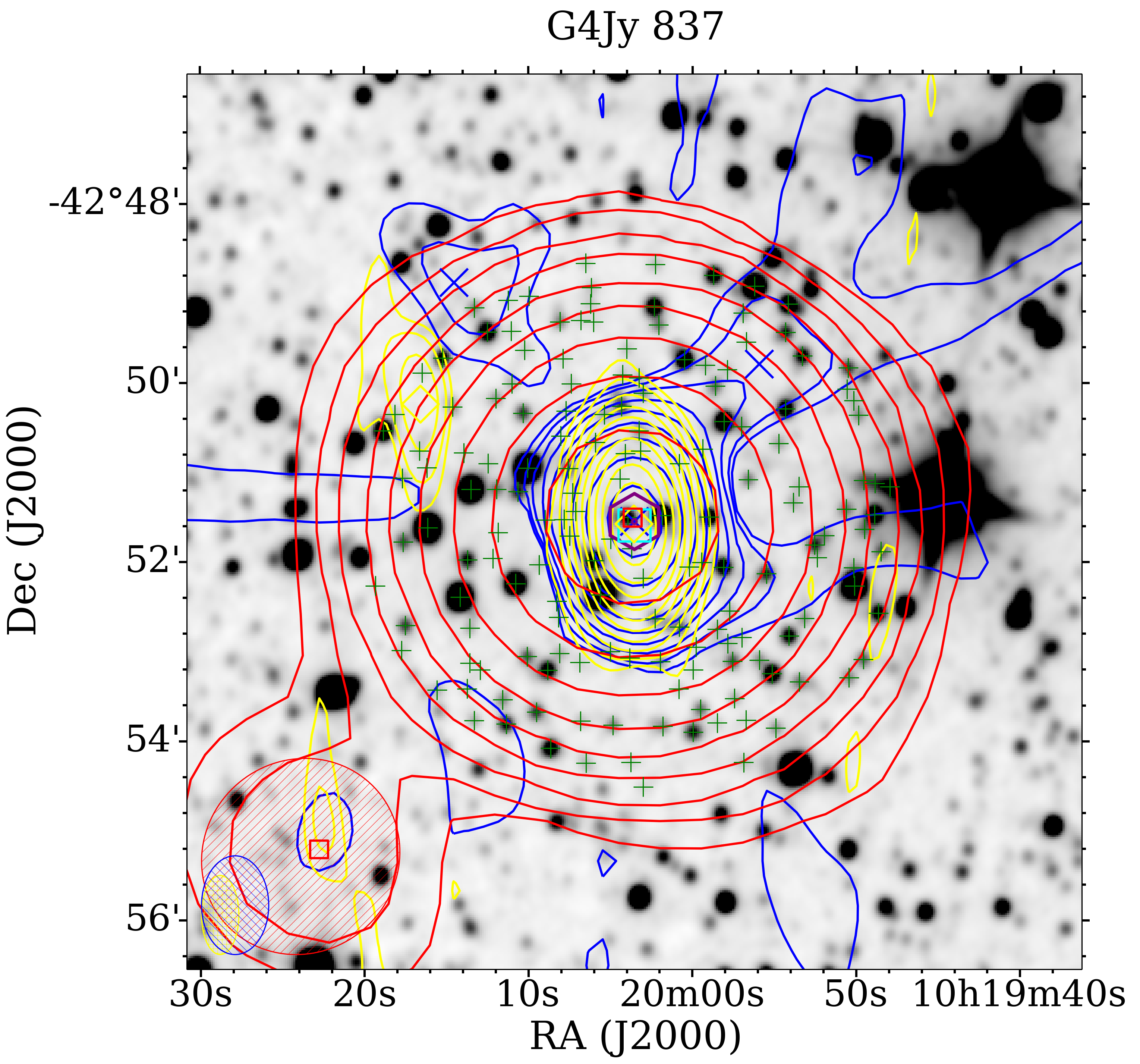}
	}
\subfigure[G4Jy~837 (RACS-low1 contours on a VHS $K$-band image)]{
	\includegraphics[width=0.44\linewidth]{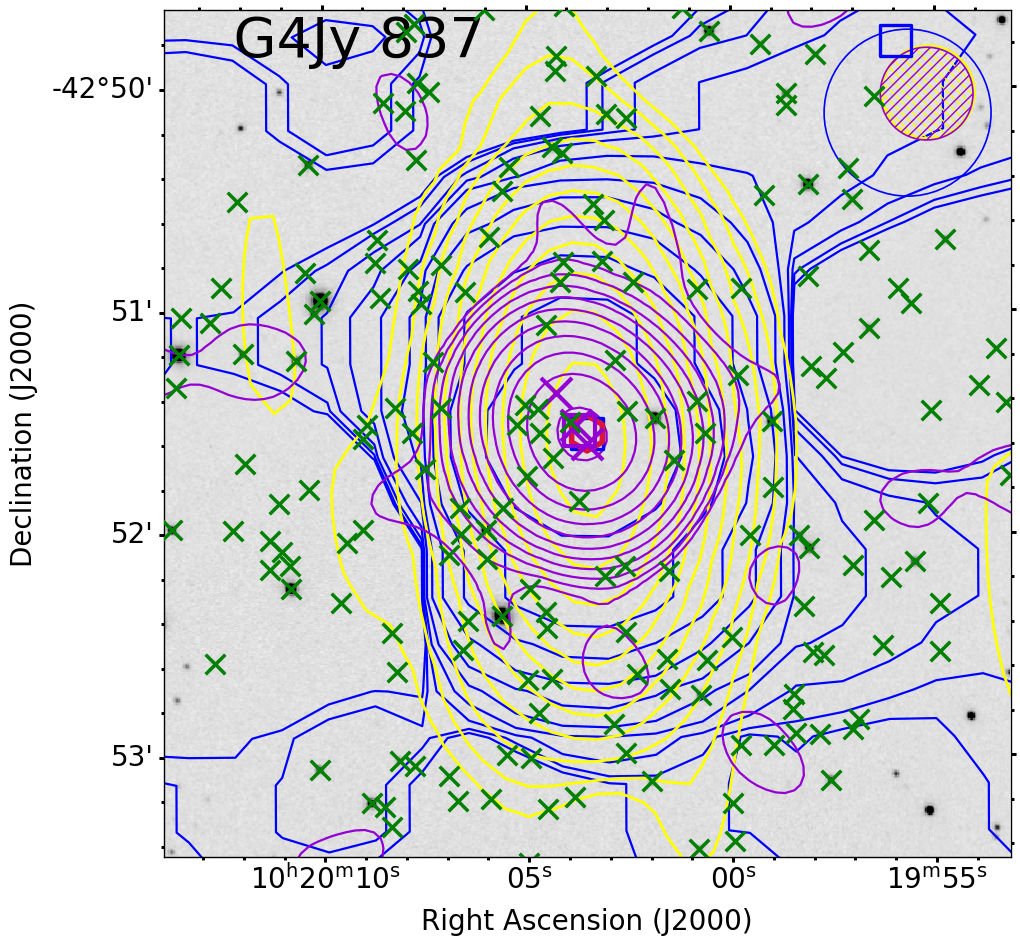} 
	} 
\caption{ Overlays of radio contours [and catalogue positions] from GLEAM (red), SUMSS (blue), TGSS (yellow), and RACS-low1 (purple) on inverted greyscale images from (a) AllWISE and (b) VHS. A hexagon indicates the radio-brightness-weighted centroid calculated by \citet{White2020b}. Also plotted is an AT20G position (cyan square). The relative synthesised-beam sizes are shown in the (a) lower-left and (b) upper-right corner of each panel, and green markers (`+'/`$\times$') indicate positions from the catalogue corresponding to the inverted greyscale image. See discussion in the text regarding the host (Section~\ref{sec:massaro}). \label{fig:possible_triple_in_RACS}}
\end{figure*}

On the assumption that {\bf G4Jy 837} is a compact radio-source, \citet{White2020a,White2020b} interpreted the AT20G position as a marker of the radio core (Figure~\ref{fig:possible_triple_in_RACS}a). As such, with no coincident source in AllWISE, it was believed that the mid-infrared data was not deep enough for host-galaxy identification, and so G4Jy 837 was assigned a host flag of `m'. Careful inspection of the new RACS-low1 overlay (Figure~\ref{fig:possible_triple_in_RACS}b) reveals three RACS-catalogue positions, which may be indicative of `triple' radio morphology. This is confirmed by a 5-GHz map \citep{Punsly2006}, which means that the AllWISE source slightly northeast of the centroid is the correct identification for G4Jy 837 (PKS B1018$-$426), as proposed by \citet{Massaro2023}. With this interpretation, the AT20G position must then mark a hotspot within the southern radio-lobe, and we update the G4Jy host-flag to `i'.

% "G4Jy837: is an extremely powerful (i.e., with a luminosity higher than all 3C radio sources at the same redshift with the exception of 3C196) FRII radio galaxy (Punsly & Tingay 2006) at z=1.28 (Murdoch et al. 1984; Stickel et al. 1993; Decarli et al. 2010) with mid-IR colors of γ-ray blazars (D’Abrusco et al. 2019)."

\setkeys{Gin}{draft=false}

\setkeys{Gin}{draft}

For {\bf G4Jy 939} (PKS B1139$-$285), we have determined the host flag to be `m', meaning that an identification in AllWISE is not possible given the depth of this dataset. Whilst a candidate host-galaxy (WISEA J114134.22$-$285048.0) is visible in the overlay produced by \citet{Sejake2023}\footnote{\url{https://archive-gw-1.kat.ac.za/public/repository/10.48479/wyab-t838/data/Overlays/GLEAM_J114134-285050_NVSS_G4Jy939_10-arcmin_24_sigma_overlay.png}}, closer inspection shows that it is not at the expected location (based upon the MeerKAT contours in purple). Therefore, we err on the side of caution and propose this source for additional follow-up in the radio and the near-infrared. This is justified by the radio structure mapped at 5\,GHz by \citet{Kapahi1998}, where two core-like features are seen southeast of the existing identification in the literature.
%%% Technically, Katlego should have looked for this in a K-band or PanSTARRS image already...

\citet{White2020a,White2020b} incorrectly identified J145235.55$-$131114.0 as the AllWISE host-galaxy of {\bf G4Jy 1197}. By inspecting VLASS contours on an $r$-band image, \citet{Massaro2023} show that the correct candidate is the adjacent source that coincides with J145235.30$-$131120.4 (marked by the radio core). We therefore update our identification for G4Jy 1197, which is also known as PKS B1449$-$129.

As shown in another 3-GHz-on-AllWISE overlay \citep[][figure 23]{Massaro2023}, the putative radio-core of {\bf G4Jy 1302} (MRC B1602$-$288) is significantly offset from the mid-infrared source that was originally identified for the G4Jy catalogue. Therefore, we record the optical identification that coincides with the `core', but seek further radio follow-up for confirmation. This is because \citet{Best1999} state that ``although the brighter object is coincident with what has the appearance of a radio core, it is the fainter object which is identified as the host radio galaxy''. Indeed, looking at their figure 31, we find confirmation that the radio contours are misleading (with respect to the core) and that the host galaxy is WISEA J160512.68$-$285912.1, as originally suggested by \citet{Best1999} and \citet{White2020a,White2020b}. The redshift remains $z = 0.482 \pm 0.001$, as obtained by \citet{Best1999} and quoted by \citet{Massaro2023}.

For {\bf G4Jy 1401} (PKS B1717$-$069), \citet{White2020a,White2020b} had noted that there are two equally-valid AllWISE sources for the host-galaxy identification, and therefore assigned this radio galaxy a host flag of `u'. We deem that the morphology in VLASS \citep{Massaro2023} is not sufficient to resolve the ambiguity, and so we retain the `u' host flag for the current version of the G4Jy catalogue.  

{\bf G4Jy 1455} shows very compact radio-emission in the MeerKAT contours\footnote{{\url{https://archive-gw-1.kat.ac.za/public/repository/10.48479/wyab-t838/data/Overlays/GLEAM_J180118-662302_SUMSS_G4Jy1455_10-arcmin_96_sigma_overlay.png}}}, with SUMSS contours in the overlay by \citet{Sejake2023} suggestive of either imaging artefacts or genuine emission that may be discernible in a lower-frequency image. The former makes cross-identification in AllWISE very straightforward, and we can confirm that this source also goes by the name of PKS B1756$-$663.

Also known as MRC B1834$-$436, {\bf G4Jy 1498} has no MIR source identified in AllWISE, but two equally-good candidates {\it are} visible in the W1-band image. The X-ray band (0.5--10\,keV) could have helped to resolve this ambiguity, but unfortunately the radio core is not detected \citep{Massaro2023b}. As such, we label the host as `u' and propose follow-up with a better-resolution radio image than TGSS (25\,arcsec) rather than adopting the optical identification of \citet{Massaro2023}.  

 {\bf G4Jy 1590} (PKS B1954$-$552) does not have sufficient resolution in the radio images to allow the correct host galaxy to be identified, in neither AllWISE nor VHS. The most-likely candidates are the two VHS sources closest to the centroid position, meaning that the AT20G position marks {\it either} the radio core or a northern hotspot close to the AGN. The latter is consistent with the image provided by \citet{Duncan1992}, for which \citet{Morganti1993} comments that ``the northern component [is] slightly extended and brighter than the southern component''.

\setkeys{Gin}{draft}

The identification of {\bf G4Jy 1843} was discussed by \citet{Sejake2023}, with the overlay supporting this AllWISE position (J233510.30$-$663655.7) available online\footnote{\url{https://archive-gw-1.kat.ac.za/public/repository/10.48479/wyab-t838/data/Overlays/GLEAM_J233511-663702_SUMSS_G4Jy1843_10-arcmin_48_sigma_overlay.png}}. However, we do accept the possibility that there may be a star \citep{Burgess2006} coinciding with the position of the radio-galaxy host. Another name for this radio source is MRC B2332$-$668, with `double' radio morphology previously being mapped by \citet{Burgess2006} using the Australia Telescope Compact Array.

For {\bf G4Jy 1854} (PKS B2348$-$252), overlays of VLASS contours on AllWISE images confirmed that the host-galaxy identification should be revised from J235050.37$-$245703.9 \citep{White2020a,White2020b} to J235049.80$-$245703.5 (this work). In addition, the new position is coincident with the optical identification proposed by \citet{Massaro2023}, and consistent with the radio core detected in 5-GHz VLA maps by \citet{Reid1999}.

Given the difficulty of host-galaxy identification (hindered by chance coincidences!), we expect the G4Jy and G4Jy-3CRE catalogues to continue to iterate between them, and so lead to more-refined and robust host-galaxy identifications for the G4Jy Sample. 

\subsubsection{ Cross-check with \citet{Maselli2022,Maselli2024}}
\label{sec:maselli}

We also thank \citet[][]{Maselli2022,Maselli2024} for paying close attention to host-galaxy identification as part of their X-ray investigations into MS4 sources \citep{Burgess2006a,Burgess2006}, which overlap with the G4Jy Sample by definition. For their Neil Gehrels {\it Swift} Observatory campaigns, they focussed on the brightest G4Jy sources (at 181\,MHz) that did not already have coverage with {\it Chandra} or XMM-{\it Newton}.

For {\bf G4Jy 446} (MRC B0420$-$625), \citet{Maselli2022} state that ``there are two viable counterparts. The given source IDs are those which include, as [the] optical counterpart, the one suggested in \citet{Burgess2006}''. For the current work we have independently identified a $K$-band counterpart (R.A. = 04:20:56.09, Dec. = $-$62:23:38.9; Figure~\ref{fig:Kband_IDs}) that is coincident with both the optical identification (J042056.09$-$622339.1) and the mid-infrared identification (J042056.03$-$622339.8) found by \citet{Maselli2022}. We note that they identified the latter in the CatWISE2020 catalogue by \citet{Marocco2021}, and we adopt it for our own catalogue. 

\citet{Maselli2022} note that there are also two viable counterparts for {\bf G4Jy 563} (B0534$-$497), these being the AllWISE sources J053613.90$-$494422.2 and J053613.56$-$494426.8. The former was chosen for the G4Jy catalogue because the alternative coincides with an optical position (R.A. = 05:36:13.61, Dec. = $-$49:44:26.7) -- identified by \citet{Burgess2006} --  that is significantly offset from the radio-jet axis that is evident in their ATCA image. Our choice is further supported by the pointlike detection in the X-ray found by \citet{Maselli2022} and \citet{Massaro2023b}, which helps the host galaxy of G4Jy 563 to be differentiated. 

%We therefore disagree with the alternative mid-infrared host (J053613.56$-$494426.8) that is suggested by \citet{Maselli2022}, and retain the AllWISE source (J053613.90$-$494422.2) that was originally chosen for the G4Jy catalogue.

% G4Jy563: is optically identified with a galaxy (Hunstead et al. 1971a; Bolton & Savage 1977) being also listed in the MS4 sample (Burgess & Hunstead 2006a), with a photometric redshift estimate of z=0.184 (Burgess & Hunstead 2006b). 
% BH2006b quote 05 36 13.61 −49 44 26.7 as the optical host, but this is offset from the radio-jet axis in the ATCA image!

%pointlike detection in the X-ray (R.A. = 05:36:13.9, Dec. = $-$49:44:20.3) <-- From Massaro's paper

Like for G4Jy 446, we have a $K$-band identification (R.A. = 09:06:52.50, Dec. = $-$68:29:40.4; Figure~\ref{fig:Kband_IDs}) for
{\bf G4Jy 752} (MRC B0906$-$682) that matches the identifications provided by \citet{Maselli2022}. These are J090652.64$-$682940.3 in the optical \citep{Burgess2006}, and CWISE J090652.71$-$682940.2 in the mid-infrared. As before, we incorporate the CatWISE2020 identification into the G4Jy catalogue.

The radio core that is seen for {\bf G4Jy 854}  (MRC B1030$-$340) in VLASS contours \citep{Massaro2023} does not coincide with any mid-infrared source in the AllWISE catalogue. Instead, \citet{Maselli2022} consult CatWISE2020 data \citep{Marocco2021} and suggest that the host galaxy is J103313.22$-$341844.9. We agree with this association, which is supported by visually inspecting the relative positions of AllWISE and CatWISE2020 sources.
% "G4Jy854: is listed in the MS4 catalog as a radio galaxy with a photometric redshift estimate z=0.5 (Burgess & Hunstead 2006b). It has an ultra steep radio spectrum (De Breuck et al. 2000). Both the mid-IR and the optical counterparts are not detected according to our analysis."
% GLEAM_J103312-341842_G4Jy_854_NVSS_RACS_DR1_1012-37A_VLASS_0_VHS_v1.png

{\bf G4Jy 1192} (B1445$-$468) {\color{black}has} many candidate host-galaxies visible in the VHS image. We believe that a higher-resolution radio-image is required to address the ambiguity in cross-identification, and confirm whether or not the AllWISE source suggested by \citet{Maselli2022}, J144828.18$-$470141.6, is the correct counterpart. To reflect this, we update the G4Jy host-flag from `m' to `u'. 
% GLEAM_J144828-470136_G4Jy_1192_SUMSS_RACS_DR1_1430-43A_VHS_v1.png

\setkeys{Gin}{draft=false}

\begin{figure}
%\vspace{-0.7cm}
\centering
	\includegraphics[width=1.0\linewidth]{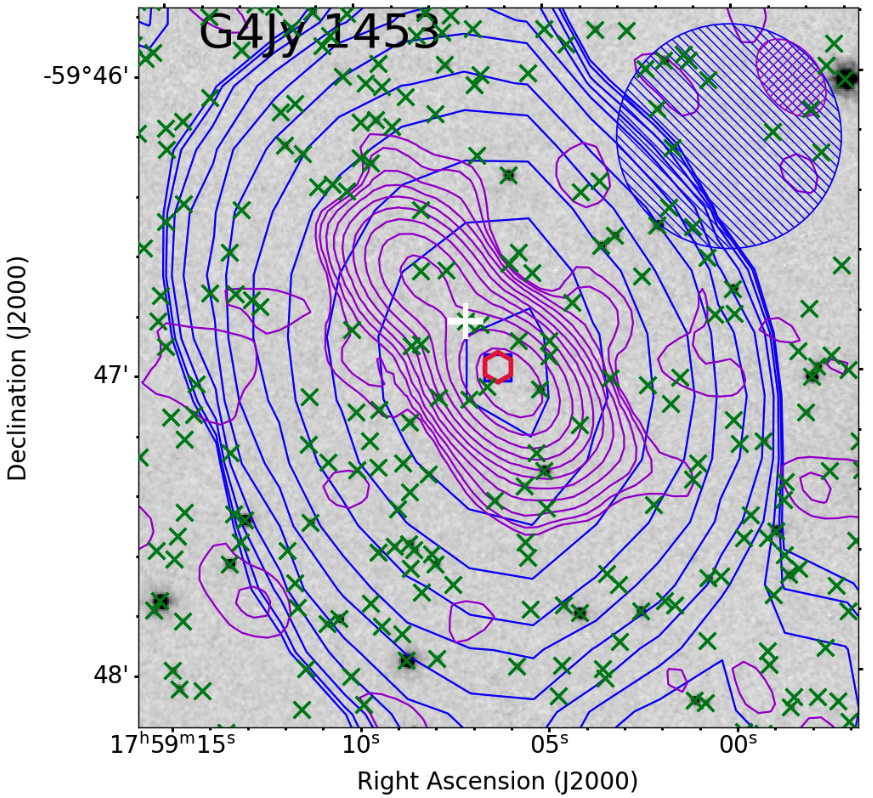} 
	
\caption{ Overlays of radio contours from SUMSS (blue) and RACS-low1 (purple) on {\color{black}an inverted greyscale image} from VHS. A red hexagon indicates the radio-brightness-weighted centroid calculated by \citet{White2020b}. The relative synthesised-beam sizes are shown in the upper-right corner of {\color{black}the} panel, and green markers (`$\times$') indicate positions from the VHS catalogue. {\color{black}The} host of G4Jy 1453 is marked with a white `+' (Section~\ref{sec:maselli}). \label{fig:uncertainty_re_m_galaxies}}
\end{figure}

\setkeys{Gin}{draft}

For {\bf G4Jy 1453} (B1754$-$597), \citet{Maselli2024} suggest that there are two viable counterparts, and favour J175907.08$-$594649.7 over J175905.73$-$594636.5. This is based upon a pointlike detection in the first scan of eROSITA data \citep{Merloni2024}\footnote{Further {\it Swift} observations have since been reduced, showing improved positional uncertainty for this G4Jy source, and further agreement with the selected AllWISE counterpart (A. Maselli, private communication).} and infrared colours in AllWISE. RACS-low1 contours show that this is nicely aligned with the radio-jet axis (Figure~\ref{fig:uncertainty_re_m_galaxies}), and so we accept this revised identification for the G4Jy catalogue.
% GLEAM_J175906-594655_G4Jy_1453_SUMSS_SB8676_cont_RACS_1812-62A_VHS_v1.png
% "G4Jy 1453: is a radio galaxy listed in the MS4 sample (a.k.a. PKS 1754-59 and AT20G J175906-594702) at the photometric redshift z=0.8 (Burgess & Hunstead 2006b) and optically identified in the literature (Hunstead et al. 1971b). The lack of a high resolution radio map prevented us confirming the host galaxy association."
% When I search VHSDR5 the best (0.19-arcsec) match is 17:59:07.09, -59:46:49.6. Yep, this is almost an exact match with the first infrared location.

As for 3 sources above, we are already in agreement regarding the host-galaxy position for the remaining 24 G4Jy sources identified in the mid-infrared by \citet{Maselli2022,Maselli2024}. Particularly satisfying is that this includes G4Jy~120, G4Jy 672, G4Jy 837, and G4Jy 1262, for which both groups (since publication of the original G4Jy catalogue) independently arrived at the same mid-infrared identification (see Section~\ref{sec:massaro} for these sources).

\subsubsection{Cases of mistaken identities}

VLASS shows that the radio core of {\bf G4Jy 175} coincides with our radio centroid, which is offset from our host galaxy. We therefore refine the position from WISEA J013716.14+092950.0 to SDSS J013716.22+092953.3 (at the position of the centroid).

For {\bf G4Jy 677}, DESI imaging \citep{Dey2019} shows that the identification made by \citet{Sejake2023} is actually a bright star. We therefore update the host-galaxy position to the adjacent optical source (PSO J118.9394+02.1757; J075545.46+021032.5), whose location is still consistent with the radio morphology shown by MeerKAT \citep{Sejake2023}.

The radio core of {\bf G4Jy 880} is detected in VLASS, at the position of SMSS J105516.80$-$272951.7 in the SkyMapper Southern Survey (SMSS DR4; \citealt{Onken2024}). This shows that the `double' radio-morphology is more asymmetric than first appreciated, and we update the host-galaxy position from WISEA J105518.07$-$272956.2.

In the case of {\bf G4Jy 964}, VLASS shows that the radio core is closer to the G4Jy centroid than our chosen position. As there is no detection in AllWISE at the location of the core, we instead update the host-galaxy identification to SDSS J115412.89+291608.4. 

VLASS is also helpful in showing that the radio morphology of {\bf G4Jy 1414} is a `double-double' (i.e. having two sets of `double' radio-brightness, likely due to intermittent AGN activity). Our original host-galaxy identification (WISEA J173034.28$-$042250.7) is almost coincident with one of the inner hotspots, and so (with the core detected in VLASS, at J173033.94$-$042242.8) we change the host-galaxy identification to PSO J262.6414$-$04.3785.

For {\bf G4Jy 1430}, \citet{White2020a, White2020b} quote the host galaxy as WISEA J174139.53+172031.5. Whilst this is nicely positioned between the two radio lobes, there are other host-galaxy candidates in the vicinity. Upon viewing the source in VLASS, the radio core appears to coincide with WISEA J174138.25+172017.4, and so we update the catalogue accordingly. 

VLASS again helps with correcting the host-galaxy identification for {\bf G4Jy 1586}. The correct host-galaxy (WISEA J195516.16$-$192535.1) is further from the centroid than the original selection (WISEA J195515.36$-$192545.4). The VLASS image also shows interesting `spatial curvature' of the extended radio emission, particularly for the southwestern lobe.

We had quoted WISEA J231956.26$-$272807.4 as the host galaxy of {\bf G4Jy 1819}, but this is a blend of two optical objects. The true host galaxy (confirmed via VLASS) is $\sim4$\,arcsec south, and so we update the catalogue position to 2MASX J23195628$-$2728117. 

In addition, we welcome revisions from the astronomical community, raised as `issues' on the GitHub repository\footnote{\url{https://github.com/svw26/G4Jy/issues}}, to ensure the continued high-quality of the G4Jy catalogue and studies of the G4Jy Sample.

\section{Towards a multiwavelength catalogue}
\label{sec:catalogue}

The original G4Jy catalogue provided detailed information for the low-frequency radio emission of all 1,863 G4Jy sources, as well  as mid-infrared data through 1,606 host-galaxy identifications in AllWISE (which itself is accompanied by near-infrared data from 2MASS). Thanks to supplementary work on identifying the host galaxies of the radio emission (described in the previous section), we now have a total of 1,733 identifications for collating additional multiwavelength data (this section) and conducting further analysis (Section~\ref{sec:results}). The work towards a multiwavelength G4Jy catalogue is completed in the accompanying paper, Paper IV \citep{White2025d}.
% 1863 - 128 - 2'n' = 1733

\subsection{Searching for photometric redshifts}
\label{sec:xmatch_photoz}

A 1-arcsec crossmatching radius is used for gathering photometric redshifts from the WISExSuperCOSMOS catalogue \citep{Bilicki2016} and the DESI Legacy Surveys DR8 photometric-redshift catalogue \citep{Duncan2022} for the southern sky. These catalogues are chosen predominantly for their large sky-coverage, noting that the average sky-density of G4Jy sources is one source per 13 deg$^2$ \citep{White2020b}. 

\citet{Bilicki2016} create their parent catalogue of $\sim20$ million sources by crossmatching two all-sky surveys, AllWISE \citep{Cutri2013} and SuperCOSMOS \citep{Hambly2001}. This is completed with a mean matching radius of $0.54 \pm 0.42$\,arcsec, noting that 2\,arcsec is the typical resolution of the SuperCOSMOS data. They then employ an artificial neural-network (ANN) algorithm, ANN$z$ \citep{Collister2004}, to determine photometric redshifts, having trained the network on Galaxy and Mass Assembly (GAMA) II \citep{Driver2011,Liske2015} spectroscopic-data. The result is a catalogue that reaches out to $z>0.4$, with a median redshift of 0.2. The normalised scatter is $\sigma_{z}=0.033$, and the outlier rate is less than 3 per cent of redshifts being beyond $3\sigma_{z}$. Due to Galactic extinction at low latitudes, the final catalogue actually covers about 70 per cent of the entire sky, and we retain their $E(B-V)$ estimate for the G4Jy catalogue. We also note that morphological cuts during pre-selection of {\it resolved} SuperCOSMOS sources largely eliminates quasars from the catalogue.

The latter is in contrast with the catalogue produced by \citet{Duncan2022}, which is able to provide accurate redshift estimates for optically-selected quasars out to at least $z = 6.0$. Based on imaging from the DESI Legacy (Imaging) Surveys DR8 \citep{Dey2019}, they catalogue $\sim900$ million sources over $19,400$\,deg$^2$, where the photometric redshift is again estimated through a machine-learning method -- this time, a combination of sparse Gaussian processes \citep{Almosallam2016a,Almosallam2016b} and Gaussian Mixture Models (GMMs). Doing so allows the data to drive the identification of regions of parameter space for individual/separate training, and so can accommodate e.g. radio-continuum-selected samples (noting that most photometric-redshift fitting is biased towards optically-selected samples). To this end, the spectroscopic training-data are from SDSS DR14 \citep{Abolfathi2018}, the SDSS DR14 Quasar catalogue \citep[DR14Q;][]{Paris2018}, the spectroscopic-redshift catalogue compiled for the {\it Herschel} Extragalactic Legacy Project \citep[HELP;][]{Shirley2021}, and the Very High-$z$ Quasar catalogue \citep[VHzQ;][]{Ross2020}. As a result, the typical scatter is $\sigma_{z}=$\,0.02--0.10, and the outlier fraction\footnote{Photometric-redshift outliers follow the common definition of satisfying |$\delta z$| / (1 + $z_{\mathrm{sp}}) > 0.15$, where $\delta z = z_{\mathrm{ph}} - z_{\mathrm{sp}}$. \label{outlier_footnote}} is $<10$ per cent for (resolved populations at) $z < 1.0 $. At the high-redshift end ($z>4.5$), the scatter is $\sigma_{z}=0.05$ and the outlier fraction is 2.5 per cent. The redshifts are also accompanied by a flag that indicates photometry that is not affected by blending or imaging artefacts (`flag\_clean'), and another that indicates the reliability of the photometric redshift (`flag\_qual') -- see table 3 of \citet{Duncan2022}. 

Whilst 110 WISExSuperCOSMOS redshifts and 768 LS-DR8 redshifts are in the G4Jy catalogue, they are superseded wherever a spectroscopic redshift is available for a particular G4Jy source. Hence, the different numbers in the `Number of $z_{\mathrm{ph}}$' column of Table~\ref{tab:z_references}.

\subsection{Searching for spectroscopic redshifts}
\label{sec:xmatch_specz}

Like for the photometric redshifts, we crossmatch the G4Jy catalogue with spectroscopic-redshift catalogues using a matching radius of 1\,arcsec on the host-galaxy positions. In order to keep track of the origin of these redshifts, we create a new column for the G4Jy catalogue: `$z$ origin', where `1' = SALT follow-up (DR1), `4' = DESI DR1, `6' = 6dFGS, `12' = SDSS DR12, and `16' = SDSS DR16. We add `0.5' to the `$z$ origin' value to represent spectra that are re-fitted by \citet[][this work]{White2025a}, and we list values that represent individual references in Table~\ref{tab:z_references}. 

\begin{table*}
\centering 
\caption{A summary of the references represented by  the `$z$ origin' flag in the G4Jy catalogue, with `0.5' added to the flag value wherever the redshift has been re-fitted. Having filtered the catalogue by this flag, `Number' refers to the number of entries in the spectroscopic-redshift ($z_{\mathrm{sp}}$; `zsp\_misc') or the photometric-redshift ($z_{\mathrm{ph}}$; `zph\_misc') column that the reference corresponds to. Similarly, the redshift range refers to sources with the corresponding flag. *The 723 redshifts from NED (the NASA/IPAC Extragalactic Database) do not have a flag to indicate whether they are spectroscopic or photometric, and so they are kept in a separate column within the multiwavelength G4Jy catalogue (see Paper IV). 164 of these redshifts are associated with z\_origin\_flag = 10.0.}
\begin{tabular}{@{}lccccc@{}}
 \hline
`$z$ origin' flag & Name & Reference  & Number of $z_{\mathrm{sp}}$ & Number of $z_{\mathrm{ph}}$ & Redshift range \\
 \hline
%Cassiopeia A & 23:23:28 & 58:48:42 \\ % Not below Dec = 30 deg
1.0 & G4Jy-SALT DR1 & \citet{White2025a}  & 42 & -- & $0.012 < z < 1.630$ \\
2.0 & G4Jy-SALT DR2 & Sejake et al. (submitted) & 127 & -- & $0.0301 < z < 3.0042$ \\
3.0 & G4Jy-3CRE (DR1) & \citet{GarciaPerez2024} & 77 & -- & $0.00075 < z < 1.62000$ \\
4.0 & DESI DR1 & \citet{AbdulKarim2025} &130 & -- & $0.00746 < z < 2.65306$ \\
% Had 201 DESI crossmatches
4.5 & DESI DR1, re-fitted & This work & 3 & -- &  $0.6337 < z < 1.7764$ \\
%5.0 & G4Jy-SALT DR3 & Sejake et al. (in prep.) & -- & -- & $? < z < ?$ \\
6.0 & 6dFGS DR3 & \citet{Jones2009} & 126 & -- &  $0.00379 < z < 0.97711$ \\
% 6dFGS reduced from 141
6.5 & 6dFGS, re-fitted & This work & 5 & -- & $0.091 < z < 1.032$ \\
7.0 & WISExSuperCOSMOS & \citet{Bilicki2016} & -- & 21 &  $ 0.16135 < z < 0.36802$ \\ % Had 46 redshifts before adding more spec-zs, with $0.11059 < z < 0.37346$
8.0 & DESI LS DR8 photo-z South & \citet{Duncan2022} & -- & 379 &  $0.010 < z < 3.030$ \\ % Previously had 556 &  $0.006 < z < 3.030$
10.0 & NED & Various references, see Table~\ref{tab:NED_references} & * & * & $0.01567 < z < 3.56990$ \\
12.0 & SDSS DR12 & \citet{Alam2015} & 66 & -- & $0.02160 < z < 2.17827$ \\
12.5 & SDSS DR12, re-fitted & \citet{White2025a} & 2 & -- & $1.196 < z < 1.706$ \\
16.0 & SDSS DR16 & \citet{Ahumada2020} & 47  & -- & $0.04385 < z < 2.12528$ \\
16.5 & SDSS DR16, re-fitted & \citet{White2025a}, this work & 6 & -- & $0.581 < z < 1.965$ \\
\hline
& & \hspace{3.2cm} Total & 631 & 400 & $0.00075 < z < 3.56990$ \\
 \hline
\label{tab:z_references}
\end{tabular}
\end{table*}

In addition, we have a `$z$-Quality flag' (Q) column to indicate the quality of the redshifts that we have compiled, as assessed through visual inspection of the optical spectroscopy and its redshift-template-fitting. The possible flag values are:
\begin{itemize}
\item Q = 1, indicating that the redshift is highly robust, with multiple spectral features detected with good signal-to-noise ratios. This flag is also assigned when a single {\it broad} emission-line\footnote{We caution that a single narrow `line' may be the result of a cosmic ray being detected, or poor background-subtraction near detector-chip edges.} (with a distinctive profile, e.g. Ly-$\alpha$) is detected with very good signal-to-noise.
\item Q = 2, indicating that the redshift is fairly robust, where multiple spectral features are detected but they are less discernible from the continuum or noise level than for Q = 1.
\item Q = 3, indicating that the redshift is a tentative measurement, as suggested by the marginal detection of multiple spectral features against the continuum/noise.
\item Q = 4, indicating that the redshift is unreliable, as it is derived from (a) a single spectral feature detected with low-to-moderate signal-to-noise, (b) one or more spectral features that have been misidentified (resulting in an incorrect redshift calculation), or (c) spectral features that are not discernible by eye.
\item Q = 5, indicating that visual inspection of the spectroscopy has not been completed. This flag is automatically given to all photometric redshifts and all redshifts acquired through automated data-mining of the NASA/IPAC Extragalactic Database (NED)\footnote{\url{https://ned.ipac.caltech.edu}}. 
\end{itemize}

`Q = 5' also allows us to flag spectroscopic redshifts that should be treated with caution for the moment. This is because some digital 6dFGS spectra are unavailable online\footnote{\url{http://www-wfau.roe.ac.uk/6dFGS/intro.html}}, or do not have visible identifications for possible emission-/absorption-lines (see Table~\ref{tab:6dfgs}). The unavailability will be addressed in future work for the G4Jy Sample. For now, we assess the quality of the redshift determination by re-fitting the 6dFGS spectra for sources listed in Table~\ref{tab:6dfgs}. We find that five are correct to within our redshift-error estimate, and we update their Q flags from `5' accordingly. Further details are in Appendix~\ref{app:6dfgs_refitting}. We also re-fit six SDSS spectra and three DESI spectra, as described in Appendices~\ref{app:sdss_refitting} and \ref{app:desi_spectra} (respectively).

  \begin{table*}%[]
    \centering
    \begin{tabular}{c|c|c|c|c|c}
    \hline
    Source name & AllWISE host-galaxy name & Counterpart in 6dFGS &  Spectroscopic  & AllWISE--6dFGS & Brief comment on
      \\ 
    & (this work)  & \citep{Jones2009} & redshift, $z_{\mathrm{sp}}$ & separation / arcsec &  the spectral fit
      \\ 
       \hline

        G4Jy 28 &  J001619.95$-$143011.0 & g0016200$-$143011 & 0.76776 & 0.2 & A single emission line (``Mg{\sc ii}'') \\
        G4Jy 229 & J021046.20$-$510101.8 & g0210462$-$510102 & 0.03154 & 0.1 & No spectral features \\
          G4Jy 411 & J040534.00$-$130813.6 &   g0405340$-$130814 & 0.57428 & 0.3 & A single emission line (``Mg{\sc ii}'') \\
G4Jy 540 & J052257.98$-$362730.8  &  g0522580$-$362731 & 0.05651 & 0.4  & A single emission line (``H$\alpha$'')  \\

       G4Jy 543 & J052527.19$-$324215.5  & g0525272$-$324216  & 0.07676 & 0.2  & No spectral features  \\
     %   G4Jy 633 & J065155.06$-$602214.8  & g0651549$-$602217  & 0.13380 & 2.108  & No spectral features  \\ % Keeping criterion to 1-arcsec matches 

     G4Jy 665 & J073104.92$-$523808.6 & g0731049$-$523808 &   0.09029 & 0.7 & No spectral features  \\ 
      G4Jy 893 & J110612.14$-$244443.8  & g1106121$-$244444  & 0.04997 & 0.4  & No spectral features  \\

   G4Jy 994 & J122343.36$-$423532.2  &  g1223434$-$423532 & 0.02658 & 0.5  & Too noisy a spectrum  \\
G4Jy 1127 &  J141331.89$-$300243.5 & g1413319$-$300244  & 0.06486 & 0.4  & Too noisy a spectrum  \\
G4Jy 1245 &  J152006.00$-$283419.9 & g1520060$-$283420  & 0.12264 & 0.2  & No spectral features  \\

G4Jy 1533 &  J192451.05$-$291430.1 &  g1924510$-$291430 & 0.35238 & 0.3  & Too noisy a spectrum  \\
G4Jy 1554 & J193252.79$-$081803.4 & g1932528$-$081804 & 0.10053 &  0.1 & Too noisy a spectrum  \\ 
G4Jy 1664 & J205604.27$-$195634.9  & g2056043$-$195635  & 0.15651 & 0.5  & No spectral features  \\
G4Jy 1708 & J213745.19$-$143255.7  & g2137452$-$143256  & 0.19979 & 0.1  & Possibly mislabelled lines \\
G4Jy 1753 &  J220156.37$-$332103.0 & g2201564$-$332103  & 0.15359 & 0.1  & No spectral features  \\
G4Jy 1842 & J233444.92$-$525119.4  & g2334449$-$525119  & 1.03189 & 0.3  & A single emission line (``Mg{\sc ii}'')  \\

    \hline
    \end{tabular}
    \caption{A list of sources for which we query the redshift that is determined through the 6dFGS spectrum (Section~\ref{sec:xmatch_specz}). This is based upon visual inspection, via the 6dFGS online database, and we download these spectra to see whether an improved redshift can be obtained (Appendix~\ref{app:6dfgs_refitting}).} 
    \label{tab:6dfgs}
\end{table*}

As for the default application of `Q = 5', this will remain in place until visual inspection of (new) data is able to confirm that these redshifts are indeed correct, noting the possibility that the existing optical/NIR identification in the literature may be incorrect \citep[e.g.][]{Burgess2006}. The latter can be explained by radio-on-optical overlays typically being used for identifying the host galaxy of the radio emission, with the limitation that optical images will miss host galaxies that are obscured by dust. Hence, Radio Galaxy Zoo \citep{Banfield2015,Alger2018,Wong2025}, \citet{White2020a,White2020b}, and \citet{Sejake2023}, for example, overlay radio contours onto {\it mid-infrared} images, which traces the underlying stellar population while being less sensitive to dust attenuation.  

As part of our crossmatching, we inspect additional 6dFGS spectra for the G4Jy Sample as a result of updated host-galaxy positions (this work), noting our previous reluctance to cross-identify 14 per cent of G4Jy sources \citep{White2020a,White2020b} without better radio-imaging being made available (such as that presented by \citealt{Sejake2023}). See Table~\ref{tab:new_6dfgs} for 16 6dFGS redshifts that are newly-added to the G4Jy catalogue, compared with those listed by \citet{White2025a}. We also remove 4 6dFGS redshifts on account of the large separation ($1 <\theta$\,/\,arcsec$ < 4$) between the original AllWISE position and the 6dFGS position.

Next we incorporate redshifts that have been obtained for the G4Jy-3CRE subset \citep{Massaro2023} by \citet{GarciaPerez2024}. This results in 15 6dFGS redshifts being replaced, which is justified by those 6dFGS spectra having low signal-to-noise or being unavailable for visual inspection (with $Q=4$ or 5; this work). % [Get this number by crossmatching 299-redshifts file with the catalogue and seeing which z-origin flags got updated from 6 to 3.] % A further 2 6dFGS spectra are replaced by DESI spectra

% Also, dropped 4 redshifts from White et al. (submitted), as a previous datafile had been created with a more lenient crossmatching radius of 3\,arcsec.] But I said in the SALT paper that it was 1 arcsec, hence removing this

  \begin{table}%[]
    \centering
    \begin{tabular}{c|c|c|c}
    \hline
    Source name &  Counterpart in 6dFGS &  Spectroscopic  & $z$-Quality
      \\ 
    & \citep{Jones2009} & redshift, $z_{\mathrm{sp}}$ & flag
      \\ 
       \hline
        G4Jy 40 &   g0021075$-$191006& 0.09551 & 1 \\
        G4Jy 47 &  g0025315$-$330246 & 0.04966 & 4  \\
          G4Jy 270 &   g0231370$-$204022 & 0.08977 & 1  \\
G4Jy 325 &  g0306527$-$120624 & 0.07870 & 2   \\
       G4Jy 531 &  g0513320$-$302850  & 0.05760 & 3  \\
   G4Jy 665 &  g0731049$-$523808 & 0.09029 & 4  \\ % Why was this 5*?
G4Jy 680 &  g0802363$-$095740 & 0.06985 & 4  \\
G4Jy 684 & g0805378$-$005818  & 0.09014 & 1    \\
G4Jy 693 &    g0816118$-$703945 & 0.03313 & 1  \\
G4Jy 1067 &  g1326106$-$272539 & 0.04369&  1   \\ 
G4Jy 1148 &  g1420037$-$493542  & 0.09136 & 3    \\
G4Jy 1166 &  g1427381$-$120350  & 0.80239 & 3   \\
G4Jy 1342 &   g1631351$-$750906  & 0.11032 & 2   \\
G4Jy 1554 &  g1932528$-$081804  & 0.10053 & 5   \\
 G4Jy 1582      &  g1952305$-$011721  & 0.05533 &  5  \\
      G4Jy 1852       &  g2347451$-$280826 &  0.02879  & 1  \\ 
    \hline
    \end{tabular}
    \caption{A list of 6dFGS redshifts that have been added to the G4Jy catalogue since the publication of \citet{White2025a}. A `$z$-Quality flag' (Section~\ref{sec:xmatch_specz}) of `5' is assigned to sources where we could not find the corresponding 6dFGS spectrum in the online database.} 
    \label{tab:new_6dfgs}
\end{table}

%\vspace{-5mm}
\subsection{Redshift summary}
\label{sec:z_summary}

\begin{figure}
\centering
\includegraphics[scale=0.52]{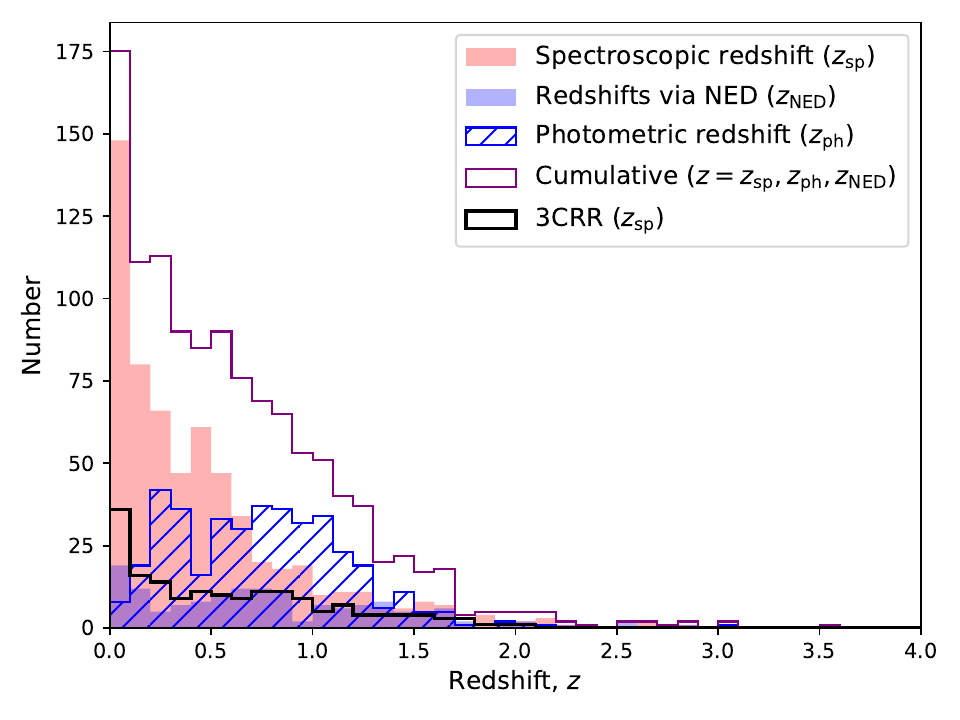}
\caption{ Redshift distributions for the whole G4Jy Sample \citep[][Section~\ref{sec:z_summary}]{White2020a,White2020b}, with that for 3CRR \citep{Laing1983} added for comparison (outlined in black; $z< 2.02$). The total numbers correspond to Table~\ref{tab:z_references}, with care taken to avoid duplication of G4Jy sources within the `cumulative' histogram (outlined in purple; $z< 3.57$). Note that `Redshifts via NED', $z_{\mathrm{NED}}$ (Table~\ref{tab:NED_references}), are separated into neither the spectroscopic category ($z_{\mathrm{sp}}$) nor the photometric category ($z_{\mathrm{ph}}$).}
\label{fig:z_histograms}
\end{figure}

We use the `z\_origin\_flag' (Table~\ref{tab:z_references}) to avoid double-counting of G4Jy sources, because some rows in the G4Jy catalogue include (for example) both a spectroscopic redshift and a photometric redshift. As such, we calculate the spectroscopic-redshift completeness to be $(631/1,863 = )$ 34 per cent, with an {\it additional} 30 per cent of the G4Jy Sample having redshifts acquired through NED (154 sources) or photometric-redshift catalogues (400 sources). The resulting redshift distributions are shown in Figure~\ref{fig:z_histograms}, with the lowest redshift being $z=0.00075$ for G4Jy~86 (aka NGC~253, a well-known star-forming galaxy), most-recently obtained by \citet{GarciaPerez2024}, and the highest redshift being $z=3.5699$ for G4Jy~1020 (aka the `triple' radio-galaxy, 4C~+03.24), obtained by \citet{1996AA...313...25V}. This demonstrates the ability of a single, complete flux-limited sample \citep{White2020a,White2020b} to probe galaxy-evolution processes across the peak in star-formation history and black-hole-accretion history at $z\sim$ 2--3 \citep[e.g.][]{Madau2014}. 

% G4Jy overlay, GLEAM_J124538+032313_NVSS_overlays_10arcmin_v9.png, shows a boring point-source for G4Jy 1020
% No room for the footnote atm: 
%\footnote{As observed at 8.3\,GHz with 0.23\,arcsec resolution \citep{1996AA...313...25V}.}

\setkeys{Gin}{draft=false}

\begin{figure}
\centering
\subfigure[$z_{\mathrm{ph}}$ values from WISExSuperCOSMOS]{\includegraphics[scale=0.52]{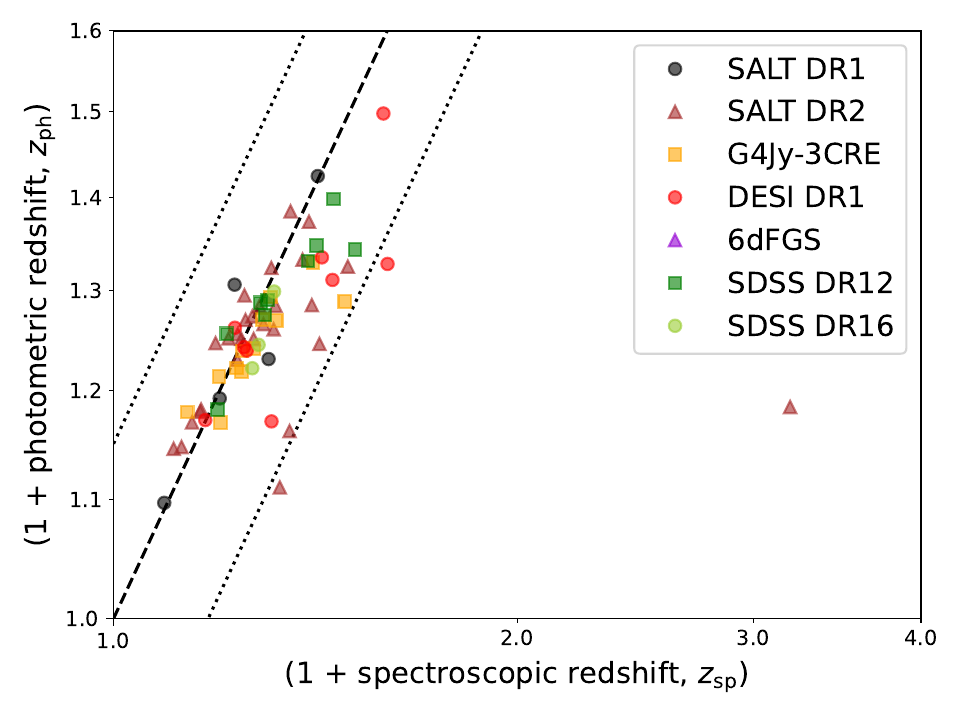}}
\subfigure[$z_{\mathrm{ph}}$ values from LS DR8 photo-$z$ South]{\includegraphics[scale=0.52]{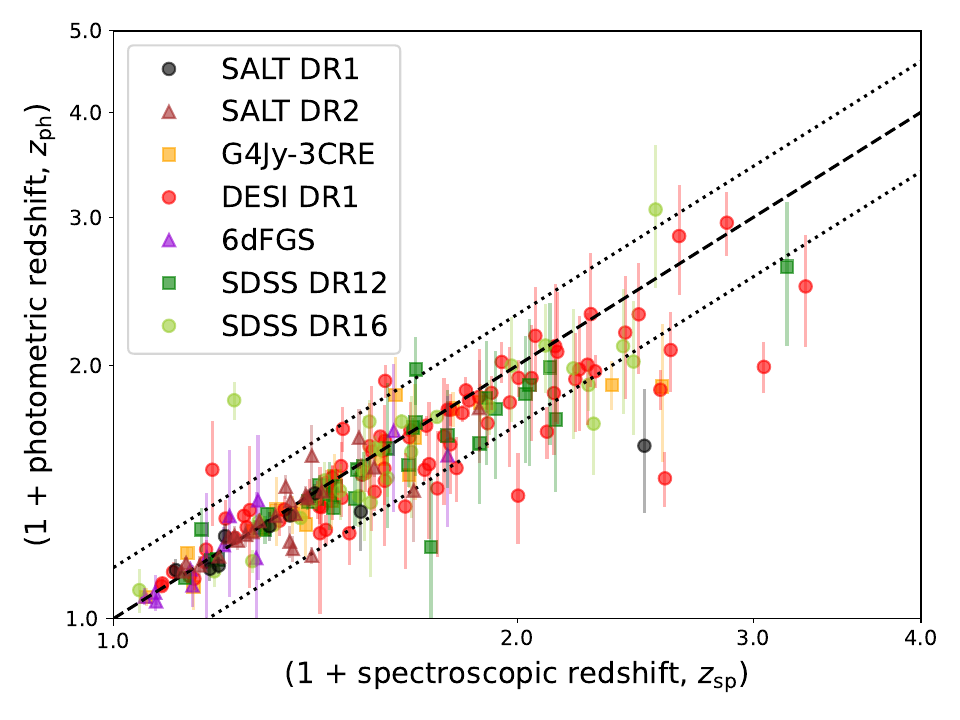}}
\caption{ The photometric redshift ($z_{\mathrm{ph}}$) versus the spectroscopic redshift ($z_{\mathrm{sp}}$), where available (Section~\ref{sec:z_summary}). The markers in the legend represent the combination of different spectroscopic datasets with the photometric dataset (a) WISExSuperCOSMOS \citep{Bilicki2016}, and (b) LS DR8 photo-$z$ South \citep{Duncan2022}. The use of Gaussian processes by the latter enables error-bars to be estimated, and all of the values appearing in panel (b) have the associated reliability flag (flag\_qual) equal to `1' (Section~\ref{sec:xmatch_photoz}). Both panels show the 1:1 relation plotted as a dashed line, with dotted lines representing the boundaries beyond which the $z_{\mathrm{ph}}$ value is deemed an `outlier' (following the same definition as in footnote~\ref{outlier_footnote}).}
\label{fig:z_comparison}
\end{figure}

Next, we use the collated spectroscopic redshifts to assess the performance of the photometric redshifts (Figure~\ref{fig:z_comparison}). We find that the outlier rate is 5 per cent for the WISExSuperCOSMOS redshifts \citep{Bilicki2016}, with one of the outliers being a spectroscopic redshift that is particularly underestimated: $z_{\mathrm{sp}}=2.19520$ (Sejake et al., submitted), compared with $z_{\mathrm{ph}} = 0.18432$ (Figure~\ref{fig:z_comparison}a). We note that estimating the redshift of quasars from their photometry is notoriously difficult, and it is unsurprising that (given the training set) any outliers are biased towards being {\it under}estimates rather than overestimates.

A similar bias is seen for the redshifts from the Legacy Surveys DR8 South catalogue \citep{Duncan2022}, although their multiple training sets (guided by deeper LS DR8 photometry) allow them to probe much higher redshifts (Figure~\ref{fig:z_comparison}b). In this case, the outlier rate is 13 per cent, which is very reasonable given their testing with LoTSS DR1 data over $1< z < 3$, where the outlier rate was 23 per cent \citep{Duncan2022}. The uncertainties in the photometric-redshifts estimates are also of interest, as they do not show the expected correspondence of larger uncertainties for the redshift outliers. This is because they are based on a combination of three components: (i) the photometric noise, with uncertainties on the input photometry scaling with optical magnitude; (ii) the training noise, which measures the degree in scatter in the {\it true} redshift for a given region of colour space; and (iii) the training density, which indicates how well-sampled is the colour/magnitude space around the source. (This means that if there is nothing similar to that source in the training sample, the uncertainty in $z_{\mathrm{ph}}$ should be larger.) These are all important considerations when making `all-purpose' catalogues that cater for rare populations (such as quasars) as well as more-common populations.

% Ken Duncan: "So for those sources that are catastrophic outliers but GPz is very confident, it could be a range of things. There might be lots of sources with similar colours and low photometric noise, but that specific source could be different for either astrophysical or observational reasons. For example, emission lines or dust mimicking colours from a different redshift. Or in some cases, it could be blending in the photometry (particularly WISE) that is biasing the observed photometry and pushing it one in direction."

\section{Results and discussion}
\label{sec:results}

Given the wealth of information available in the updated G4Jy catalogue, we provide a glimpse of the intrinsic radio properties of the G4Jy Sample in this paper, and present further analysis (based on additional data) in the accompanying Paper IV \citep{White2025d}. We encourage the reader to compare the `results' sections of both papers, for a more-complete understanding of the sample and its significance for galaxy-evolution studies.

\subsection{Radio luminosities}
\label{sec:luminosities}

\begin{figure}
\centering
\includegraphics[scale=0.52]{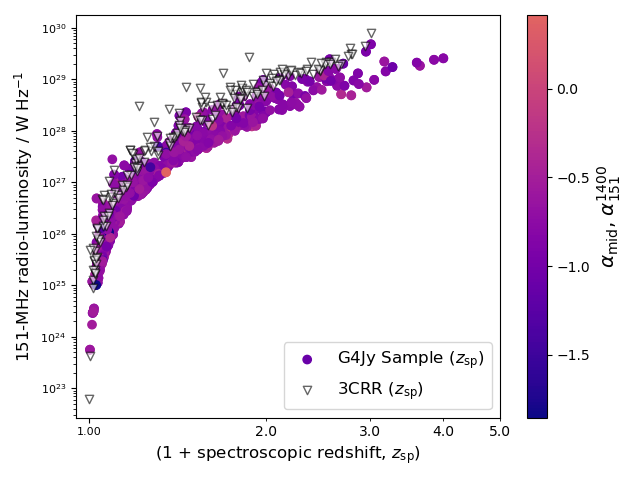} 
\caption{ Distribution of 151-MHz radio luminosties as a function of redshift (Section~\ref{sec:luminosities}), with the 3CRR sources \citep{Laing1983} added for comparison. Note that the G4Jy sources shown here are only those with a verified spectroscopic redshift ($z_{\mathrm{sp}}$), which applies to 34 per cent of the sample. The datapoints are colour-coded by a two-point spectral index ($\alpha$) between 151\,MHz and 1400\,MHz, assuming that the radio emission can be described by a power-law function ($S_{\nu} \propto \nu^{\alpha}$). }
\label{fig:luminosities}
\end{figure}

Thanks to the collation of redshifts for the G4Jy catalogue (Section~\ref{sec:catalogue}), we can now calculate intrinsic properties of the G4Jy Sample, following on from \citet{White2025a}. Of foremost importance are the radio luminosities (Figure~\ref{fig:luminosities}), which reveal just how powerful these radio galaxies are, although modulated by environmental properties such as gas density. As expected, the lower flux-density threshold of the G4Jy Sample compared with 3CRR \citep{Laing1983} allows the former to probe higher redshifts, as well as lower radio-luminosities at almost all redshifts. For interest, we include (as the third/colour axis) the spectral index ($\alpha$) between 151\,MHz and 1400\,MHz (assuming that the radio emission follows a power-law description, $S_\nu \propto \nu^{\alpha}$). This shows that following up ultra-steep-spectrum ($\alpha < -1.3$; \citealt{DeBreuck2000}) sources does not necessarily lead to high redshifts ($z > 1$) being probed (see also Sejake et al., submitted). Furthermore, this result continues the debate by \citet{Pinjarkar2025}, as we see no trend in the spectral index with either redshift or radio luminosity. However, we recognise that their finding -- that the $\alpha$--luminosity relation underpins the apparent $\alpha$--$z$ correlation -- is more appropriate for a wider sampling of luminosity (to lower values). 

\begin{figure}
\centering
\includegraphics[scale=0.52]{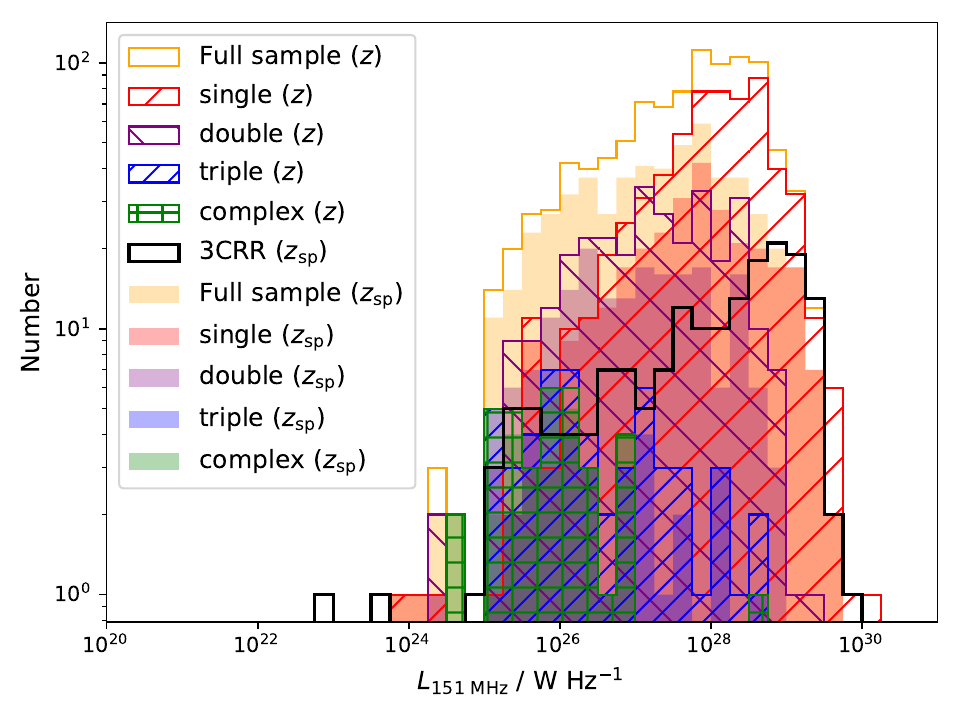} 
\caption{ Distributions of the 151-MHz radio-luminosities (Section~\ref{sec:luminosities}) for the G4Jy Sample, colour-coded by the radio morphology of the source \citep{White2020a,White2020b} in the 45-arcsec imaging from NVSS and SUMSS. `Cumulative redshifts' ($z = z_{\mathrm{sp}}$, $z_{\mathrm{ph}}$, $z_{\mathrm{NED}}$) are used to determine the outlined histograms, and spectroscopic redshifts ($z_{\mathrm{sp}}$) are used to determine the shaded histograms. The luminosity distribution for 3CRR sources \citep{Laing1983} is added for comparison (black, unfilled histogram). }
\label{fig:luminosities_by_morphology}
\end{figure}

We present a histogram of the radio luminosities in Figure~\ref{fig:luminosities_by_morphology} and see that they break the $10^{30}$\,W\,Hz$^{-1}$ luminosity threshold via a photometric-redshift estimate of $z_{\mathrm{ph}} = 3.030$ \citep{Duncan2022} for G4Jy~27. However, we note that the associated quality flag deems this to be an `unreliable' redshift, and so should be treated with caution. If we consider just the spectroscopic redshifts (shaded histograms in Figure~\ref{fig:luminosities_by_morphology}) then the highest radio luminosity (so far) is still an impressive $L_{\mathrm{151\,MHz}} = 4.8 \times 10^{29}$\,W\,Hz$^{-1}$ (also visible in Figure~\ref{fig:luminosities}). This source, G4Jy 38, also happens to be the source with the highest radio luminosity in the 3CRR sample, 3C~9 \citep[at $z_{\mathrm{sp}} = 2.0158$;][]{AbdulKarim2025}.

In addition, we inspect the distribution in radio luminosities as a function of radio morphology (Figure~\ref{fig:luminosities_by_morphology}). `Single' morphology refers to compact radio-sources, but may also include extended radio-sources whose angular-size measurements are limited by the resolution of the NVSS/SUMSS data (at 45\,arcsec). Like the `double' (double-lobed) radio-sources and the `triple's (sources where the radio-core is detected in addition to the two radio lobes), they span a wide range in luminosities. In contrast, the G4Jy sources with `complex' morphology have a relatively narrow range at lower luminosities. {\color{black}This is likely an observational bias}: the limited resolution of the radio data means that the `complex' label tends to be assigned to low-redshift systems, which are typically at lower luminosity than their high-redshift counterparts (due to Malmquist bias). We note that the radio-luminosity distribution will be investigated further once higher-resolution radio images for the full sample enable more-complete (and more-precise) identification of different radio morphologies -- for example, `core-brightened' FR-I radio-galaxies and `edge-brightened' FR-II radio-galaxies, following the morphological classification of \citet{Fanaroff1974}.

\subsection{Linear sizes} 
\label{sec:linearsizes}

\begin{figure}
\centering
\includegraphics[scale=0.5]{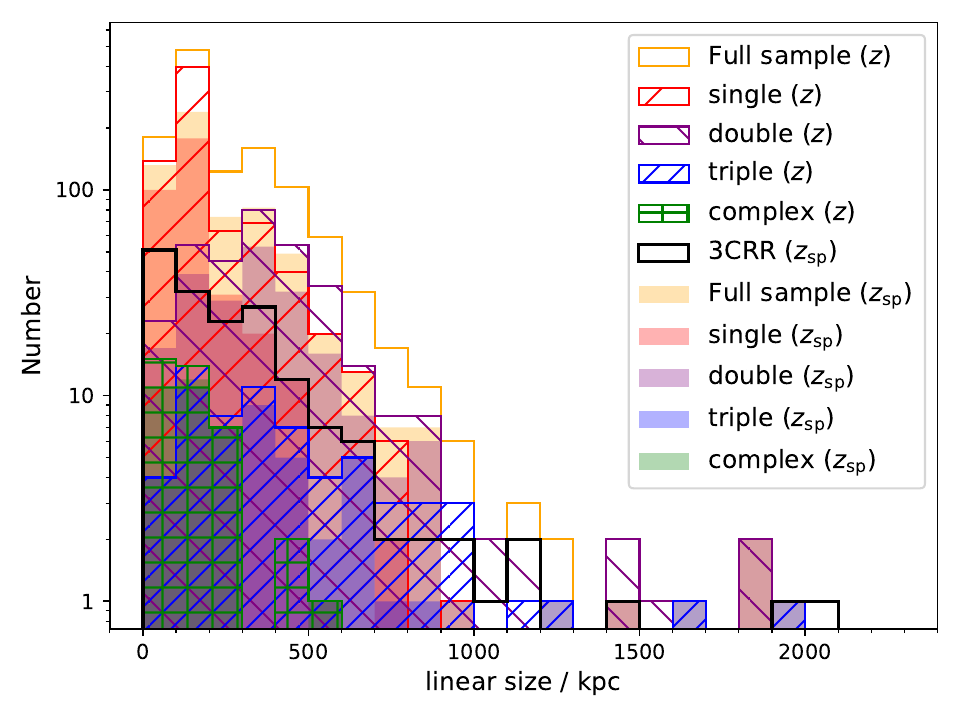}
\caption{ Distributions in the linear sizes of the G4Jy Sample, colour-coded by the morphology of the source \citep{White2020b,White2020a} in the 45-arcsec radio-imaging from NVSS and SUMSS. The size distribution for 3CRR sources \citep{Laing1983} is added for comparison (black histogram), with one source (3C\,236) having a linear size that is beyond the plot range (at 4530\,kpc). The outlined histograms represent linear sizes calculated from the `cumulative redshifts' ($z = z_{\mathrm{sp}}$, $z_{\mathrm{ph}}$, $z_{\mathrm{NED}}$), whilst the shaded histograms represent linear sizes calculated from spectroscopic redshifts alone ($z_{\mathrm{sp}}$). }
\label{fig:linearsizes}
\end{figure}

Another intrinsic property that we can now calculate are the linear sizes of G4Jy sources, which are presented in Figure~\ref{fig:linearsizes}. {\color{black}These are derived from the angular sizes calculated for the G4Jy catalogue \citep{White2020a,White2020b}, based on the NVSS/SUMSS components associated with each source. This association was initiated by cross-correlating GLEAM contours with the NVSS- and SUMSS-catalogue positions, with refinement informed by visual inspection of the NVSS and SUMSS contours (subject to 45-arcsec spatial resolution). For single-component sources, we inherit the deconvolved angular-size from the respective catalogue. In the case of the NVSS catalogue, this value is accompanied by an upper-limit flag (`--' or '$<$'), whilst for SUMSS-based G4Jy sources, that remain unresolved, an upper-limit ('$<$') is associated with the fitted major-axis measurement. (For the current work, we note that 64 per cent of linear-size measurements for `single' morphologies are upper limits.) For estimating the angular size of sources that are multi-component in NVSS/SUMSS, \citet{White2020b} calculated the maximum separation of the NVSS/SUMSS components associated with a particular source. For further details, we direct the reader to Section 6.3.1 of Paper I.

A consequence of the 45-arcsec spatial resolution provided by NVSS/SUMSS is that the number of sources with `double' or `triple' radio morphology will be underestimated. This is exacerbated for higher-redshift radio sources where much higher resolution is required to determine the appropriate morphology. Nevertheless, the angular-size distribution is `modulated' by the redshift distribution, resulting in the linear-size distribution shown in Figure~\ref{fig:linearsizes}.} When considering the radio morphology of the source, we see a noticeable `tail' towards larger sizes in the `double' and `triple' distributions. This could be because these FR-II-like morphologies are easier to characterise in terms of angular size, since the latter is still calculated based on the NVSS/SUMSS components (i.e. from a position of peak emission to another position of peak emission). Meanwhile, FR-I radio-galaxies are likely to be classified as having `single' morphology despite showing extended emission. Again, as in Figure~\ref{fig:luminosities_by_morphology}, the `complex' sources are more-restricted in their distribution {\color{black} as a consequence of selection effects, since more-extended complex sources do not exceed the surface-brightness limits necessary for inclusion.  Furthermore, such morphology needs to be spatially resolved, and hence there is a dependence on the angular size of the source. } 

% (NGC\,6251, 3C\,326 = G4Jy\,1282, and 3C\,236) having linear sizes that are beyond the plot range (i.e. 1900\,kpc to 4530\,kpc)

For space reasons, we discuss giant radio-galaxies (GRGs) in {\color{black}Appendix C of} the accompanying Paper IV \citep{White2025d}.

\section{Conclusions}
\label{sec:conclusions}

The G4Jy Sample \citep{White2020a,White2020b} is a complete, flux-limited ($S_{\mathrm{151\,MHz}}> 4$\,Jy) compilation of the `brightest' radio-sources in the southern sky, as defined in G4Jy Paper I. The accompanying paper, Paper II in the series, detailed the procedures that went into host-galaxy identification for 1,606 out of the 1,863 sources -- with the note that such identification is not required for G4Jy~571 (the Flame Nebula), nor for G4Jy~1605 (the northern cluster-relic of Abell 3667). Whilst the G4Jy catalogue provided comprehensive radio and mid-infrared information, via GLEAM \citep{HurleyWalker2017}, NVSS \citep{Condon1998}, SUMSS \citep{Mauch2003,Murphy2007}, and AllWISE \citep{Cutri2013}, only brief analysis was presented in Paper I. The focus there was, instead, on comparing the flux-density scale between GLEAM and 3CRR \citep{Laing1983} for the 67 3CRR sources that appear in the G4Jy Sample.

For the current work, Paper III, we update the G4Jy catalogue by crossmatching it with multiple optical datasets (via a 1-arcsec matching radius, throughout). This allows us to perform further analysis of the sample, as summarised within the following conclusions:

\begin{enumerate}
    \item{ Thanks to visual inspection of radio contours [from MeerKAT \citep{Sejake2023}, VLASS \citep{Lacy2020}, and RACS-low1 \citep{Hale2021}] on AllWISE-W1, $K$-band, and optical images -- accompanied by thorough literature checks -- it has been possible to identify the galaxy hosting the radio emission for a further 127 G4Jy sources. This does not include revised host-galaxy identifications, which were made for 16 sources (i.e. 1 per cent of the sample). } % 1,590 of the hosts in the original catalogue remain in the updated catalogue, to within *0.1* arcsec. 
    \item{ We document a tentative identification for G4Jy 1289 (WISEA J155900.91$-$213935.5), because of its special interest as a candidate giant radio-galaxy (GRG) that has particularly-diffuse radio emission. However, the location of the radio `core' remains unconfirmed, and so this identification is not provided in the G4Jy catalogue. Similarly (discounting G4Jy~571 and G4Jy~1605, as mentioned above), 128 other sources remain unidentified in the catalogue. Of these, 68 have their host-galaxy identification limited by the depth of the available mid-infrared/near-infrared/optical catalogues. (Those that are not affected by imaging artefacts are likely to be at high redshift.) }
 %   \item{ [Cross-checks and mistaken identifications] }
    \item{ Our eventual aim is to obtain spectroscopic completeness for the sample, where all of the spectroscopy is available online in digitised form (e.g. via \url{ https://zenodo.org/communities/g4jy/records}). This means re-observing some G4Jy sources with modern instruments, even if a spectroscopic redshift already exists in the literature. We visually inspect spectroscopy from the following datasets in order to assign a `$z$-Quality' (Q) flag to the redshift that has been fitted: 6dFGS DR3 \citep{Jones2009}, SDSS DR12 \citep{Alam2015}, SDSS DR16 \citep{Ahumada2020}, G4Jy-3CRE \citep{GarciaPerez2024}, DESI DR1 \citep{AbdulKarim2025}, G4Jy-SALT DR1 \citep{White2025a}, and G4Jy-SALT DR2 (Sejake et al., submitted). Following \citet{White2025a}, we re-fit 16 6dFGS spectra, four SDSS DR16 spectra, and three DESI spectra. The result is 631 G4Jy sources having {\it verified} spectroscopic redshifts (see Table~\ref{tab:z_references}). }
    \item{ We also gather photometric redshifts for the sample, from WISExSuperCOSMOS (\citealt{Bilicki2016}; 110 sources) and from DESI LS DR8 photo-$z$ South (\citealt{Duncan2022}; 768 sources). These show outlier rates of 5 per cent and 13 per cent, respectively, when compared to the spectroscopic redshifts that we have collected for the sample.}
    \item{ Data-mining the NASA/IPAC Extragalactic Database (NED) allows us to acquire a mixture of photometric and spectroscopic redshifts for 723 sources, 164 of which do not have redshifts via other means. The references associated with the redshifts from NED are provided in the G4Jy catalogue and expanded upon in Table~\ref{tab:NED_references}. }
    \item{ In summary, we have gathered redshift information for 64 per cent of the G4Jy Sample, with 34 per cent of these redshifts being from visually-inspected spectroscopy. Furthermore, this work has enabled new redshift associations for 473 sources. We find that, as it stands, the G4Jy Sample spans redshifts from $z=0.00075$ to $z=3.56990$, demonstrating the ability of a single, flux-limited sample to probe galaxy-evolution processes across the peak in star-formation and AGN activity at $z\sim2$--3.  }
    \item{ The highest 151-MHz radio-luminosity in the sample, based on a spectroscopic redshift, is $L_{\mathrm{151\,MHz}} = 4.8 \times 10 ^{29}$\,W\,Hz$^{-1}$ for G4Jy~38 (aka 3C~9). Meanwhile, the lower luminosities seen for `complex' radio morphologies is deemed to be the result of observational bias.} 
    \item{ Redshift information also allows us to calculate the linear sizes for the sample, which span from 1\,kpc (for G4Jy~288) to 1961\,kpc (for G4Jy~923). {\color{black}We again note the more-restricted values for `complex' sources, but caution that these are likely due to surface-brightness limits rather than a consequence of particularly-dense environments.} }
\end{enumerate}

We note that the above science results are a glimpse of the more-in-depth multiwavelength analysis that is conducted in the accompanying paper, Paper IV \citep{White2025d}. With respect to spectroscopic completeness, we expect this to be completed in stages, with the rate of progress dictated by how much telescope time is awarded to observe the optically-fainter sources. This is critical for avoiding observational bias in the conclusions drawn from sample studies, and taking full advantage of the fact that the G4Jy Sample is {\it unbiased with respect to the radio-jet axis orientation}. Furthermore, being defined across the Southern sky means that the sample is perfect for AGN research using (for example) the Atacama Large Millimeter/submillimeter Array (ALMA), the Square Kilometre Array (SKA), and its precursor telescopes.

\section*{Acknowledgements}

{\color{black} We thank the referee for their comments, which helped to improve the manuscript. In addition, we} thank Kenneth Duncan and Dustin Lang for helpful discussions, and those involved in the Astro Data Lab Science Platform (\url{https://datalab.noirlab.edu/}). We also thank Philip Best for providing overlays from their \citet{Best1999} paper.

This work is based on the research supported in part by the National Research Foundation of South Africa (Grant Number 151060). The financial assistance of the South African Radio Astronomy Observatory (SARAO) towards this research is also hereby acknowledged. CJR acknowledges support from the DFG via the Collaborative Research Center SFB1491, \textit{Cosmic Interacting Matters -- From Source to Signal} (project no. 445052434).

Some of the observations reported in this paper were obtained with the Southern African Large Telescope (SALT), under program 2020-1-MLT-008 (PI: White), and the MeerKAT telescope (via proposal SCI-20190418-SW-01). The latter is operated by the South African Radio Astronomy Observatory, which is a facility of the National Research Foundation, an agency of the Department of Science and Innovation.

Based on observations obtained at the Southern Astrophysical Researchntelescope, which is a joint project of the Minist\'{e}rio da Ci\^{e}ncia, Tecnologia e Inova\c{c}\~{o}es (MCTI/LNA) do Brasil, the US National Science Foundation’s NOIRLab, the University of North Carolina at Chapel Hill (UNC), and Michigan State University (MSU). Based upon observations carried out at the Observatorio Astron{\' o}mico Nacional on the Sierra San Pedro M{\' a}rtir (OAN-SPM), Baja California, M{\' e}xico. This research uses services or data provided by the Astro Data Lab, which is part of the Community Science and Data Center Program of NSF NOIRLab. NOIRLab is operated by the Association of Universities for Research in Astronomy (AURA), Inc. under a cooperative agreement with the U.S. National Science Foundation.

We acknowledge the use of the ilifu cloud computing facility -- www.ilifu.ac.za, a partnership between the University of Cape Town, the University of the Western Cape, Stellenbosch University, Sol Plaatje University, the Cape Peninsula University of Technology and the South African Radio Astronomy Observatory. The ilifu facility is supported by contributions from the Inter-University Institute for Data Intensive Astronomy (IDIA - a partnership between the University of Cape Town, the University of Pretoria and the University of the Western Cape), the Computational Biology division at UCT and the Data Intensive Research Initiative of South Africa (DIRISA).

This research has used the NASA/IPAC Extragalactic Database, which is funded by the National Aeronautics and Space Administration and operated by the California Institute of Technology.

The National Radio Astronomy Observatory is a facility of the National Science Foundation operated under cooperative agreement by Associated Universities, Inc. We also make use of data products from 2MASS, which is a joint project of the University of Massachusetts and the Infrared Processing and Analysis Center/California Institute of Technology, funded by the National Aeronautics and Space Administration and the National Science Foundation. 

This scientific work uses data obtained from Inyarrimanha Ilgari Bundara / the Murchison Radio-astronomy Observatory. We acknowledge the Wajarri Yamaji People as the Traditional Owners and native title holders of the Observatory site. CSIRO’s ASKAP radio telescope is part of the Australia Telescope National Facility (https://ror.org/05qajvd42). Operation of ASKAP is funded by the Australian Government with support from the National Collaborative Research Infrastructure Strategy. ASKAP uses the resources of the Pawsey Supercomputing Research Centre. Establishment of ASKAP, Inyarrimanha Ilgari Bundara, the CSIRO Murchison Radio-astronomy Observatory and the Pawsey Supercomputing Research Centre are initiatives of the Australian Government, with support from the Government of Western Australia and the Science and Industry Endowment Fund. This paper includes archived data obtained through the CSIRO ASKAP Science Data Archive, CASDA (https://data.csiro.au).

Funding for the Sloan Digital Sky Survey IV has been provided by the Alfred P. Sloan Foundation, the U.S. Department of Energy Office of Science, and the Participating Institutions. SDSS-IV acknowledges support and resources from the Center for High Performance Computing  at the University of Utah. The SDSS website is www.sdss.org. SDSS-IV is managed by the Astrophysical Research Consortium for the Participating Institutions of the SDSS Collaboration including the Brazilian Participation Group, the Carnegie Institution for Science, Carnegie Mellon University, Center for Astrophysics | Harvard \& Smithsonian, the Chilean Participation Group, the French Participation Group, Instituto de Astrof\'isica de Canarias, The Johns Hopkins University, Kavli Institute for the Physics and Mathematics of the Universe (IPMU) / University of Tokyo, the Korean Participation Group, Lawrence Berkeley National Laboratory, Leibniz Institut f\"ur Astrophysik Potsdam (AIP), Max-Planck-Institut f\"ur Astronomie (MPIA Heidelberg), Max-Planck-Institut f\"ur Astrophysik (MPA Garching), Max-Planck-Institut f\"ur Extraterrestrische Physik (MPE), National Astronomical Observatories of China, New Mexico State University, New York University, University of Notre Dame, Observat\'ario Nacional / MCTI, The Ohio State University, Pennsylvania State University, Shanghai Astronomical Observatory, United Kingdom Participation Group, Universidad Nacional Aut\'onoma de M\'exico, University of Arizona, University of Colorado Boulder, University of Oxford, University of Portsmouth, University of Utah, University of Virginia, University of Washington, University of Wisconsin, Vanderbilt University, and Yale University.

%%%%%%%%%%%%%%%%%%%%%%%%%%%%%%%%%%%%%%%%%%%%%%%%%%
%\vspace{-7mm}

\section*{Data Availability}

Overlays for the full sample, amongst other G4Jy data (such as optical spectroscopy), can be downloaded from the Zenodo repository: \url{https://zenodo.org/communities/g4jy/}. We encourage others to also make their G4Jy data available here, in the interest of Open Access, discoverability, and general best practices \citep{Chen2022}. We also wish to encourage further work on the G4Jy Sample that is beyond that conducted by our research group. Furthermore, G4Jy names are now fully-resolvable in many astronomical databases, including the NASA/IPAC Extragalactic Database, which helps to speed up literature searches for this complete sample.

%\vspace{-7mm}

%%%%%%%%%%%%%%%%%%%% REFERENCES %%%%%%%%%%%%%%%%%%

% The best way to enter references is to use BibTeX:

\bibliographystyle{mnras}
\bibliography{G4Jy_Paper_III}

% Alternatively you could enter them by hand, like this:
% This method is tedious and prone to error if you have lots of references
%\begin{thebibliography}{99}
%\bibitem[\protect\citeauthoryear{Author}{2012}]{Author2012}
%Author A.~N., 2013, Journal of Improbable Astronomy, 1, 1
%\bibitem[\protect\citeauthoryear{Others}{2013}]{Others2013}
%Others S., 2012, Journal of Interesting Stuff, 17, 198
%\end{thebibliography}

%%%%%%%%%%%%%%%%%%%%%%%%%%%%%%%%%%%%%%%%%%%%%%%%%%

%%%%%%%%%%%%%%%%% APPENDICES %%%%%%%%%%%%%%%%%%%%%

\appendix

\section{Fornax A}
\label{app:FornaxA}

For (the original) G4Jy Paper II, G4Jy~1110 was mistakenly cross-identified as Fornax A, due to its similar radio-morphology and orientation on the sky. Instead, it is also known as PKS B1358$-$113, and is the brightest cluster galaxy of ACO~1836. Furthermore, with a counterpart in the 6dFGS (g1401419$-$113625) at a redshift of $z = 0.03797$, G4Jy~1110 has a linear extent of 277\,kpc \citep{White2025d}. 

Meanwhile, Fornax~A (NGC~1316) is a famous merger remnant with a linear size of 389\,kpc \citep{McKinley2015} at $z = 0.006$ \citep{Humason1956}. A detailed optical study by \citet{Schweizer1980} -- showing the presence of an outer envelope, loops and `ripples' of stars -- allowed them to infer that the system has undergone multiple merger events. [See also the flickering activity and feedback studied by \citet{Maccagni2020,Maccagni2021} with MeerKAT.]
In fact, using a sample of 43 radio galaxies, \citet{Heckman1986} found that between one quarter and one third of the sources are associated with merger remnants. That is, their host galaxies show highly-disturbed optical morphologies (e.g. shells, tails, fans) that are suggestive of recent collisions. 

In the case of Fornax~A, the brightness of the radio lobes has permitted detailed imaging at both high \citep{Anderson2018} and low frequencies \citep{McKinley2015,Wayth2018}, using the Australia Telescope Compact Array (ATCA) and the MWA, respectively. It is excluded from the GLEAM Extragalactic Catalogue \citep{HurleyWalker2017} upon which the G4Jy Sample is based, as are other `A-team' sources listed in Table~\ref{tab:GLEAMexclusions}. As such, we revise the completeness estimate for the G4Jy Sample (section 7.4 of \citealt{White2020b}) from $>$95.5 per cent to 95.5 per cent.
%1863 = 0.955 * total
%1863/0.955 = total = 1950
%Now: 1863/1951 = 0.95489

\begin{table}
\centering 
\caption{An updated list of the brightest sources in the Southern sky (Dec.~$< 30^{\circ}$, $|b| > 10^{\circ}$) that do not appear in the G4Jy Sample. The flux densities ($S_{\mathrm{151\,MHz}}$) and spectral indices ($\alpha$) shown are approximate values \citep{HurleyWalker2017}, based on measurements (spanning 60--1400\,MHz) from the NASA/IPAC Extragalactic Database (NED)\protect\footnotemark. The exceptions are for *Fornax~A and *Orion~A (the Orion Nebula), with the values for Fornax A being based upon work by \citet{McKinley2015}. For Orion~A, the flux-density and spectral-index values are determined via the method described in appendix~A of \citet{White2020b}. Note that its spectral index is valid only very locally at 151\,MHz, due to the high degree of spectral curvature. }
\begin{tabular}{@{}lcccc@{}}
 \hline
Source & R.A.  & Dec. & $S_{\mathrm{151\,MHz}}$ & $\alpha$ \\
 &  (h:m:s) &  (d:m:s) & /\,Jy &  \\
 \hline
%Cassiopeia A & 23:23:28 & 58:48:42 \\ % Not below Dec = 30 deg
Centaurus A & 13:25:28 & $-$43:01:09 & 1577 & $-$0.50 \\
*Fornax A & 03:22:41 & $-$37:12:29 & 761 & $-$0.77 \\
Hercules A & 16:51:08 & +04:59:33 & 509 & $-$1.07 \\
Hydra A & 09:18:06 & $-$12:05:44 & 367 & $-$0.96 \\
*Orion A & 05:35:17 & $-$05:23:23 & 67 & $+1.1$\\
Pictor A & 05:19:50 & $-$45:46:44 & 515 & $-$0.99 \\
Taurus A & 05:34:32 & +22:00:52& 1425 & $-$0.22 \\
Virgo A & 12:30:49 & +12:23:28 & 1096 & $-$0.86 \\

\hline
\label{tab:GLEAMexclusions}
\end{tabular}
\end{table}

\section{New overlays}
\label{app:overlays}

As detailed in Section~\ref{sec:new_IDs}, the G4Jy catalogue contains new host-galaxy identifications based on radio/optical/NIR images, with improved spatial resolution {\color{black}for the radio contours. Overlays that are not already shown in the main text are provided here,} in Figure~\ref{fig:Kband_IDs}.

\setkeys{Gin}{draft=false}

\begin{figure*}
%\vspace{-0.7cm}
\centering

\subfigure[G4Jy~6 ($K$-band image from VIKING DR2; \citealt{Edge2016})]{
	\includegraphics[width=0.4\linewidth]{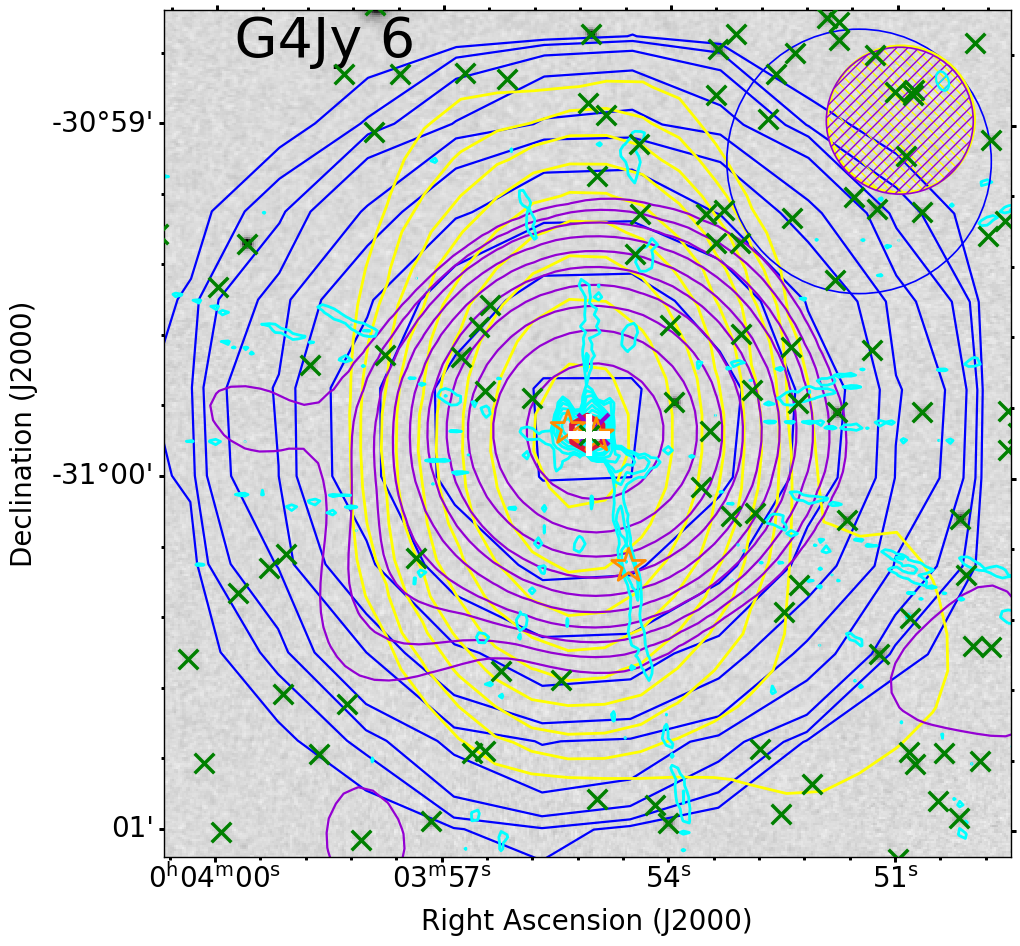}
	}
\subfigure[G4Jy~39]{
	\includegraphics[width=0.4\linewidth]{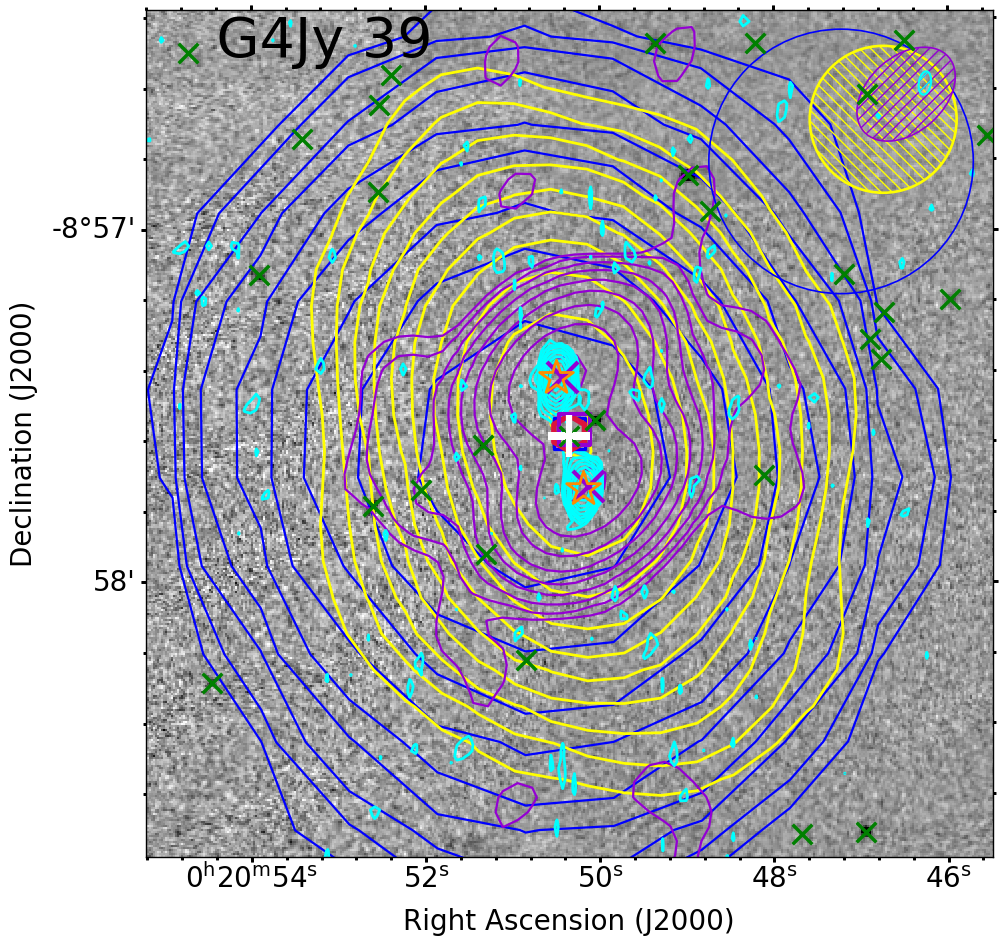} 
	}  \\[-0.05cm]
\subfigure[G4Jy~117 (MIR image from AllWISE)]{
	\includegraphics[width=0.4\linewidth]{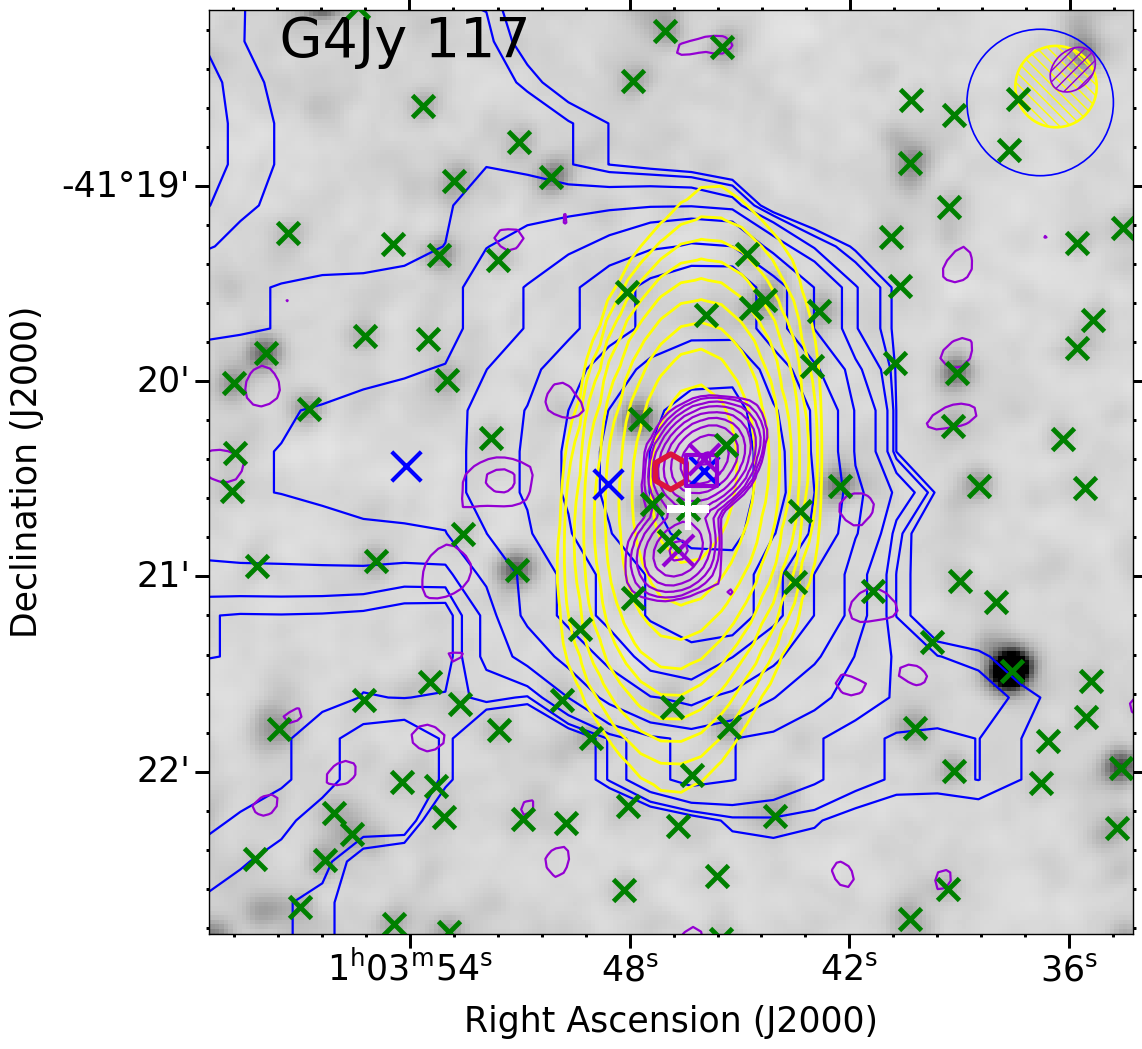} 
	} 
\subfigure[G4Jy~225 ($g$-band image from PanSTARRS)]{
	\includegraphics[width=0.4\linewidth]{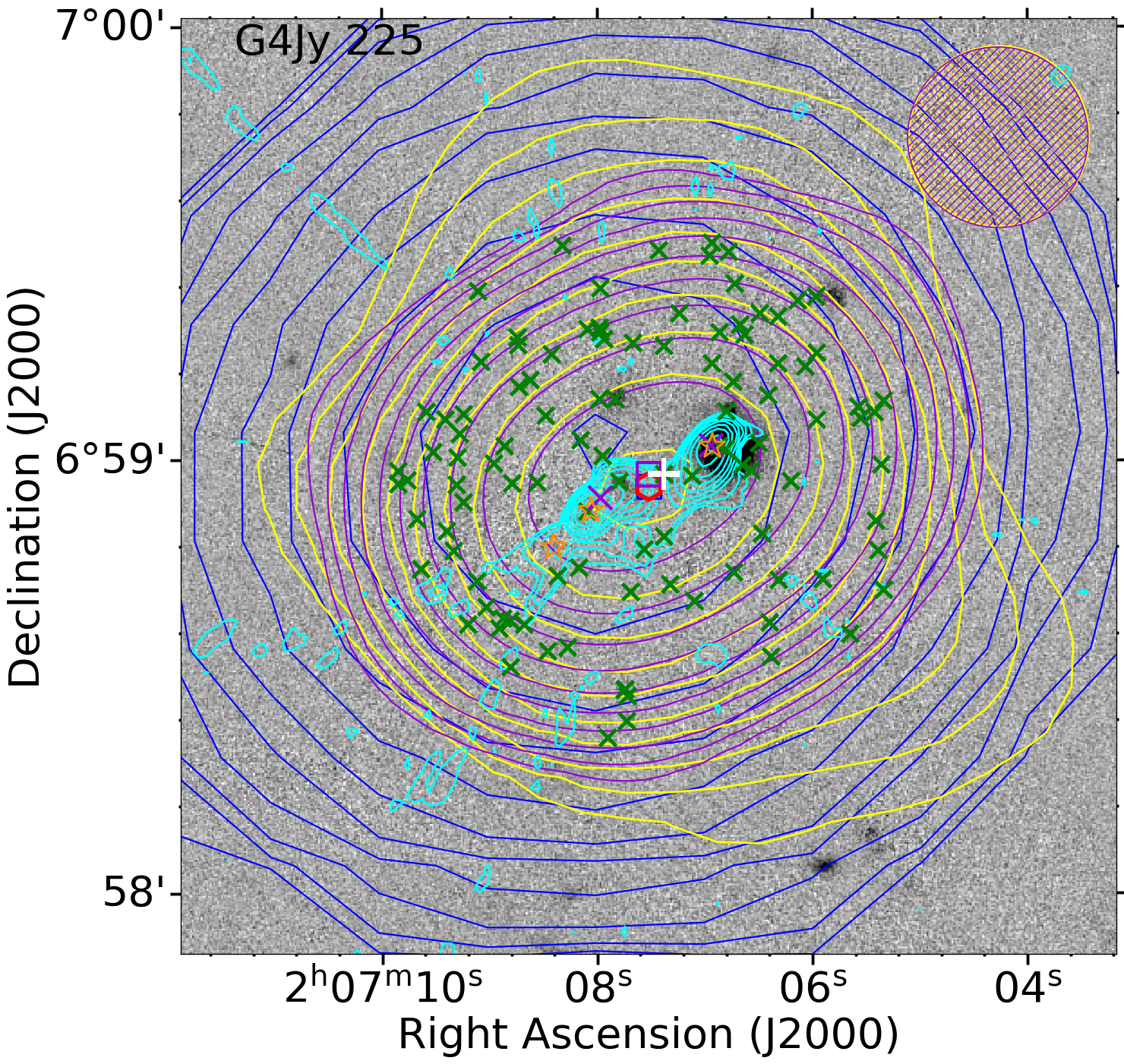} 
	}  \\[-0.05cm]
 \subfigure[G4Jy~446]{
	\includegraphics[width=0.4\linewidth]{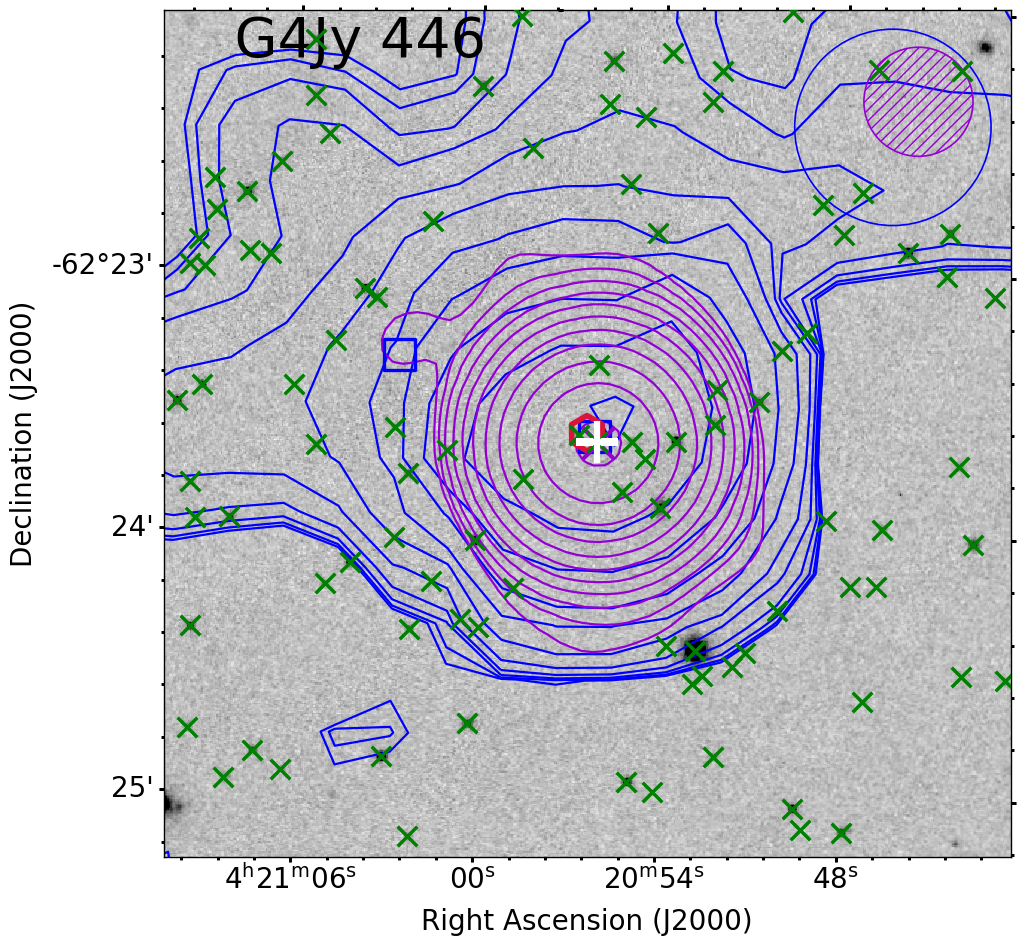} 
	} 
     \subfigure[G4Jy~471]{
	\includegraphics[width=0.4\linewidth]{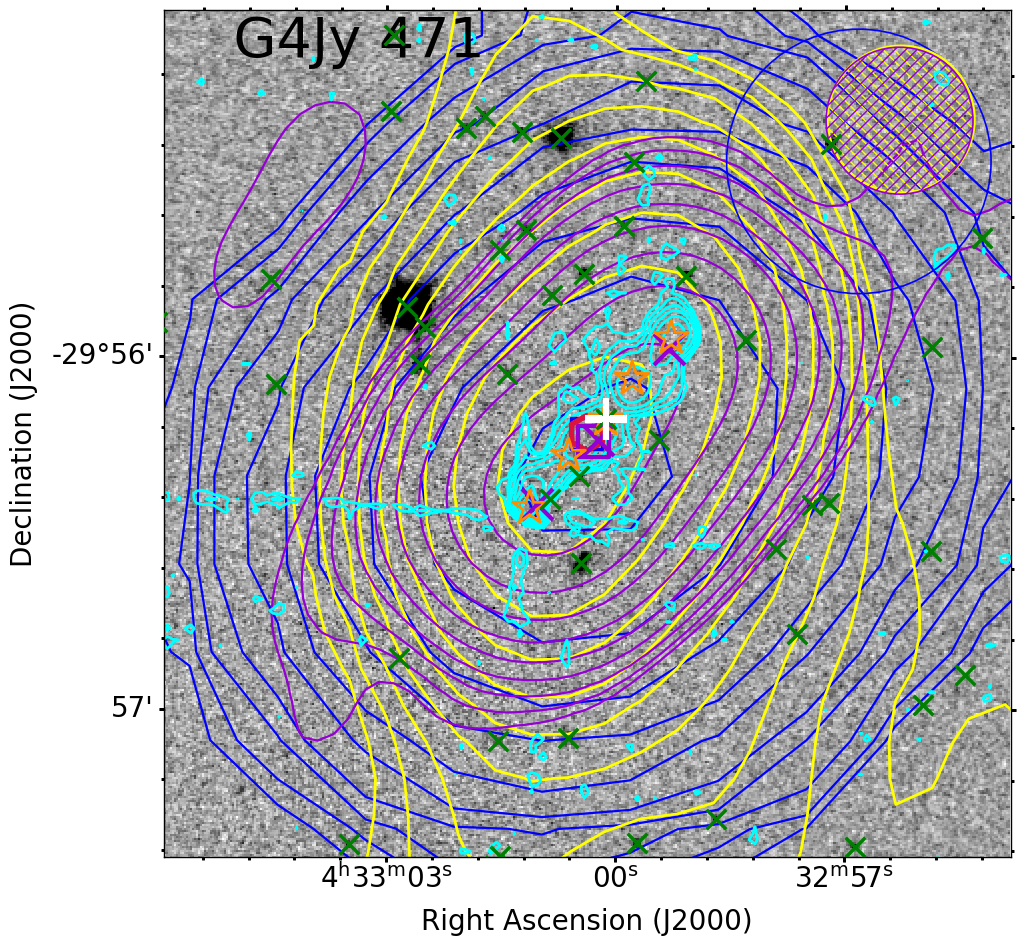} 
	} 

\caption{ Overlays of radio contours [and catalogue positions] from SUMSS/NVSS (blue), RACS-low1 (purple), and TGSS (yellow) on $K$-band images from VHS (unless otherwise stated). A red hexagon indicates the radio-brightness-weighted centroid calculated by \citet{White2020b}. Also plotted, where available, are cyan contours based on radio images from either MeerKAT-2019 follow-up \citep{Sejake2023} or VLASS, with their peak-brightness positions indicated by orange stars. The relative synthesised-beam sizes are shown in the upper-right corner of each panel. Green crosses (`$\times$') indicate positions from the catalogue corresponding to the inverted grey-scale image, and a white `+' marks the position of the galaxy hosting the radio emission. \label{fig:Kband_IDs}}
\end{figure*}

\setcounter{figure}{0}

\begin{figure*}
%\vspace{-0.7cm}
\centering
\subfigure[G4Jy~565]{
	\includegraphics[width=0.4\linewidth]{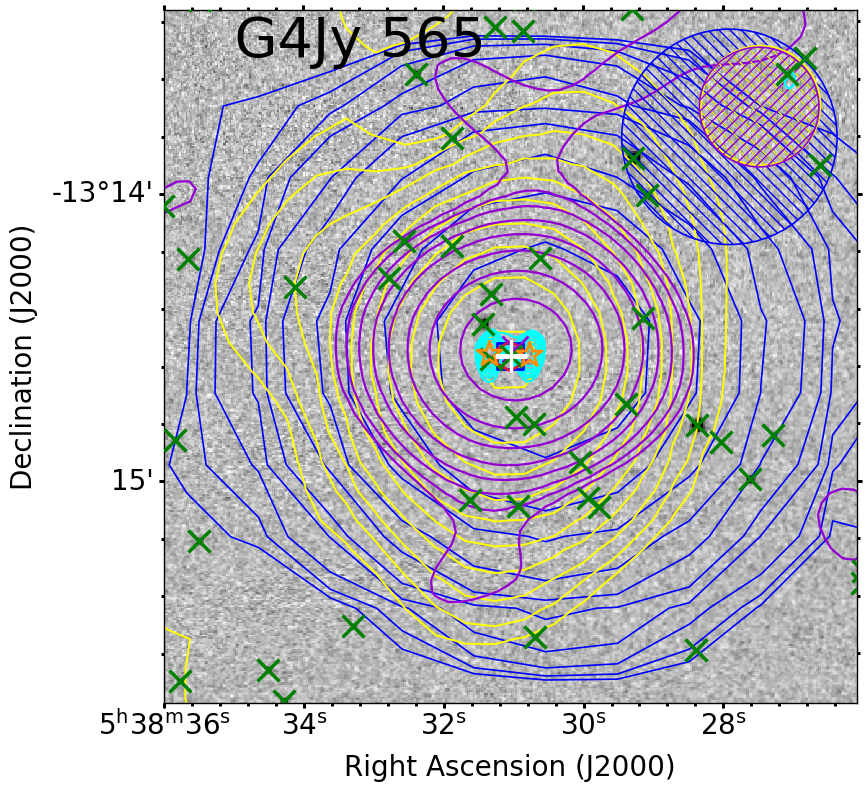} 
	}  
\subfigure[G4Jy~593 ]{
	\includegraphics[width=0.4\linewidth]{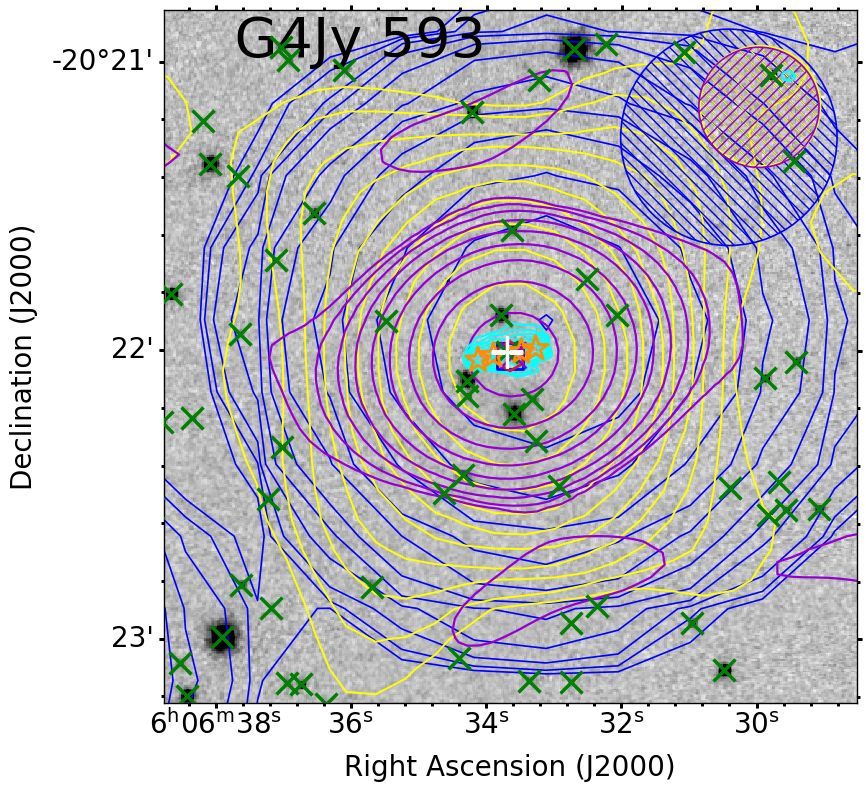} 
	} \\[-0.05cm]
\subfigure[G4Jy~752 ]{
	\includegraphics[width=0.4\linewidth]{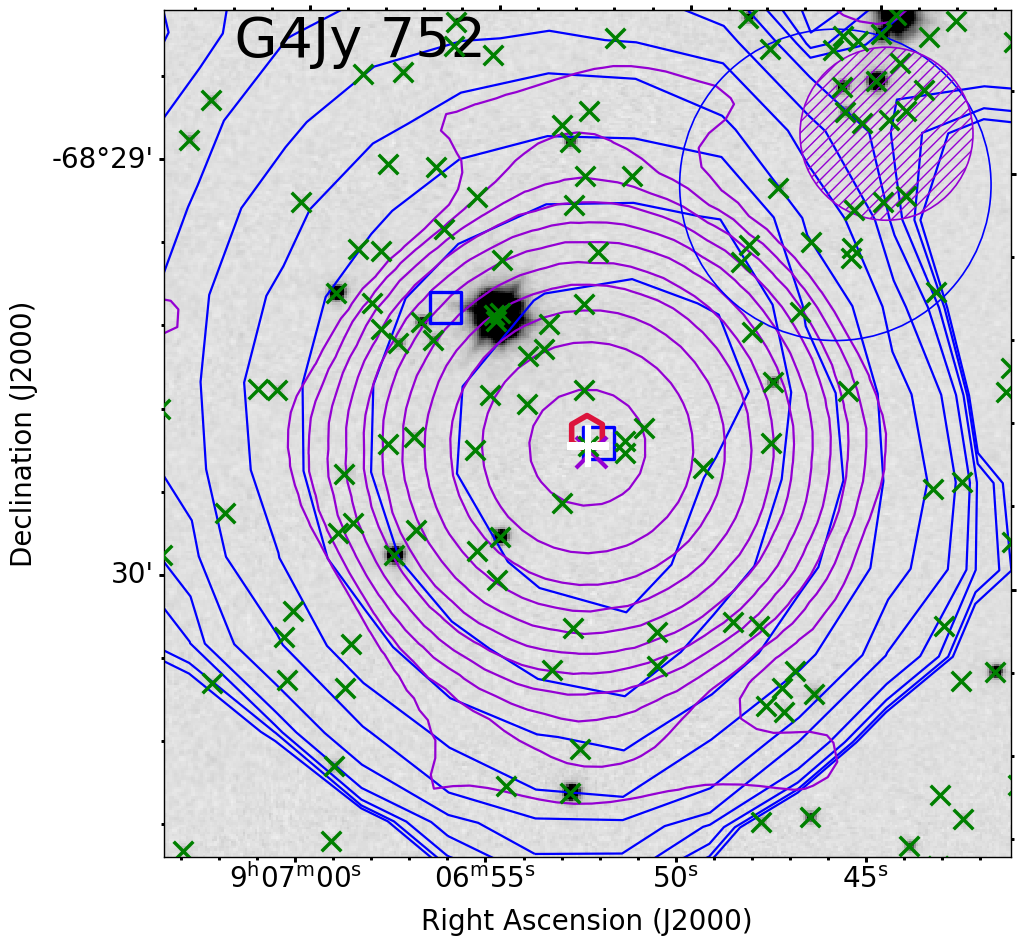} 
	}  
 \subfigure[G4Jy~759]{
	\includegraphics[width=0.4\linewidth]{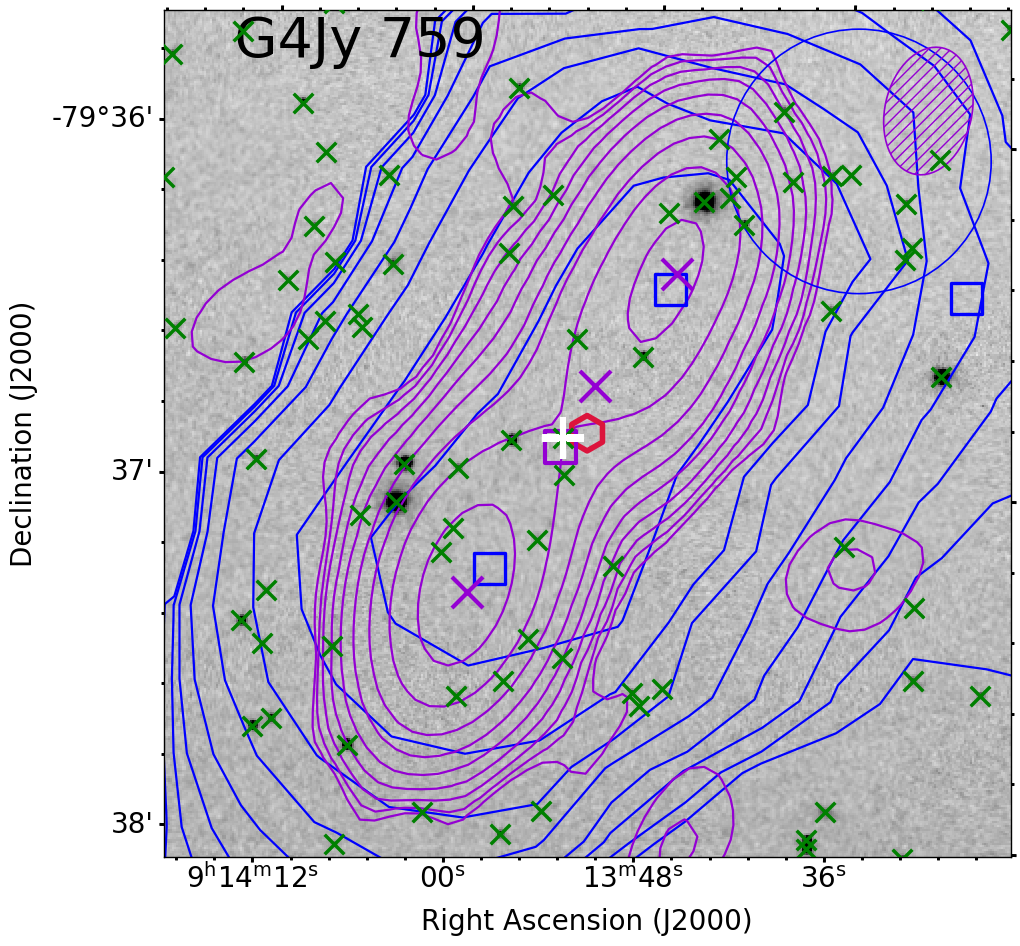} 
	} \\[-0.05cm]
\subfigure[G4Jy~961]{
	\includegraphics[width=0.4\linewidth]{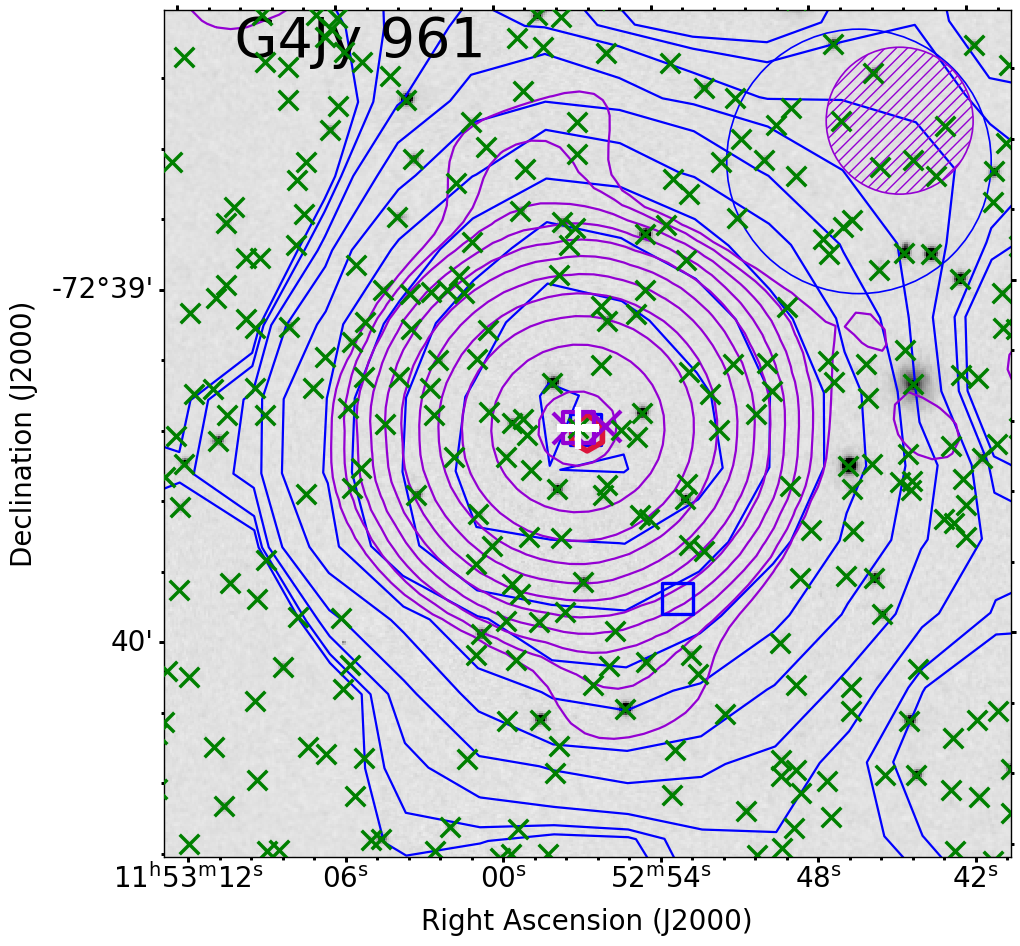}
	}
    \subfigure[G4Jy~1116 ]{
	\includegraphics[width=0.4\linewidth]{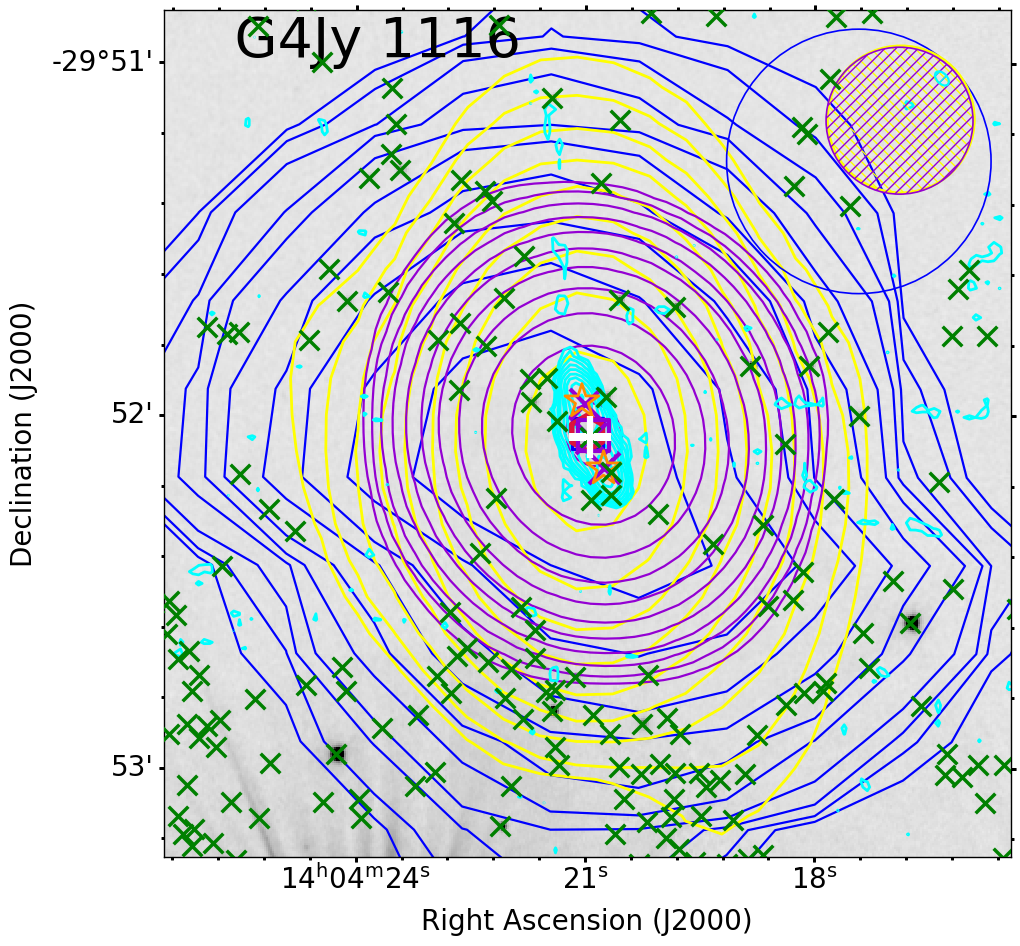} 
	} 

\caption{ {\it Continued} -- Overlays of radio contours [and catalogue positions] from SUMSS/NVSS (blue), RACS-low1 (purple), TGSS (yellow) and MeerKAT/VLASS (cyan) on inverted grey-scale $K$-band images from VHS (unless otherwise stated). A white `+' marks the position of the galaxy hosting the radio emission.}
\end{figure*}

\setcounter{figure}{0}

\begin{figure*}
%\vspace{-0.7cm}
\centering

\subfigure[G4Jy~1121 ]{
	\includegraphics[width=0.4\linewidth]{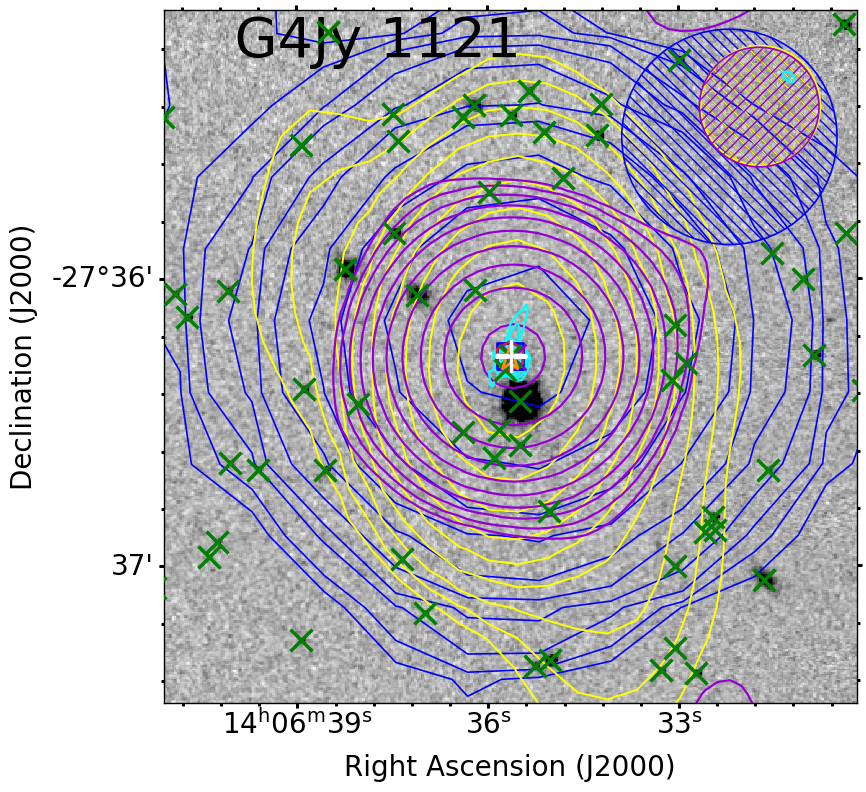} 
	}  
 \subfigure[G4Jy~1134 ($g$-band image from PanSTARRS)]{
	\includegraphics[width=0.41\linewidth]{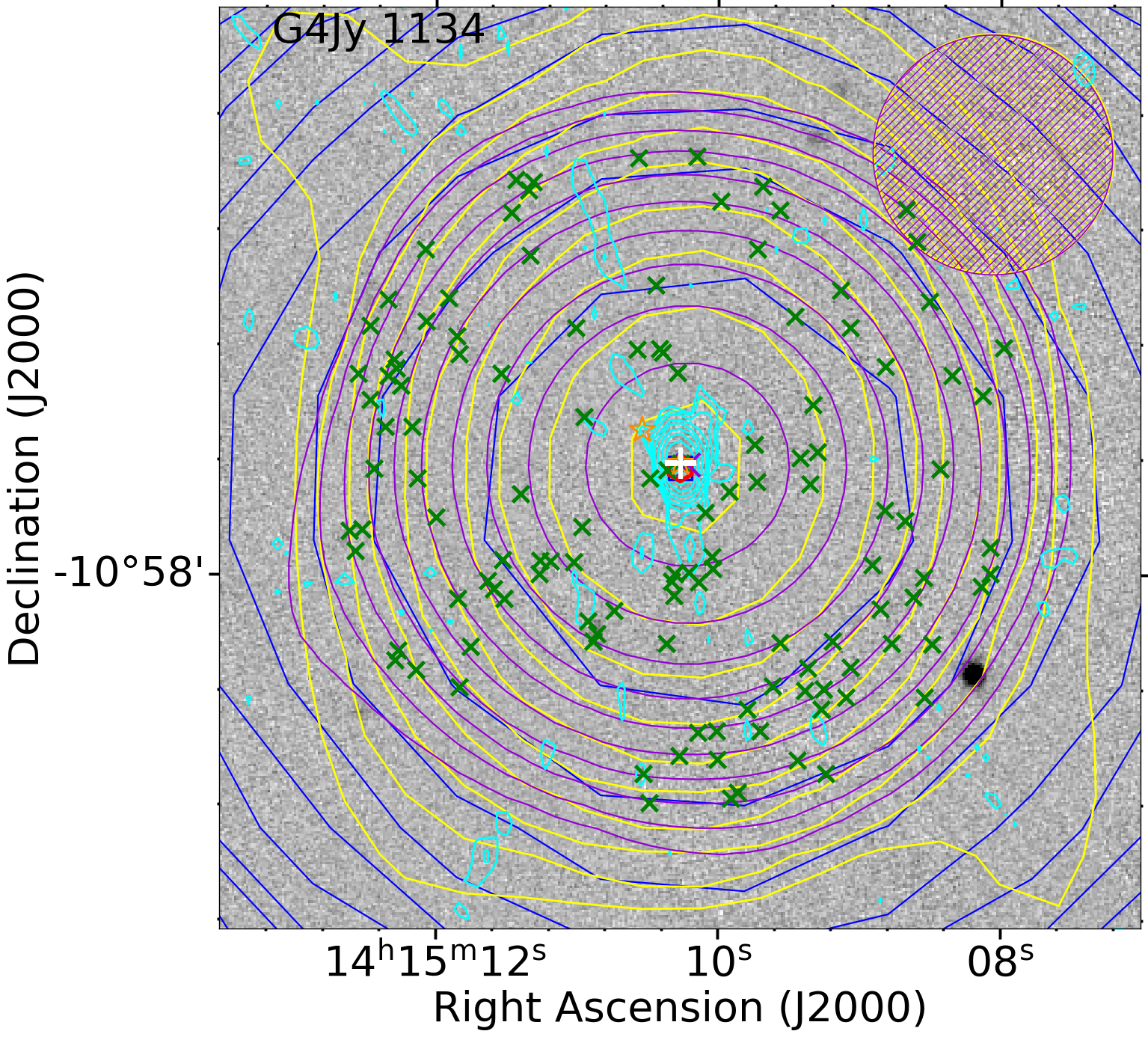} 
	} \\[-0.05cm]
     \subfigure[G4Jy~1159]{
	\includegraphics[width=0.4\linewidth]{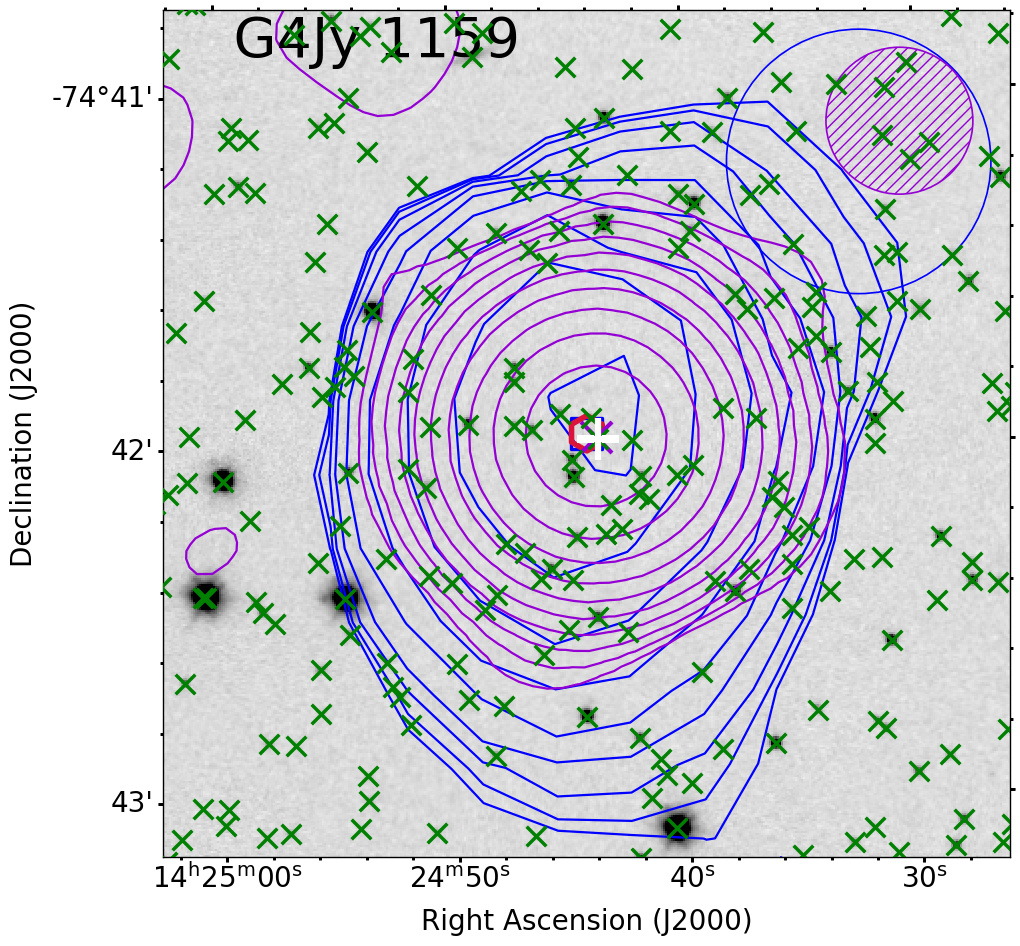} 
	} 
    \subfigure[G4Jy~1183 (MIR image from AllWISE)]{
	\includegraphics[width=0.4\linewidth]{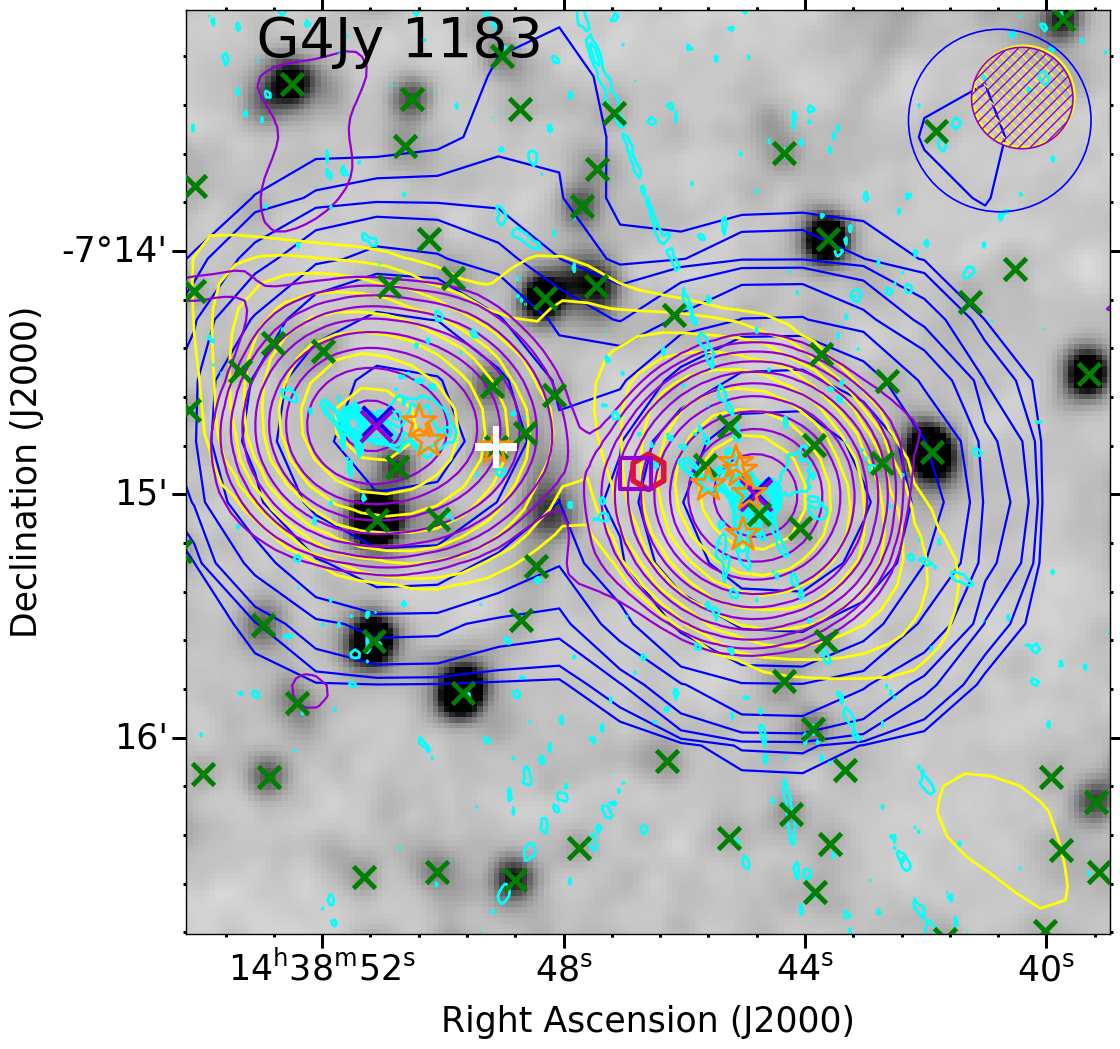}
	}\\[-0.05cm]
\subfigure[G4Jy~1229 ]{
	\includegraphics[width=0.4\linewidth]{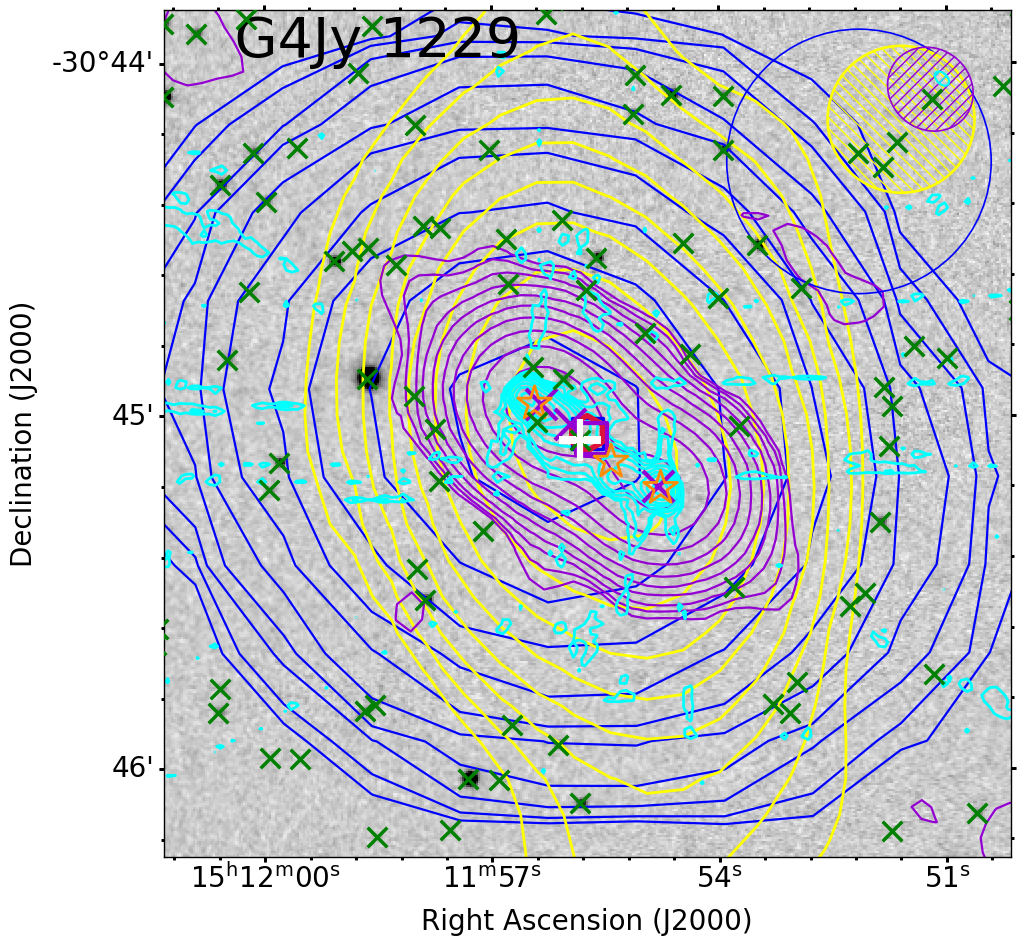} 
	}  
\subfigure[G4Jy~1241]{
	\includegraphics[width=0.4\linewidth]{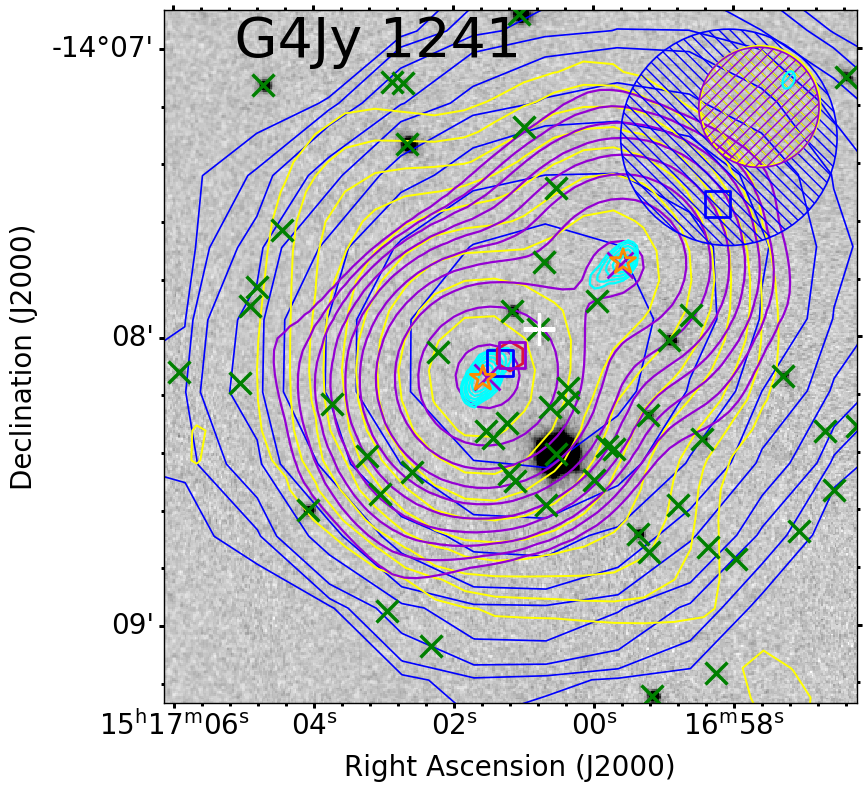} 
	} 

\caption{ {\it Continued} -- Overlays of radio contours [and catalogue positions] from SUMSS/NVSS (blue), RACS-low1 (purple), TGSS (yellow) and MeerKAT/VLASS (cyan) on inverted grey-scale $K$-band images from VHS (unless otherwise stated). A white `+' marks the position of the galaxy hosting the radio emission.}
\end{figure*}

\setcounter{figure}{0}

\begin{figure*}
%\vspace{-0.7cm}
\centering

\subfigure[G4Jy~1256]{
	\includegraphics[width=0.4\linewidth]{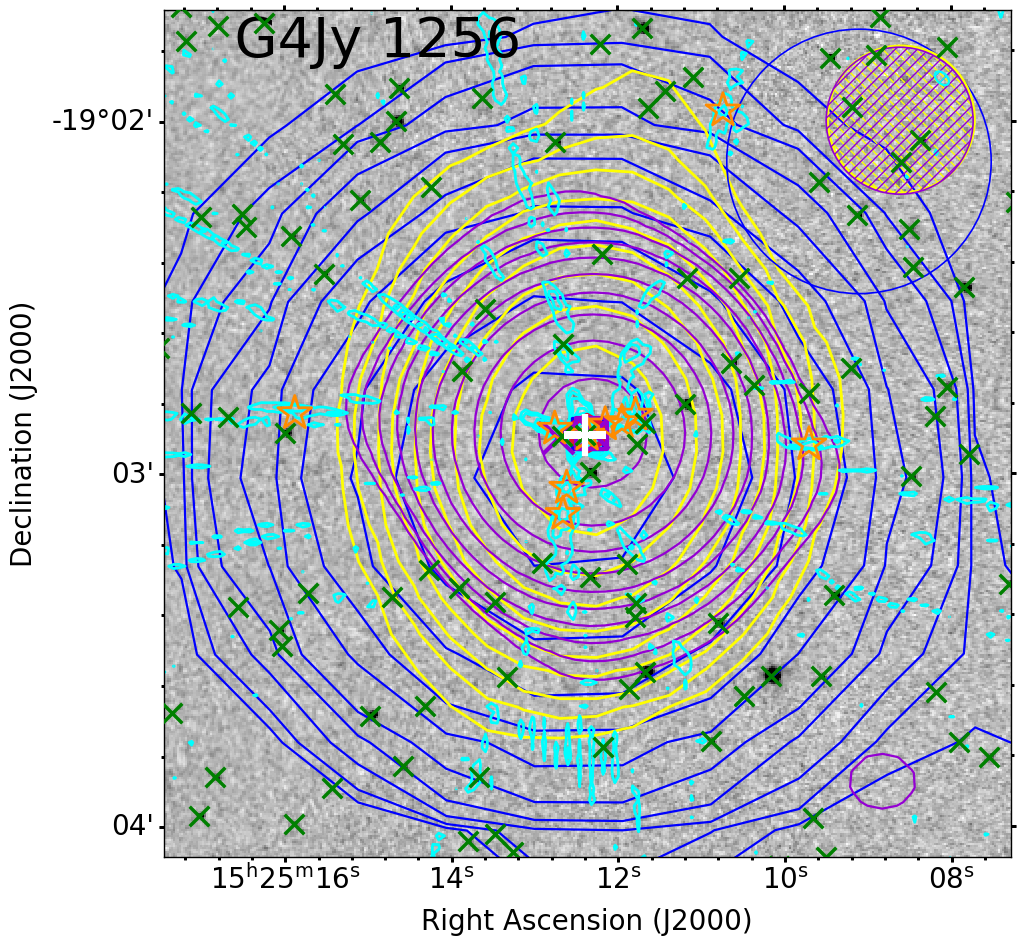} 
	}  
 \subfigure[G4Jy~1292]{
	\includegraphics[width=0.4\linewidth]{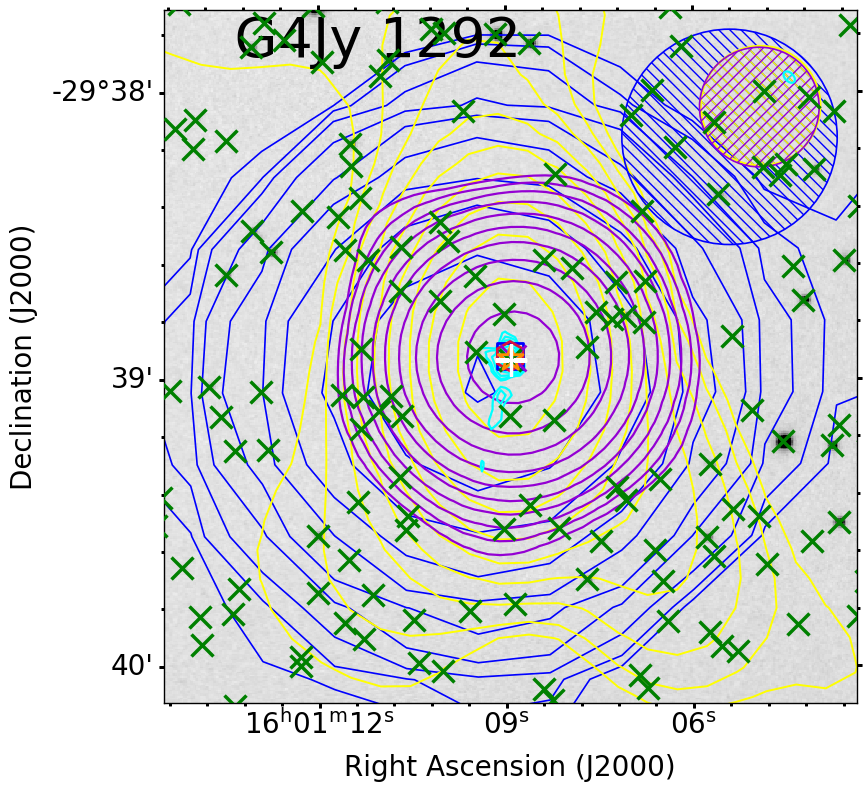} 
	} \\[-0.05cm]
     \subfigure[G4Jy~1295]{
	\includegraphics[width=0.4\linewidth]{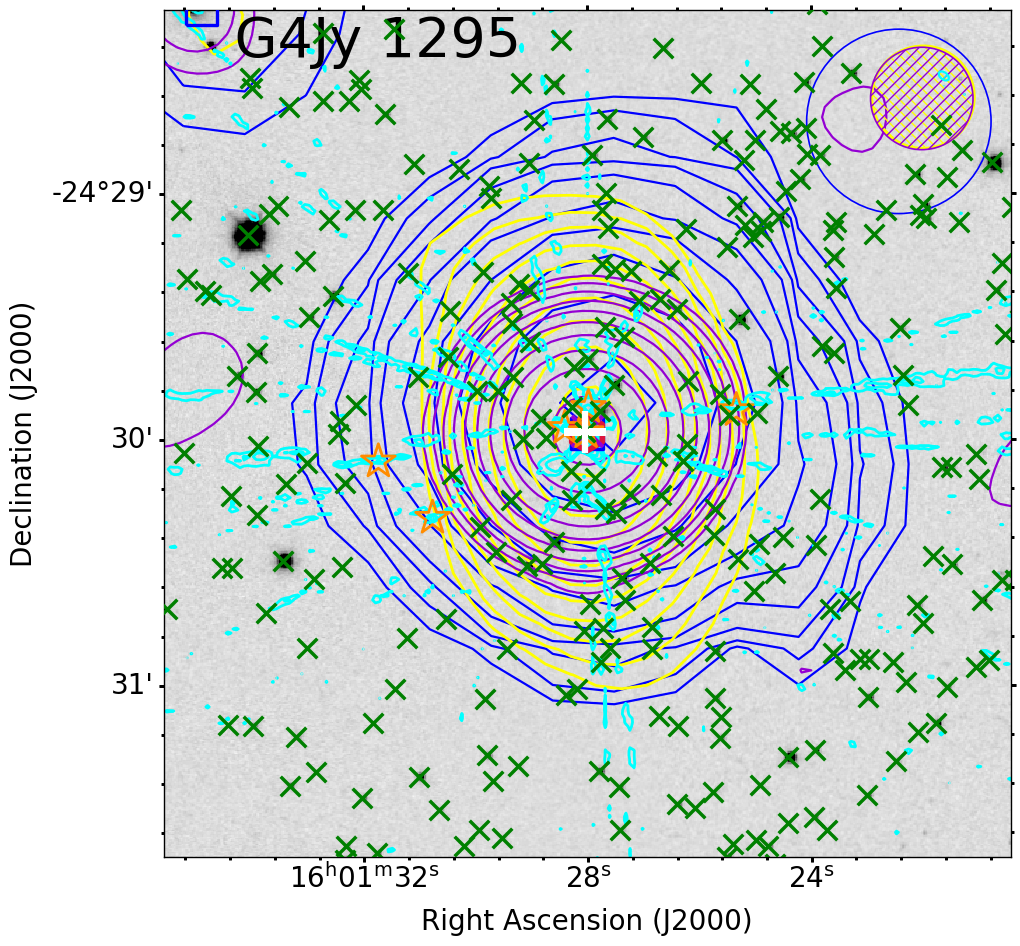} 
	} 
    \subfigure[G4Jy~1301 ($H$-band image from UKIDSS; \citealt{Lawrence2013})]{
	\includegraphics[width=0.425\linewidth]{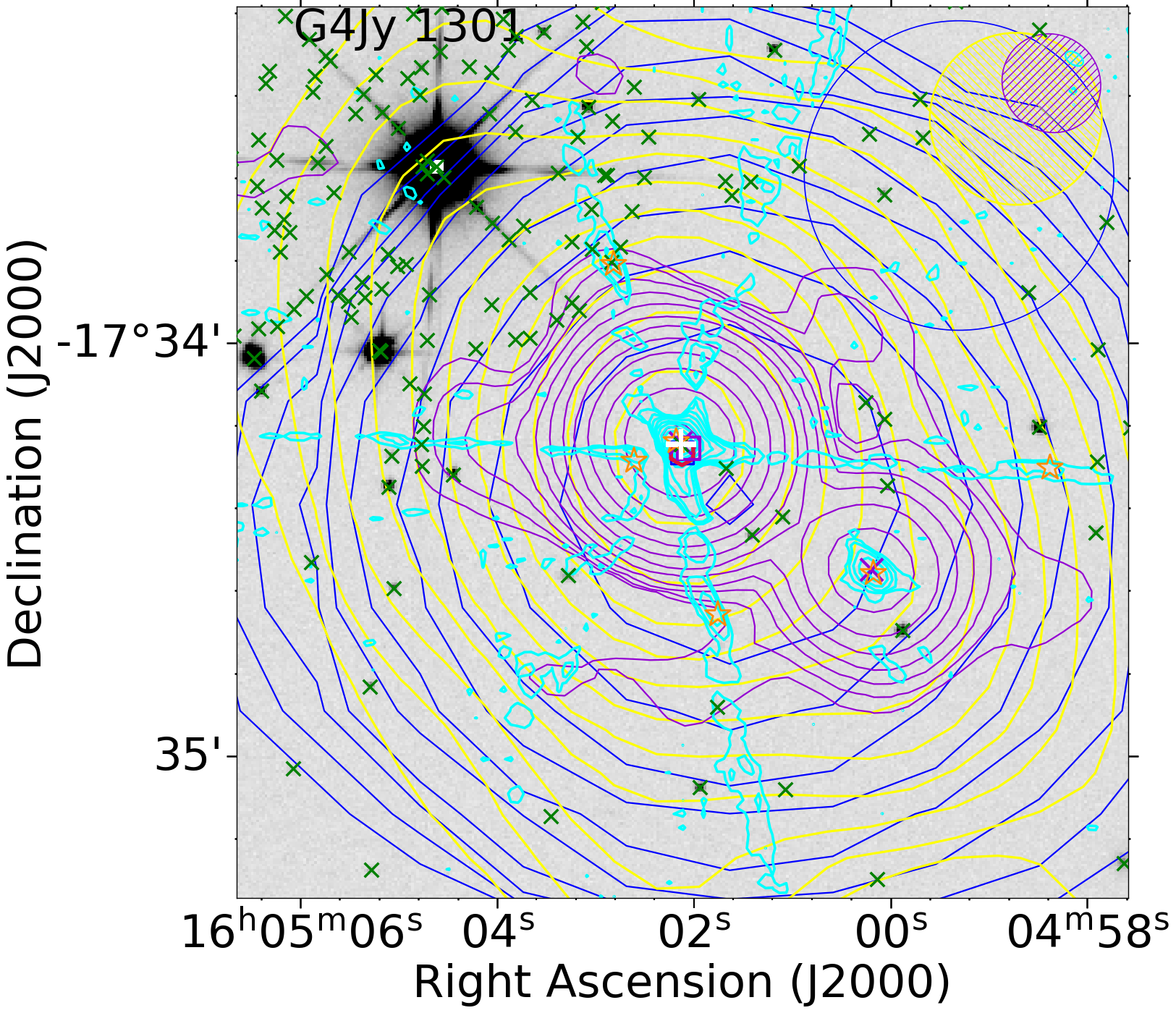}
	} \\[-0.05cm]
\subfigure[G4Jy~1305 ]{
	\includegraphics[width=0.4\linewidth]{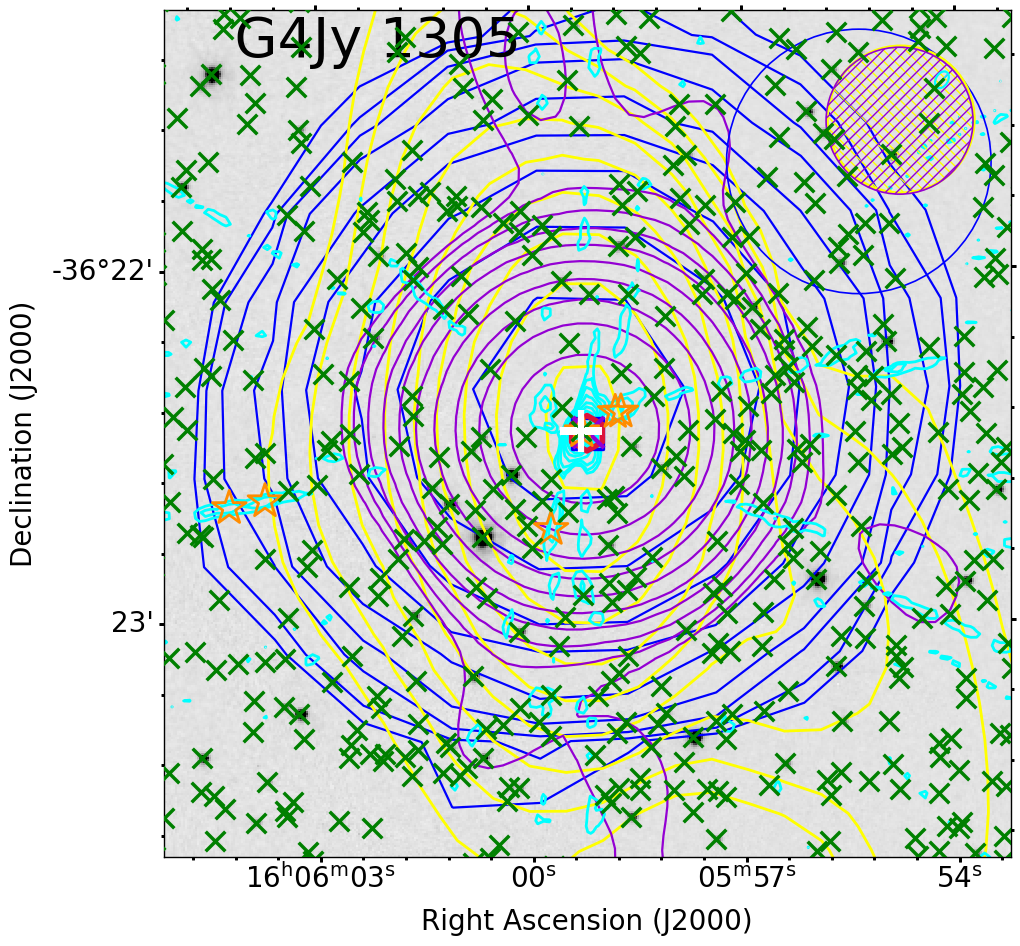} 
	} 
\subfigure[G4Jy~1310]{
	\includegraphics[width=0.4\linewidth]{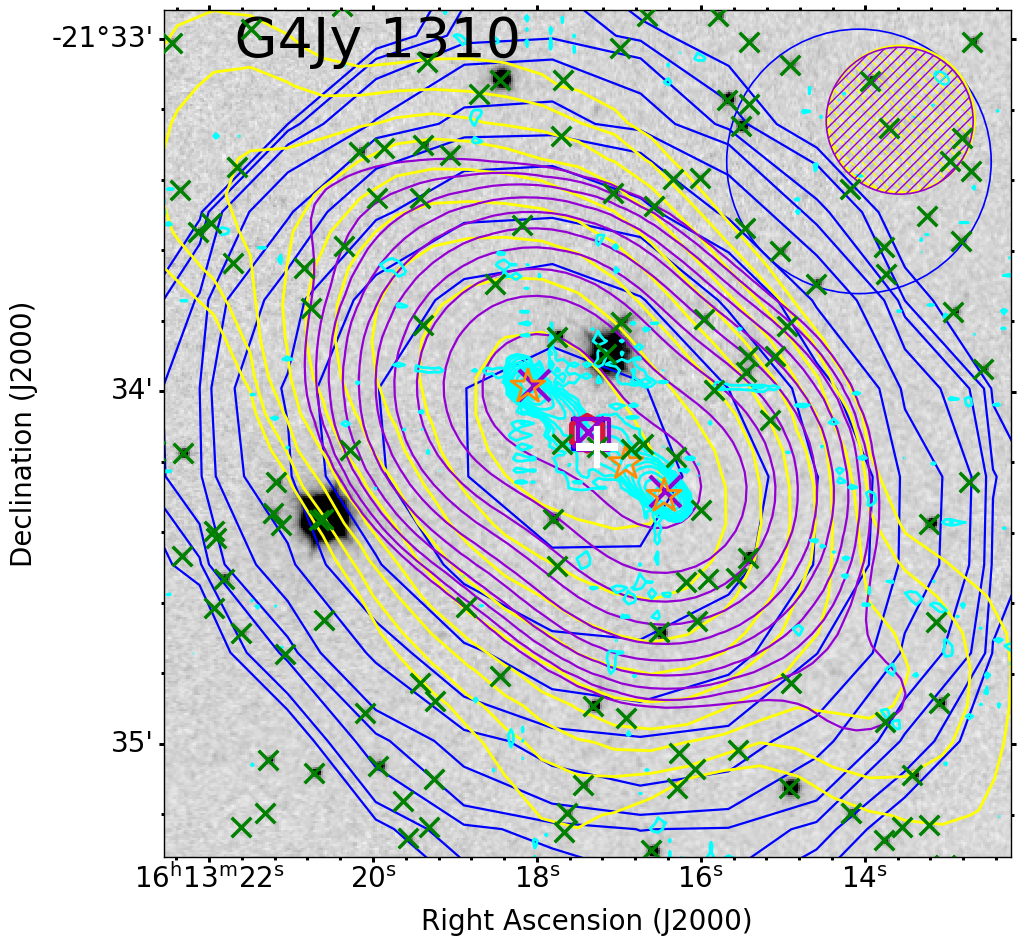} 
	}

\caption{ {\it Continued} -- Overlays of radio contours [and catalogue positions] from SUMSS/NVSS (blue), RACS-low1 (purple), TGSS (yellow) and MeerKAT/VLASS (cyan) on inverted grey-scale $K$-band images from VHS (unless otherwise stated). A white `+' marks the position of the galaxy hosting the radio emission.}
\end{figure*}

\setcounter{figure}{0}

\begin{figure*}
%\vspace{-0.7cm}
\centering

 \subfigure[G4Jy~1329]{
	\includegraphics[width=0.4\linewidth]{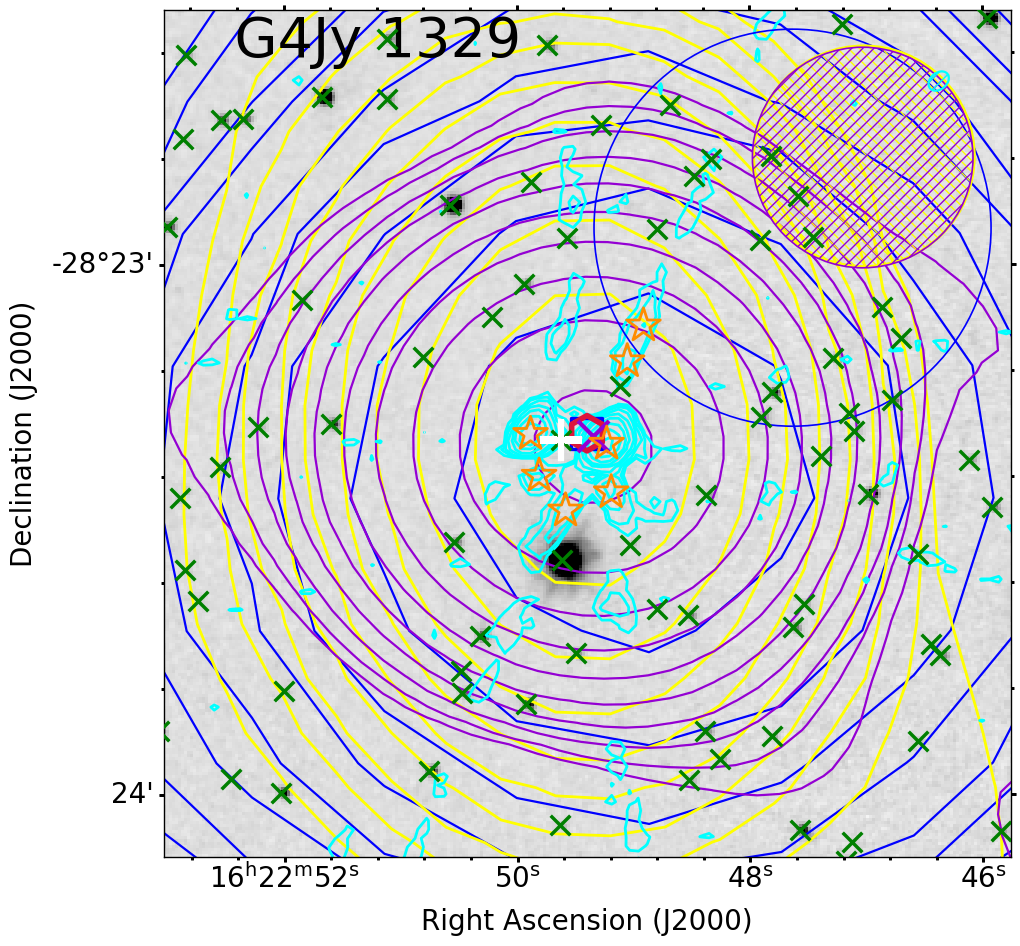} 
	} 
     \subfigure[G4Jy~1330]{
	\includegraphics[width=0.4\linewidth]{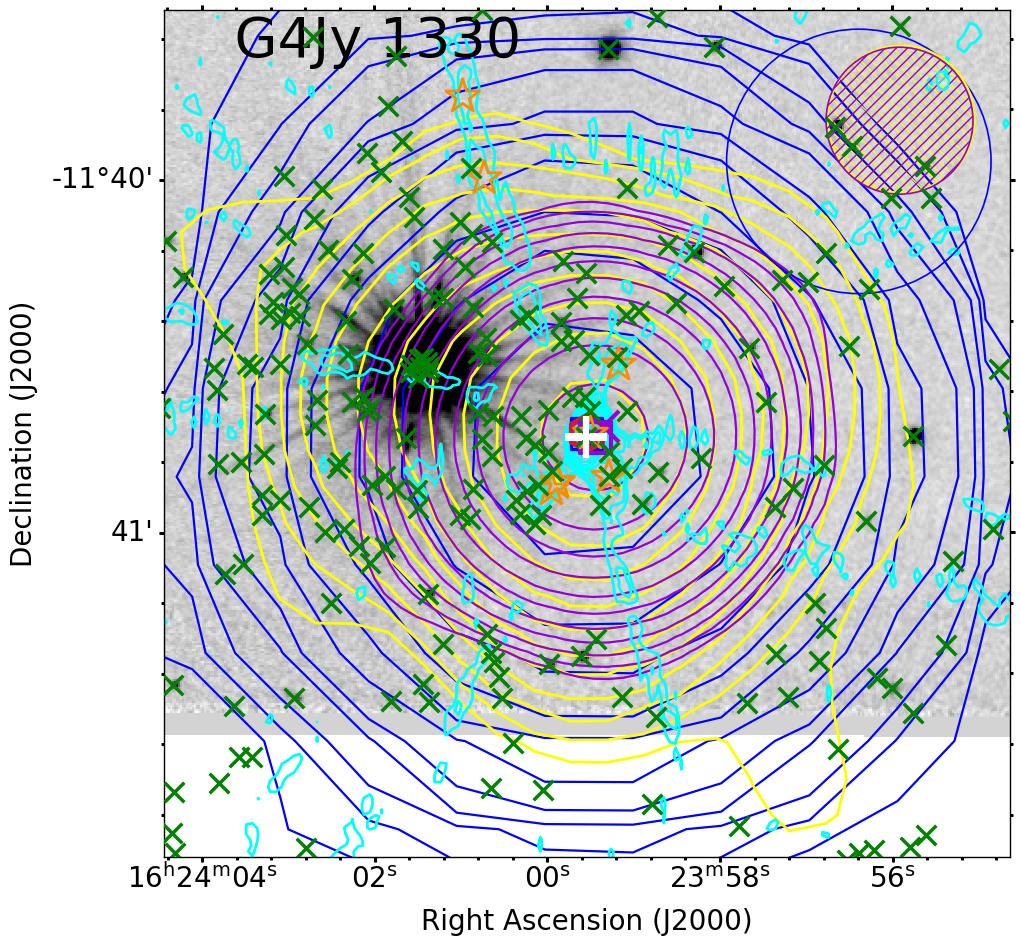} 
	} \\[-0.05cm]
    \subfigure[G4Jy~1390 (MIR image from AllWISE)]{
	\includegraphics[width=0.42\linewidth]{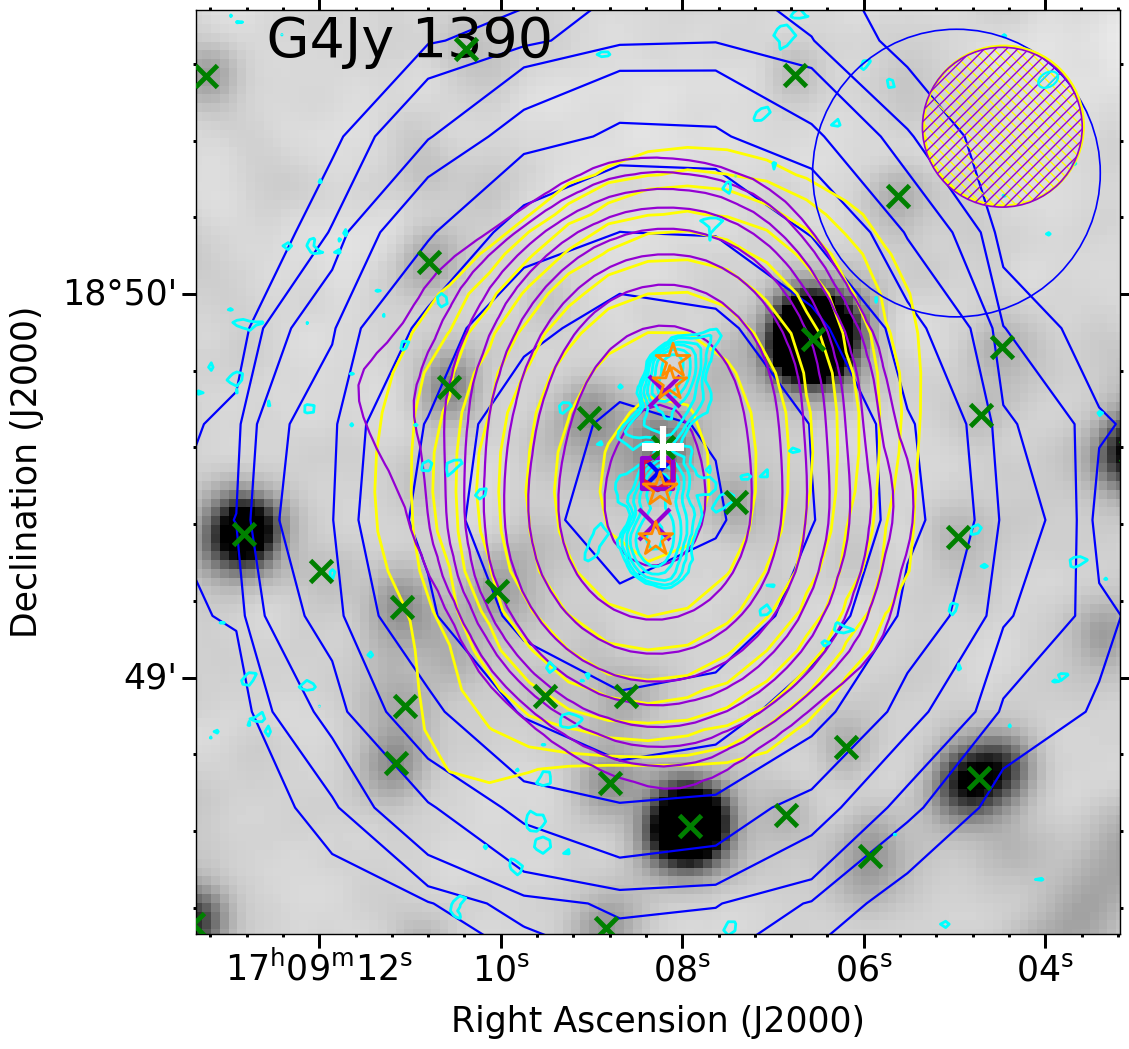}
	}
\subfigure[G4Jy~1418 (MIR image from AllWISE)]{
	\includegraphics[width=0.42\linewidth]{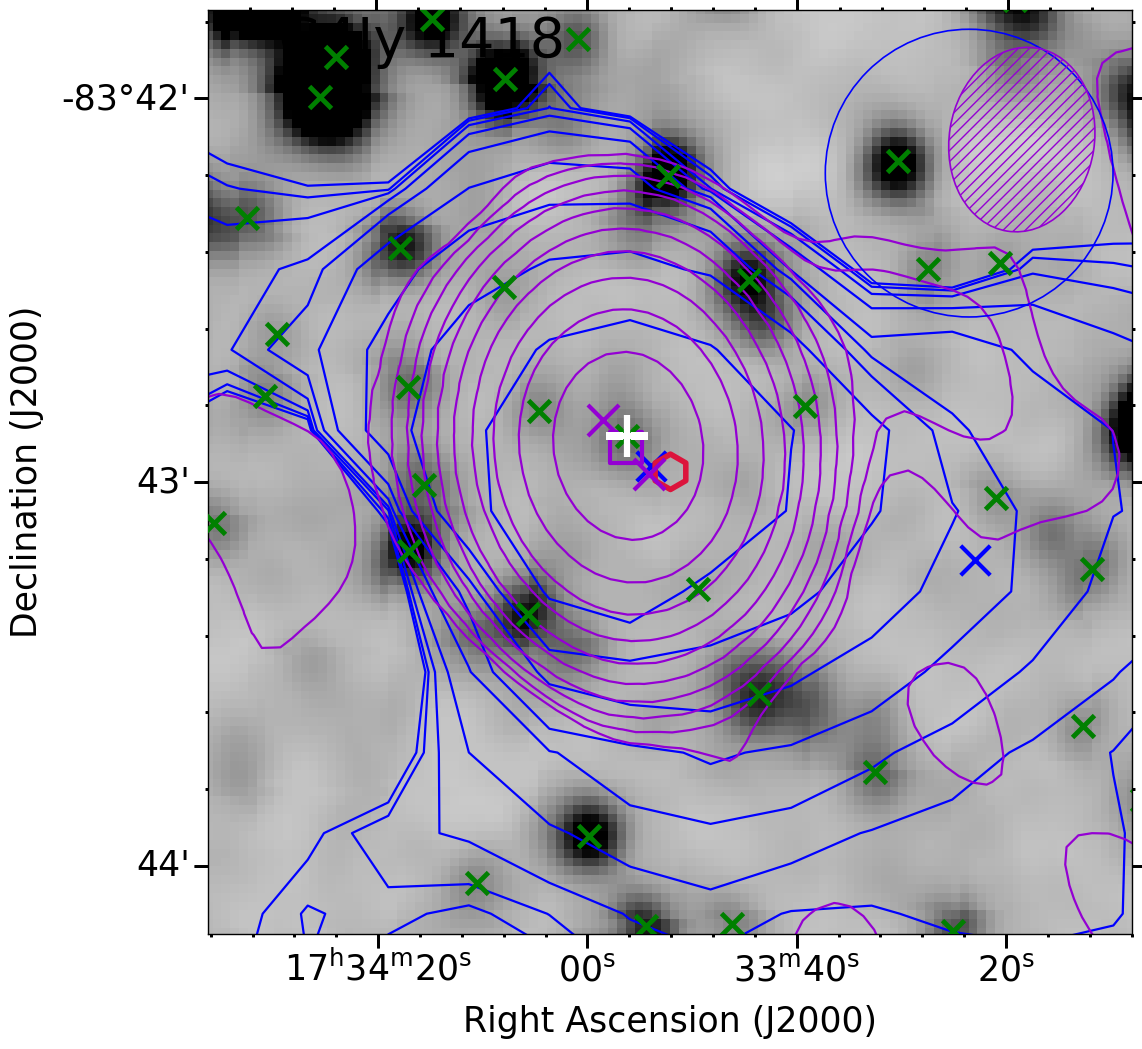}  
	}\\[-0.05cm]
    \subfigure[G4Jy~1441 ($g$-band image from PanSTARRS)]{
	\includegraphics[width=0.42\linewidth]{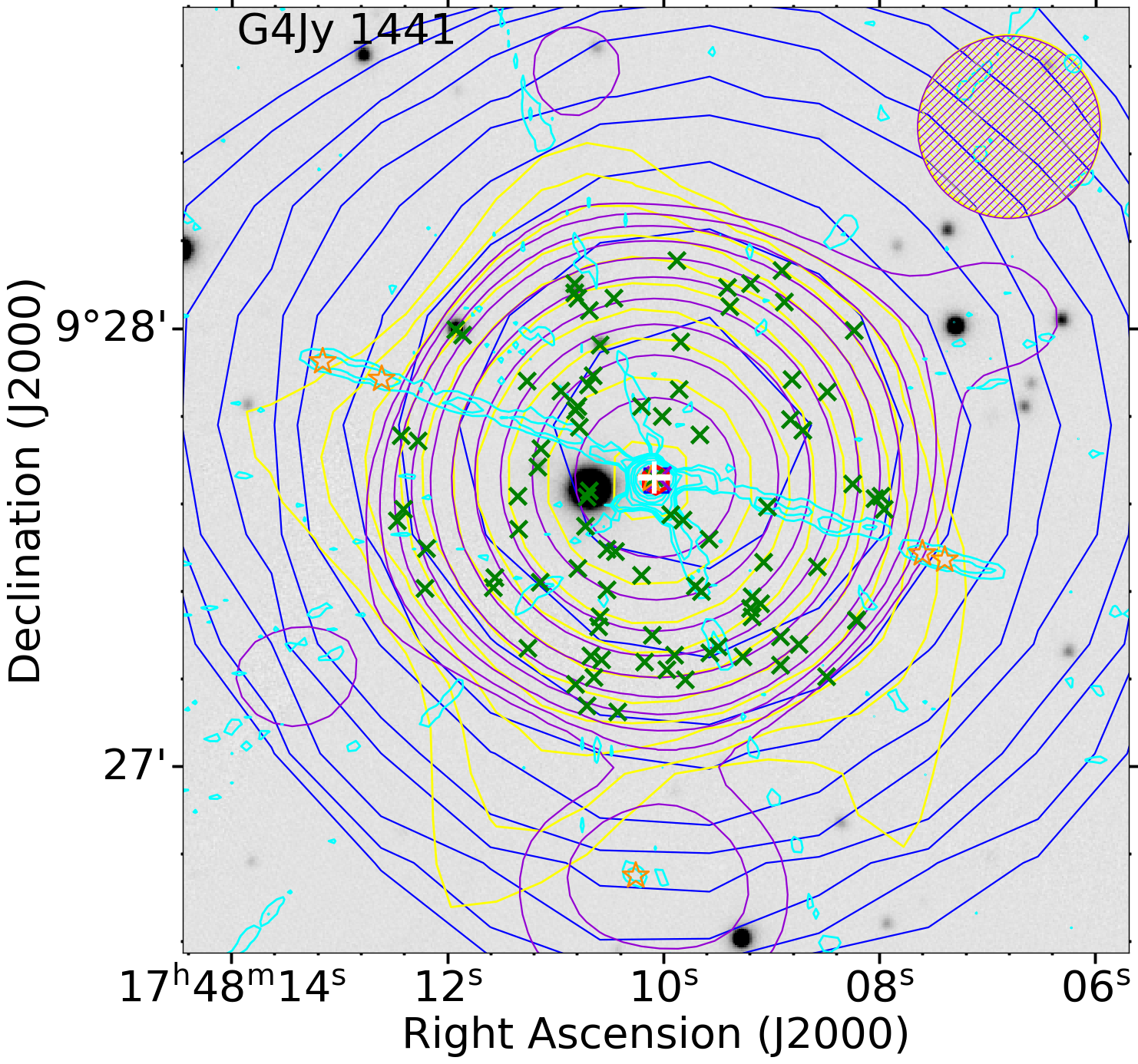} 
	}  
     \subfigure[G4Jy~1449 ($g$-band image from PanSTARRS)]{
	\includegraphics[width=0.42\linewidth]{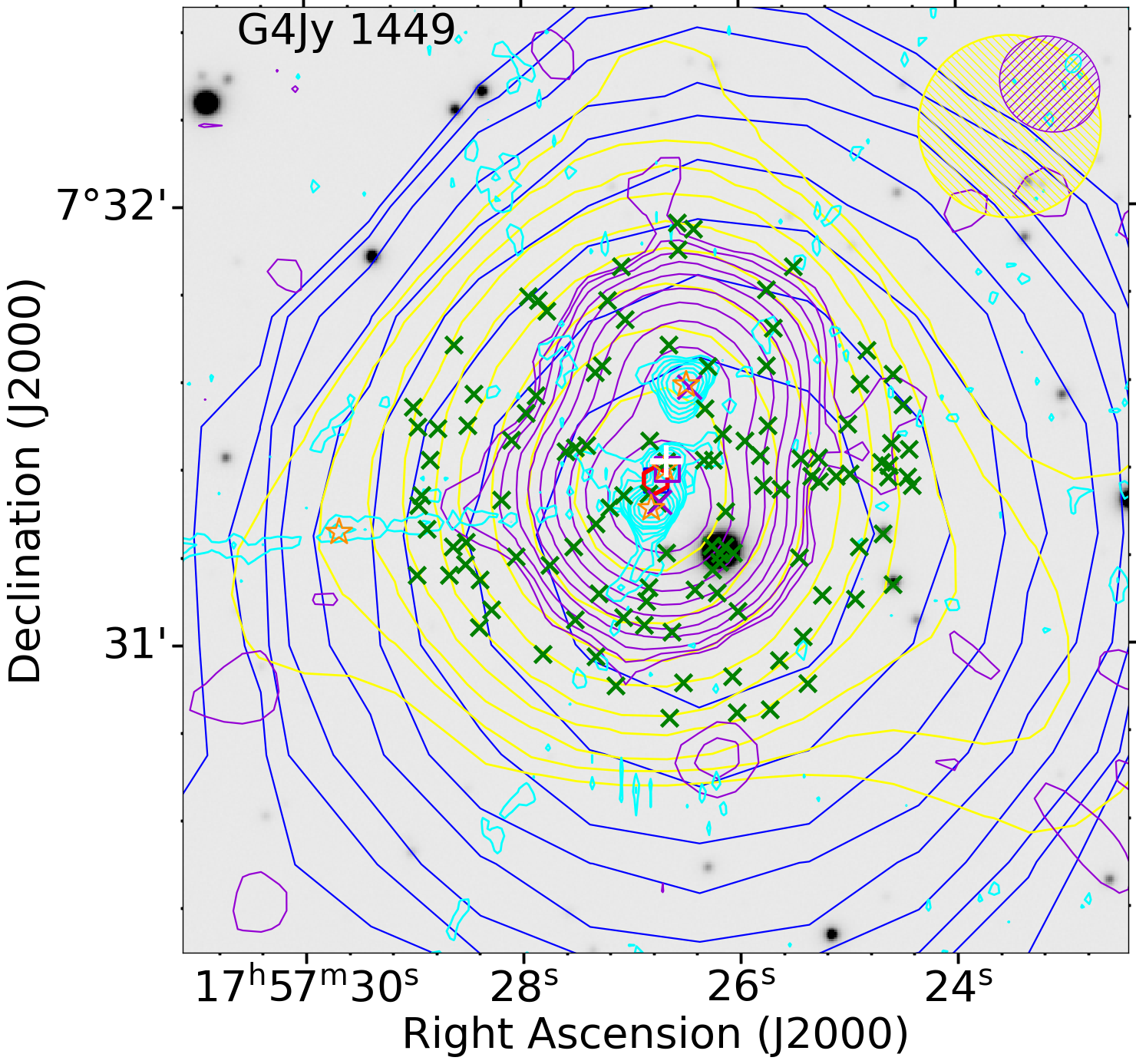} 
	} 

\caption{ {\it Continued} -- Overlays of radio contours [and catalogue positions] from SUMSS/NVSS (blue), RACS-low1 (purple), TGSS (yellow) and MeerKAT/VLASS (cyan) on inverted grey-scale $K$-band images from VHS (unless otherwise stated). A white `+' marks the position of the galaxy hosting the radio emission.}
\end{figure*}

\setcounter{figure}{0}

\begin{figure*}
%\vspace{-0.7cm}
\centering

\subfigure[G4Jy~1458]{
	\includegraphics[width=0.41\linewidth]{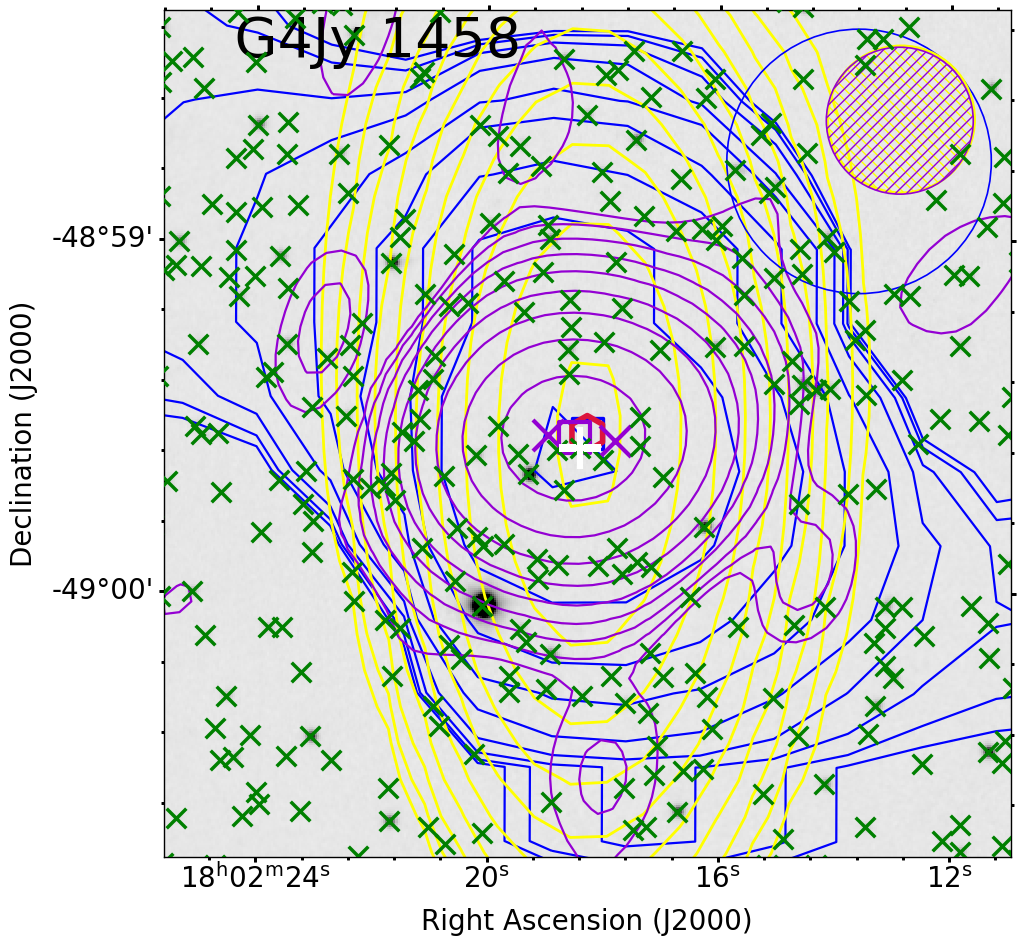}
	}
\subfigure[G4Jy~1464 ]{
	\includegraphics[width=0.41\linewidth]{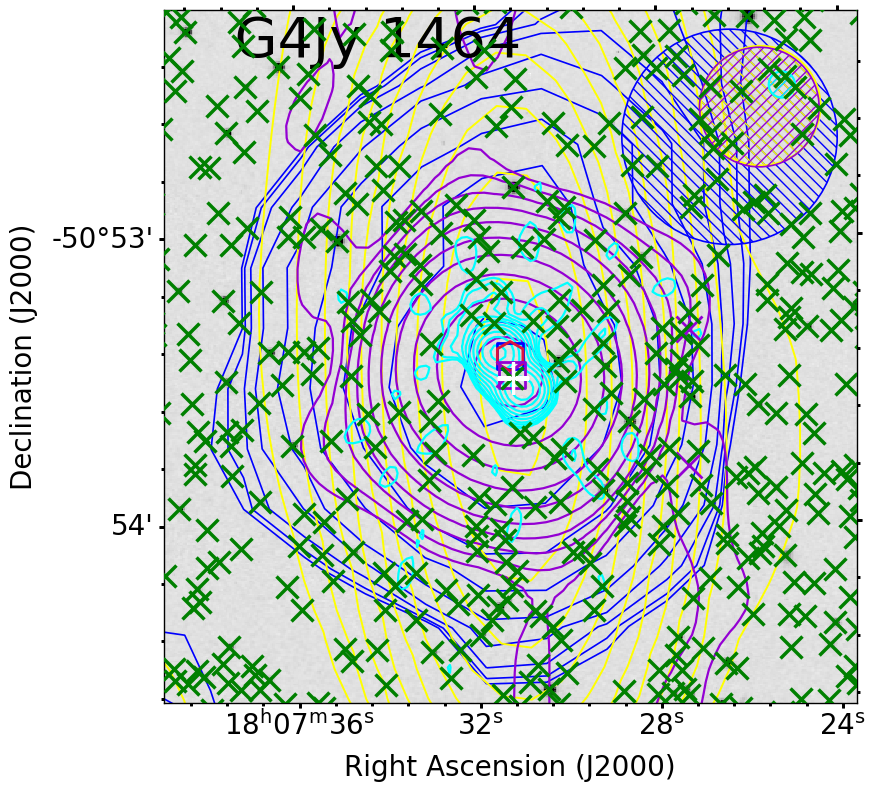} 
	}  \\[-0.05cm]
\subfigure[G4Jy~1465 ($g$-band image from PanSTARRS)]{
	\includegraphics[width=0.41\linewidth]{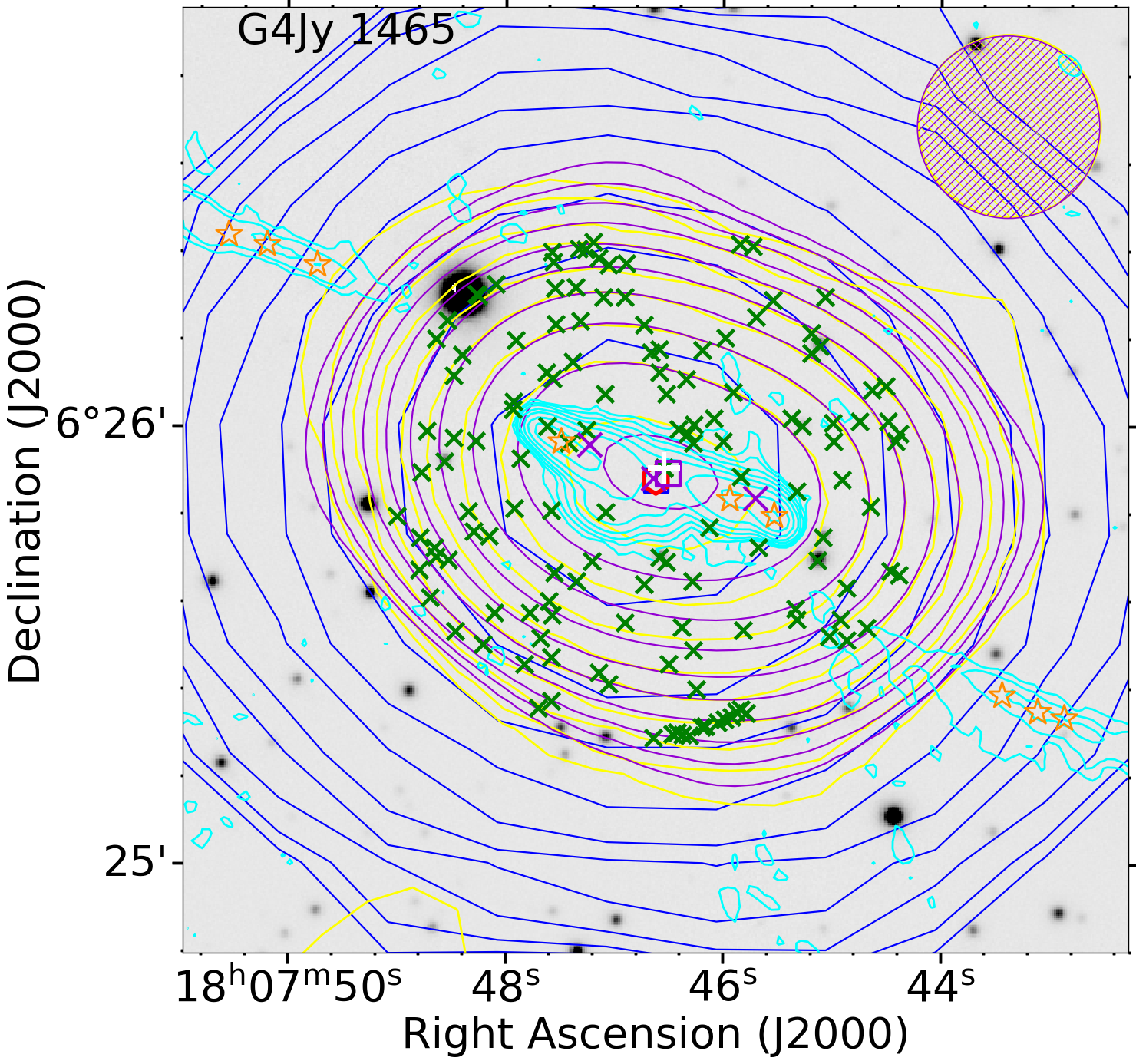} 
	} 
\subfigure[G4Jy~1513 ($g$-band image from PanSTARRS)]{
	\includegraphics[width=0.42\linewidth]{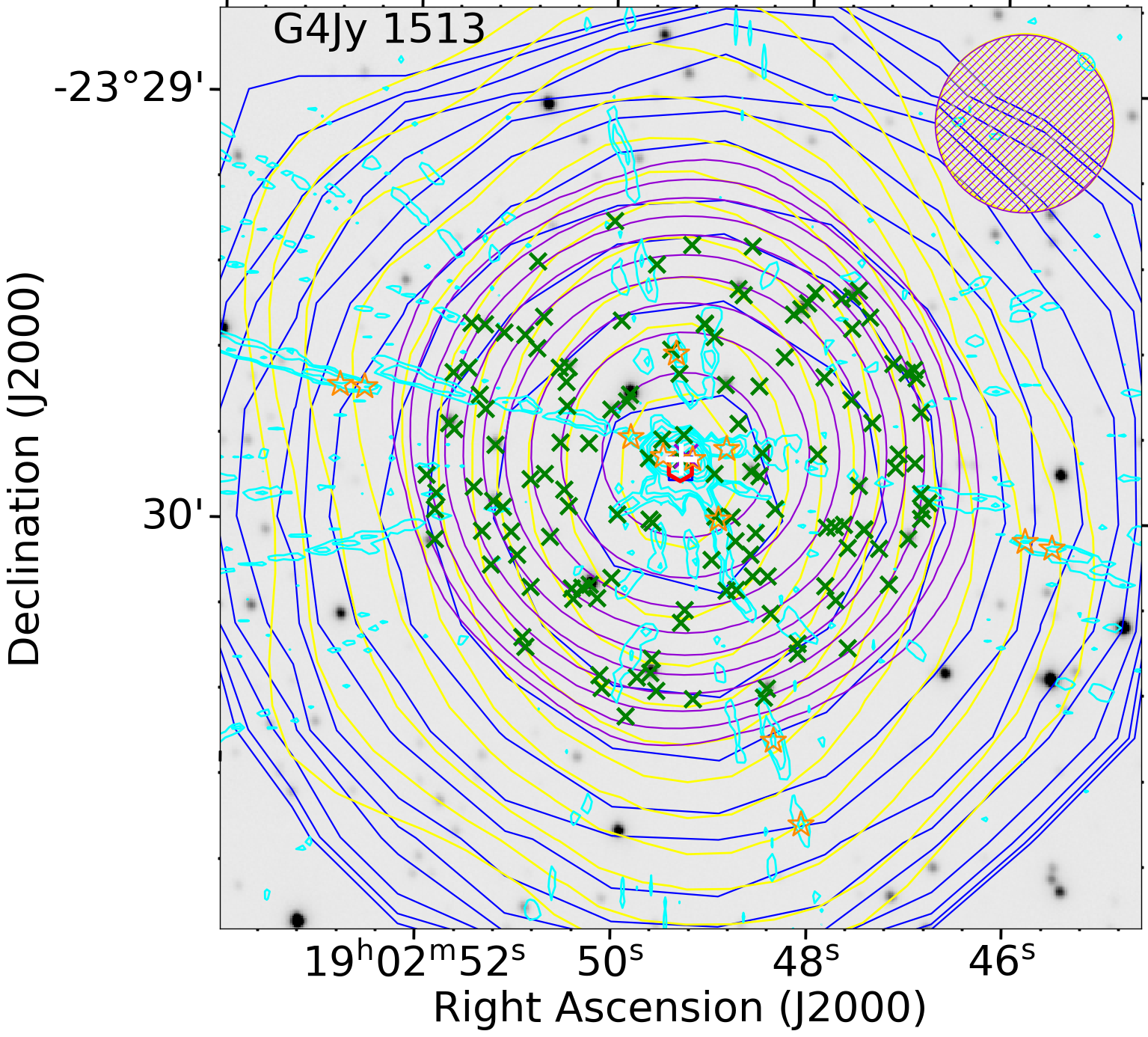} 
	}  \\[-0.05cm]
 \subfigure[G4Jy~1520]{
	\includegraphics[width=0.42\linewidth]{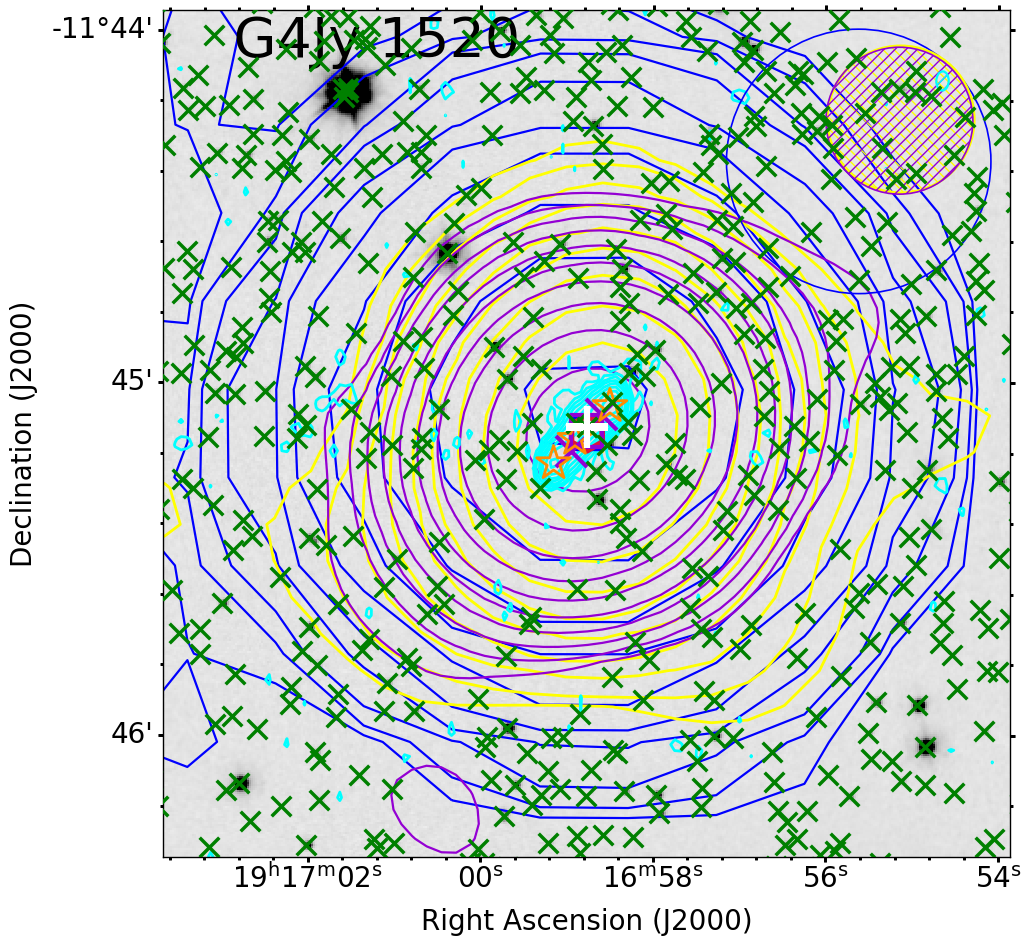} 
	} 
\subfigure[G4Jy~1529]{
	\includegraphics[width=0.41\linewidth]{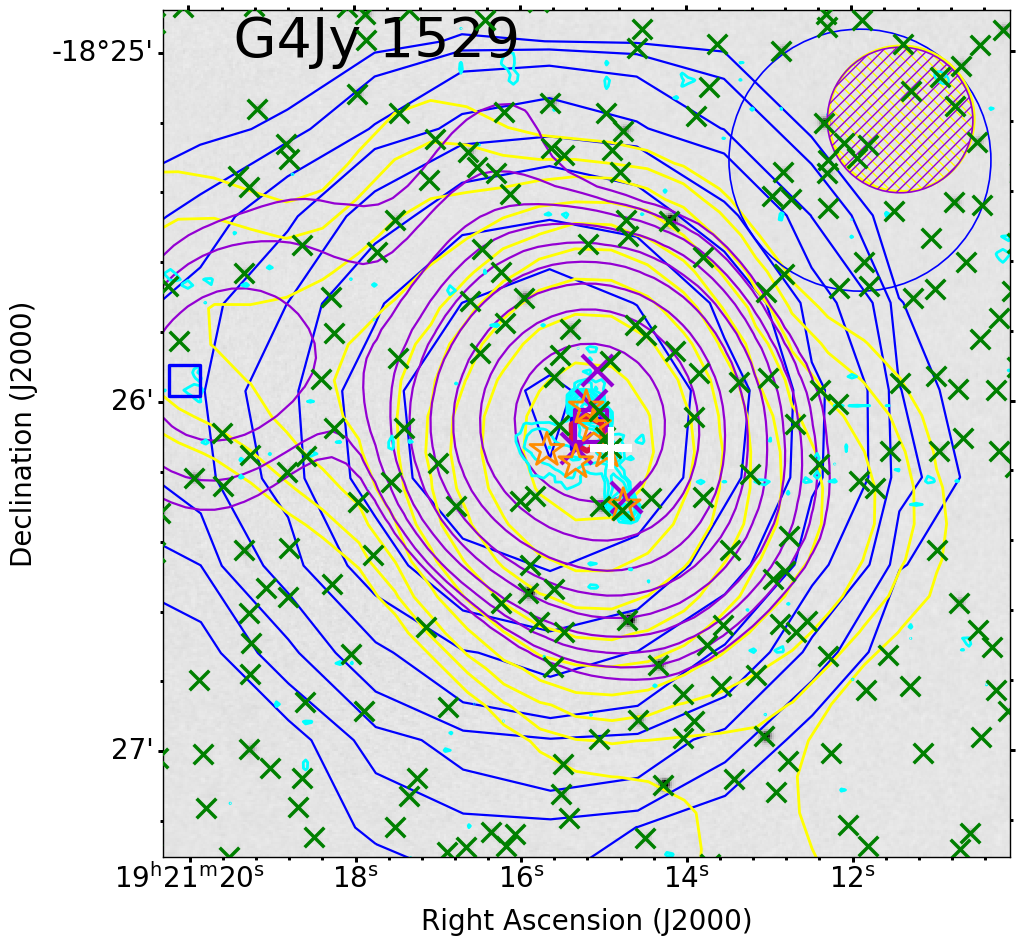}
	}

\caption{ {\it Continued} -- Overlays of radio contours [and catalogue positions] from SUMSS/NVSS (blue), RACS-low1 (purple), TGSS (yellow) and MeerKAT/VLASS (cyan) on inverted grey-scale $K$-band images from VHS (unless otherwise stated). A white `+' marks the position of the galaxy hosting the radio emission.}
\end{figure*}

\setcounter{figure}{0}

\begin{figure*}
%\vspace{-0.7cm}
\centering
\subfigure[G4Jy~1532 ]{
	\includegraphics[width=0.42\linewidth]{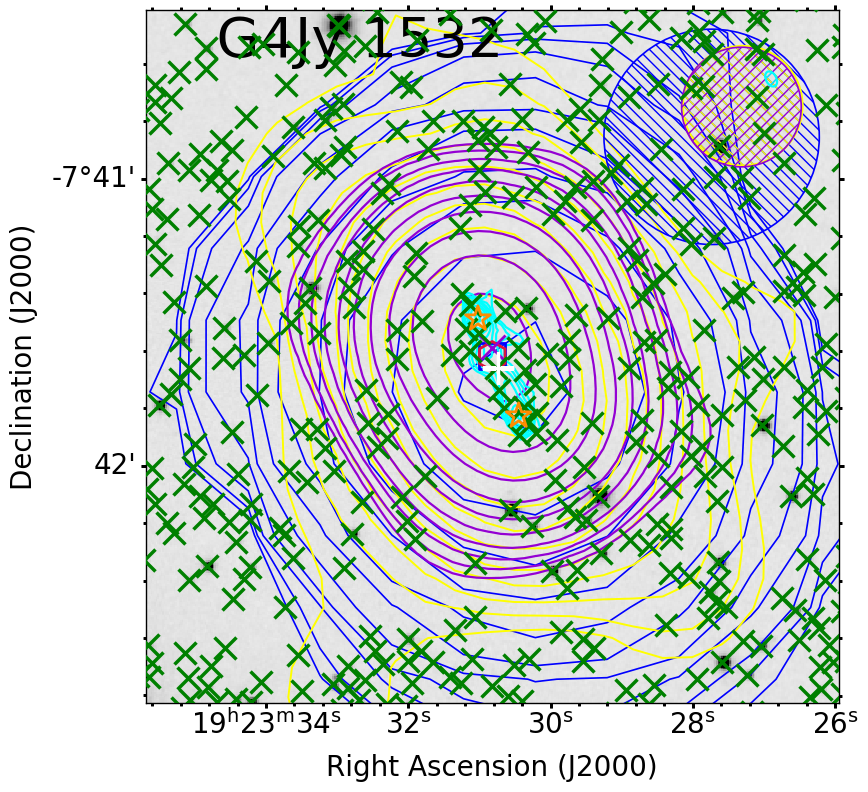} 
	}  
\subfigure[G4Jy~1575 (MIR image from AllWISE)]{
	\includegraphics[width=0.4\linewidth]{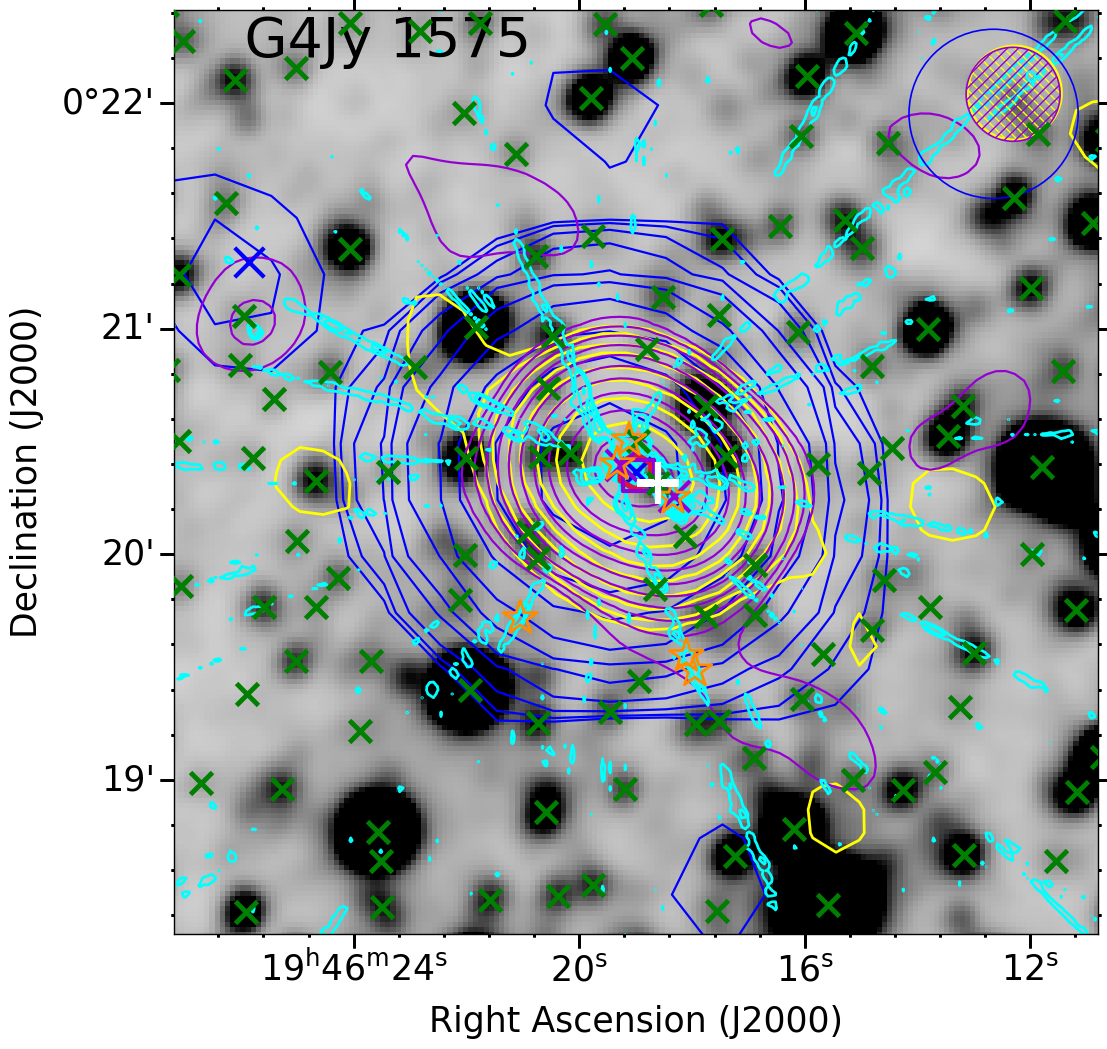} 
	} \\[-0.05cm]
\subfigure[G4Jy~1587 ]{
	\includegraphics[width=0.41\linewidth]{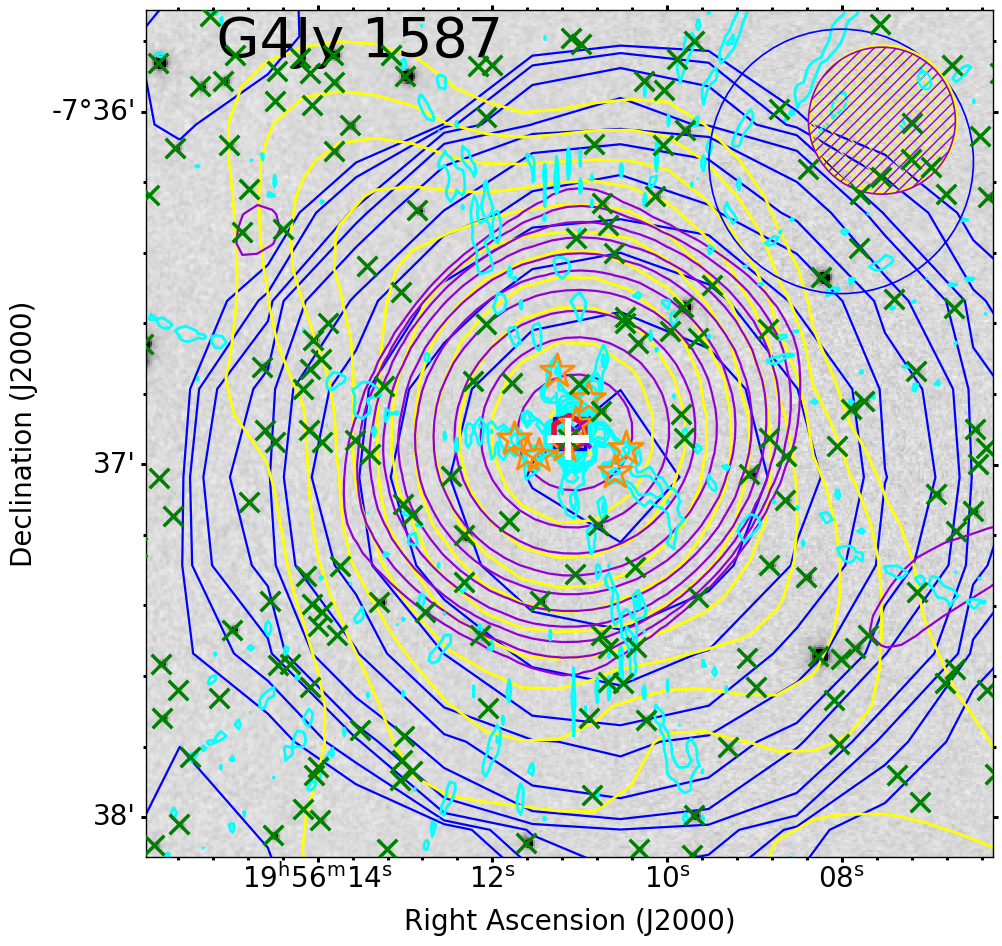} 
	}  
 \subfigure[G4Jy~1595 (MIR image from AllWISE)]{
	\includegraphics[width=0.42\linewidth]{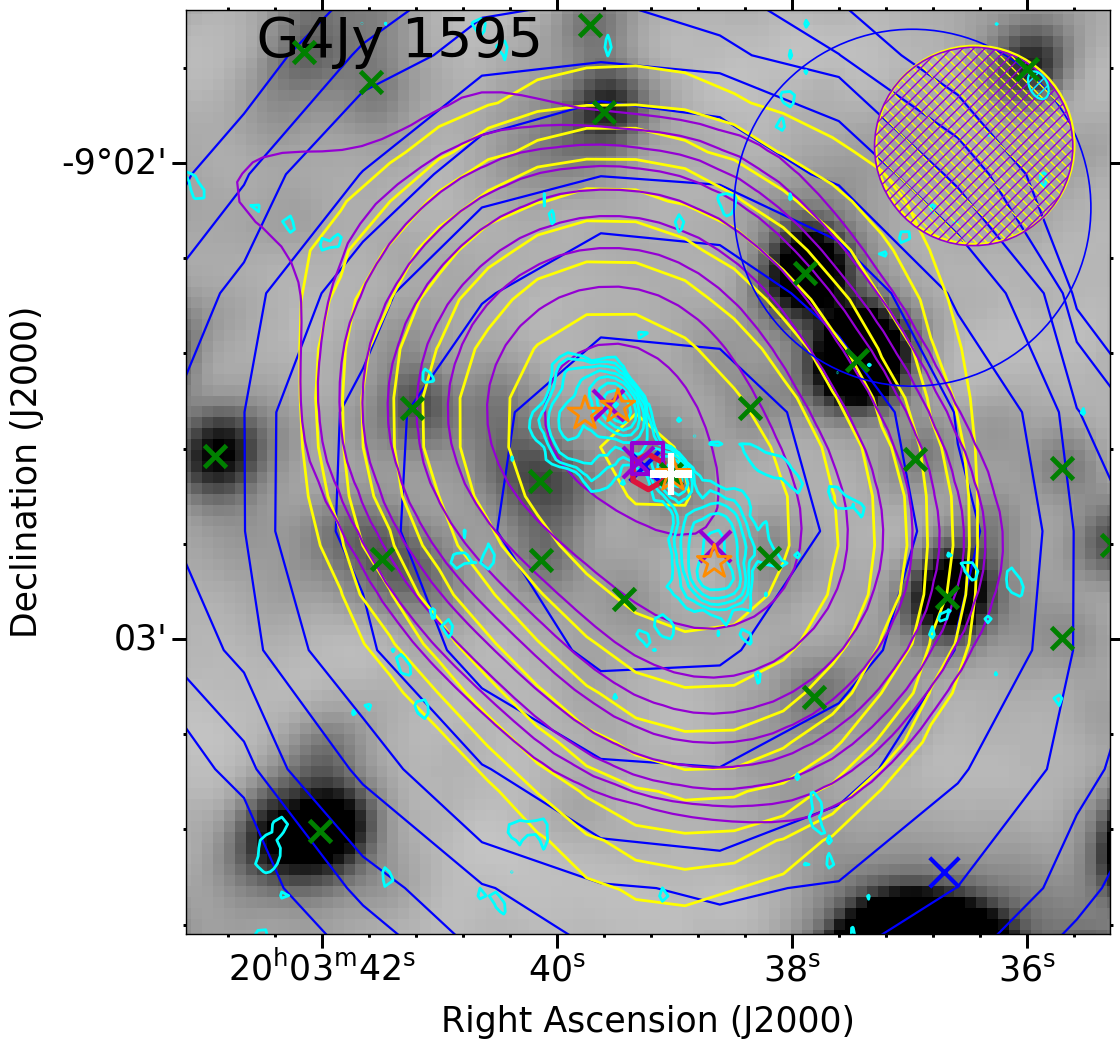} 
	} \\[-0.05cm]
     \subfigure[G4Jy~1603 ]{
	\includegraphics[width=0.41\linewidth]{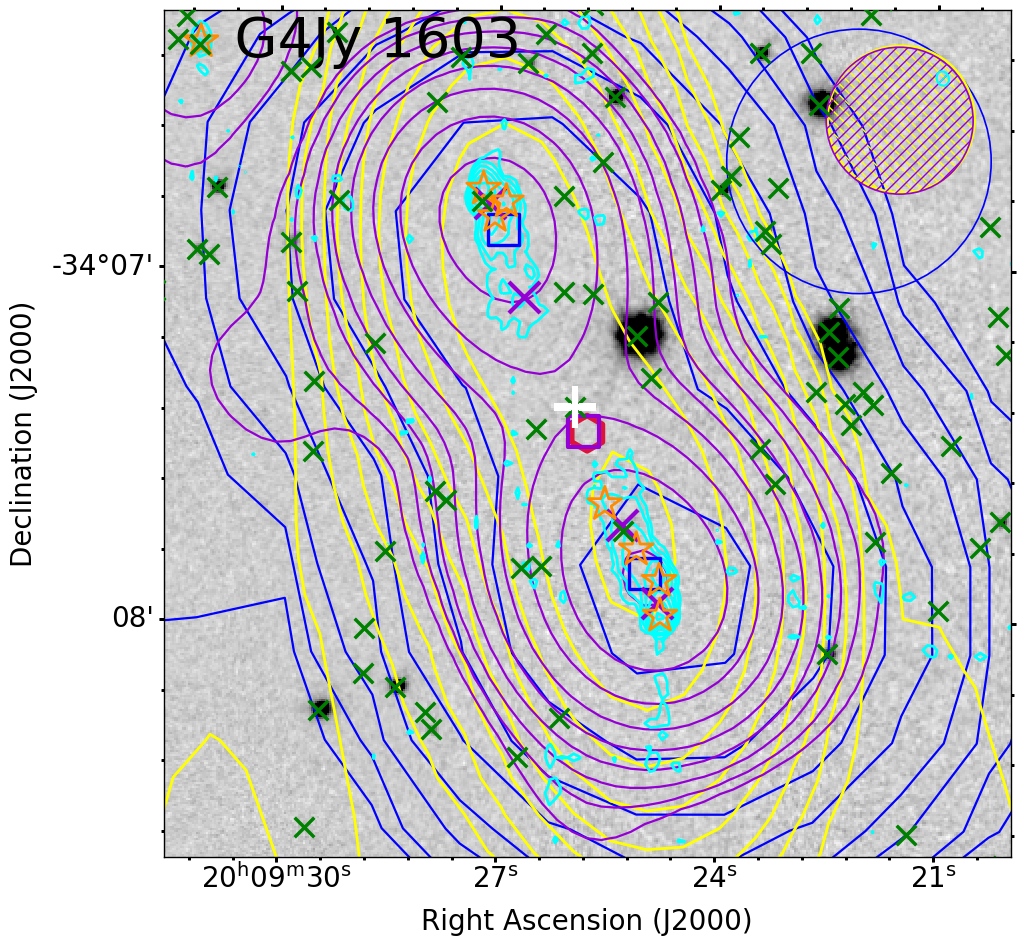} 
	} 
         \subfigure[G4Jy~1814 ]{
	\includegraphics[width=0.42\linewidth]{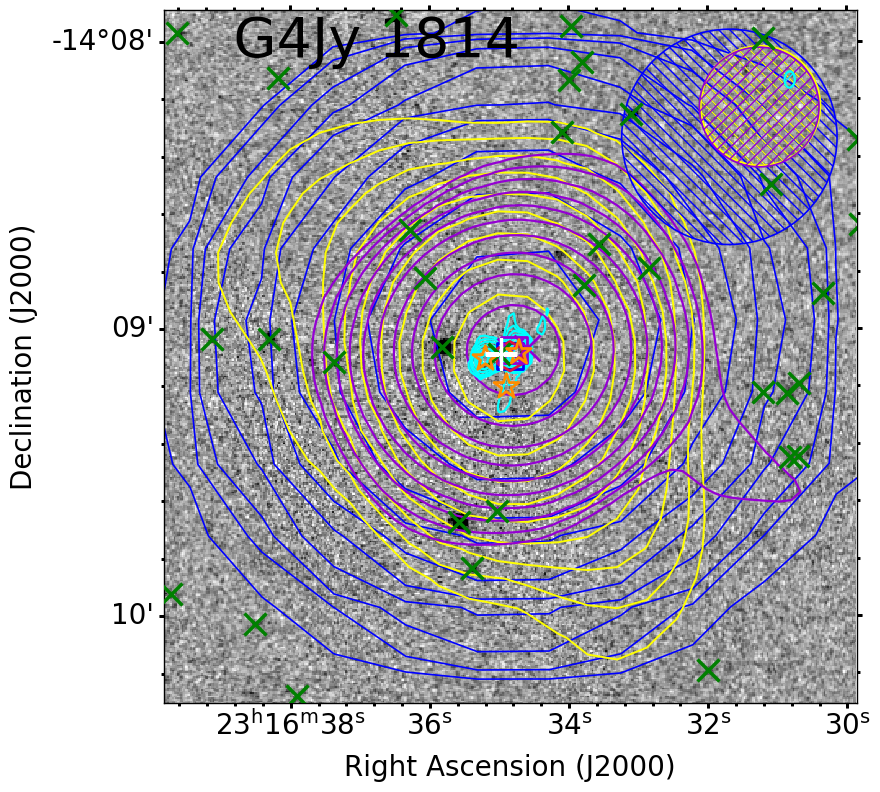} 
	} 

\caption{ {\it Continued} -- Overlays of radio contours [and catalogue positions] from SUMSS/NVSS (blue), RACS-low1 (purple), TGSS (yellow) and MeerKAT/VLASS (cyan) on inverted grey-scale $K$-band images from VHS (unless otherwise stated). A white `+' marks the position of the galaxy hosting the radio emission.}
\end{figure*}

\setkeys{Gin}{draft}

\section{Redshift references via NED}

Following on from Table~\ref{tab:z_references}, Table~\ref{tab:NED_references} provides a guide to the Bibcodes that can be found in the G4Jy catalogue where the `z\_origin\_flag' is equal to `10'. This indicates that the redshift was semi-automatically `data-mined' from the NASA/IPAC Extragalactic Database (NED) and has not been superseded by either a more-recent $z_{\mathrm{sp}}$ or $z_{\mathrm{ph}}$ value. Note that additional Bibcodes exist in the catalogue (where z\_origin\_flag does not equal 10), and these can be looked up via \url{https://ned.ipac.caltech.edu/reflookup}. % Due to repetition of Bibcodes, can't say how many more there are in the catalogue

\begin{table}
\centering 
\caption{ A guide to redshift references that have been `data-mined' through the NASA/IPAC Extragalactic Database (NED). Each of these are associated with a `z origin' flag of `10' in the G4Jy catalogue, as indicated in Table~\ref{tab:z_references}. The asterisk, *, indicates references that were added manually, having found that the automated result had cited the incorrect publication. (Please note that not all of the references have undergone the same level of scrutiny, and so we apologise in advance to authors for any remaining misreferences.) }
\begin{tabular}{@{}lcr@{}}
 \hline
 &   & No. G4Jy  \\
Bibcode & Reference  & sources \\
 \hline

  1976ApJS...31..143W    & \citet{1976ApJS...31..143W} & 1 \\
  1978MNRAS.185..149H    & \citet{1978MNRAS.185..149H} & 3 \\ 
 % 1979ApJ...229...73W    & \citet{1979ApJ...229...73W} & 1 \\
  1984ApJ...286..498J    & \citet{1984ApJ...286..498J}* & 1 \\
  1984MNRAS.209..159P    & \citet{1984MNRAS.209..159P} & 1 \\
  1984PASA....5..341M    & \citet{1984PASA....5..341M}* & 1 \\
  1985PASP...97..932S    & \citet{Spinrad1985} & 2 \\
  1988AJ.....96.1775O    & \citet{1988AJ.....96.1775O} & 1 \\
  1988ApJ...327..561W    & \citet{1988ApJ...327..561W} & 1 \\
  1988PASP..100.1423M    & \citet{1988PASP..100.1423M} & 1 \\
  1989MNRAS.238.1171D    & \citet{1989MNRAS.238.1171D} & 2 \\
  1989QSO...M...0000H    & \citet{1989QSO...C......0H} & 4 \\ %8
  1990AJ....100.1014M    & \citet{1990AJ....100.1014M} & 1 \\
  1991ApJS...75..297H    & \citet{1991ApJS...75..297H} & 1 \\
  1991MNRAS.253..287A    & \citet{1991MNRAS.253..287A} & 1 \\
  1991RC3.9.C...0000d    & \citet{1991rc3..book.....D} & 1 \\
  1992ApJS...83...29S    & \citet{1992ApJS...83...29S} & 1 \\
  1992NED11.R......1N    & The NED Team  & 1 \\
  1993MNRAS.263..999T    & \citet{1993MNRAS.263..999T}* & 1 \\
%  1994A\&A...281..355B    & \citet{1994AA...281..355B} & 1 \\
%  1994ApJ...428...65H    & \citet{1994ApJ...428...65H} & 1 \\
%  1994MNRAS.269..998D    & \citet{1994MNRAS.269..998D}* & 1 \\
  1995A\&A...299...17H    & \citet{1995AA...299...17H} & 1 \\
  1995MNRAS.277..553B    & \citet{1995MNRAS.277..553B} & 2 \\ %3
  1996A\&A...313...25V    & \citet{1996AA...313...25V}* & 1 \\
  1996ApJS..107...19M    & \citet{1996ApJS..107...19M} & 15 \\ %17
  1996MNRAS.279L..13R    & \citet{1996MNRAS.279L..13R} & 1 \\
  1998ApJS..118..327K    & \citet{1998ApJS..118..327K} & 1 \\
%  1998MNRAS.299L..25B    & \citet{1998MNRAS.299L..25B} & 1 \\
  1998MNRAS.300..269I    & \citet{1998MNRAS.300..269I} & 1 \\
    1999ApJS..122...29B    & \citet{1999ApJS..122...29B}* & 1 \\
  1999MNRAS.303..565T    & \citet{1999MNRAS.303..565T} & 1 \\
  1999MNRAS.310..223B    & \citet{Best1999} & 10 \\
  2000MNRAS.315...21B & \citet{2000MNRAS.315...21B} & 2 \\
  2001AJ....121.1241D    & \citet{2001AJ....121.1241D} & 1 \\
  2002A\&A...386...97J    & \citet{2002AA...386...97J} & 1 \\
  2002AJ....123.3018M    & \citet{2002AJ....123.3018M} & 1 \\
 % 20032dF...C...0000C    & \citet{2003astro.ph..6581C}* & 1 \\
  2003MNRAS.346.1021B    & \citet{2003MNRAS.346.1021B} & 1 \\
%  2004AJ....128.1558S    & \citet{2004AJ....128.1558S} & 1 \\
  2005AJ....130..896S    & \citet{2005AJ....130..896S} & 1 \\
  2005ApJS..158..161H    & \citet{2005ApJS..158..161H} & 1 \\
  2006AJ....131..114B    & \citet{Burgess2006} & 17 \\ %16
  2006MNRAS.366.1067B    & \citet{2006MNRAS.366.1067B} & 1 \\
  2008MNRAS.387..639H    & \citet{2008MNRAS.387..639H} & 1 \\
%  20096dF...C...0000J    & \citet{Jones2009} & 1 \\
  2009ApJS..184..398H    & \citet{2009ApJS..184..398H} & 2 \\
  2009MNRAS.395.1099B    & \citet{2009MNRAS.395.1099B} & 1 \\
%  2010MNRAS.401..633D    & \citet{2010MNRAS.401..633D} & 1 \\
%  2011AJ....142..165T    & \citet{2011AJ....142..165T} & 1 \\
  2011MNRAS.415.2245K    & \citet{2011MNRAS.415.2245K} & 1 \\
  2012ApJS..199...26H    & \citet{2012ApJS..199...26H} & 3 \\ % 5
  2014ApJS..210....9B    & \citet{Bilicki2014} & 11 \\ % 14
  
   2014MNRAS.440..696A    & \citet{2014MNRAS.440..696A} & 2 \\
  2015ApJS..218...10V    & \citet{2015ApJS..218...10V} & 2 \\
  2015PASA...32...10F    & \citet{2015PASA...32...10F}* & 4 \\
  2016AJ....152...25M    & \citet{2016AJ....152...25M} & 1 \\
%  2016ApJ...818..113N    & \citet{2016ApJ...818..113N} & 1 \\

  2016SDSSD.C...0000:    & \citet{2017ApJS..233...25A}* & 1 \\ % 2
  2017A\&A...599A.123N    & \citet{Nesvadba2017} & 1 \\
      2017AJ....153..157T    & \citet{2017AJ....153..157T} & 2 \\
  2017ApJ...836..174C    & \citet{Callingham2017} & 2 \\
    2018A\&A...610A...1M    & \citet{2018AA...610A...1M} & 1 \\
  2018A\&A...618A..80G    & \citet{2018AA...618A..80G} & 2 \\ %3

  2018ApJ...859...38N    & \citet{2018ApJ...859...38N} & 1 \\
   % 2018ApJS..238....9K    & \citet{2018ApJS..238....9K} & 1 \\
  2018ApJS..239...33Y    & \citet{2018ApJS..239...33Y} & 1 \\ % 4
\hline
{\it Continued} \\
\hline
\label{tab:NED_references}
\end{tabular}
\end{table}

\setcounter{table}{0}

\begin{table}
\centering 
\caption{ {\it Continued -- } A guide to the redshift references that have been `data-mined' through the NASA/IPAC Extragalactic Database (NED). }
\begin{tabular}{@{}lcr@{}}
 \hline
 &   & No. G4Jy  \\
Bibcode & Reference  & sources \\
 \hline

  2019ApJ...873..132M    & \citet{2019ApJ...873..132M} & 1 \\
  2019ApJS..240...34M    & \citet{2019ApJS..240...34M} & 2 \\

  2019ApJS..245....3G    & \citet{2019ApJS..245....3G} & 1 \\
  2020A\&A...642A.153D    & \citet{2020AA...642A.153D} & 1 \\
  2022ApJ...929..108G    & \citet{2022ApJ...929..108G} & 10 \\
  2022ApJ...936..146X    & \citet{2022ApJ...936..146X} & 2 \\
  2022ApJS..259...37S    & \citet{2022ApJS..259...37S} & 3 \\
  2022ApJS..261....5M    & \citet{2022ApJS..261....5M} & 1 \\
  2022ApJS..261....6K    & \citet{2022ApJS..261....6K} & 2 \\
  2022ApJS..262...51M & \citet{Maselli2022} & 1 \\
  2022MNRAS.510..786S    & \citet{2022MNRAS.510..786S} & 1 \\ % 2
  2022MNRAS.512..874T    & \citet{2022MNRAS.512..874T} & 1 \\
  2023A\&A...671A..32C    & \citet{2023AA...671A..32C} & 4 \\
  2023ApJS..264....4M    & \citet{2023ApJS..264....4M} & 1 \\
  2023ApJS..265....8R    & \citet{2023ApJS..265....8R} & 1 \\ %2
  2023ApJS..265...25J    & \citet{2023ApJS..265...25J} & 1 \\
  2023ApJS..265...32M & \citet{Massaro2023}
& 3 \\
2023ApJS..268...17K    & \citet{2023ApJS..268...17K} & 1 \\
2023MNRAS.518.4290S & \citet{Sejake2023} & 1 \\
  2024A\&A...683A..20C    & \citet{2024AA...683A..20C} & 1 \\
  2024ApJS..271....8G & \citet{GarciaPerez2024} & 1 \\
  
  \hline
& \hspace{2cm} Total & 164 \\

\hline
\label{tab:NED_references}
\end{tabular}
\end{table}

\section{Re-fitting 6dFGS spectra}
\label{app:6dfgs_refitting}

We visually inspected the spectra for G4Jy host-galaxies that are in the 6dFGS, and found that (of those with spectra available online; Section~\ref{sec:xmatch_specz}) 16 had questionable redshift determinations. We therefore downloaded the 6dFGS spectra for these sources (see Table~\ref{tab:6dfgs}) and re-fitted them. Our findings are as follows.

\setkeys{Gin}{draft=false}

\begin{figure*}
%\vspace{-0.7cm}
\centering
\subfigure[G4Jy 28 at the 6dFGS-catalogue redshift, $z = 0.768$]{
	\includegraphics[width=0.47\linewidth]{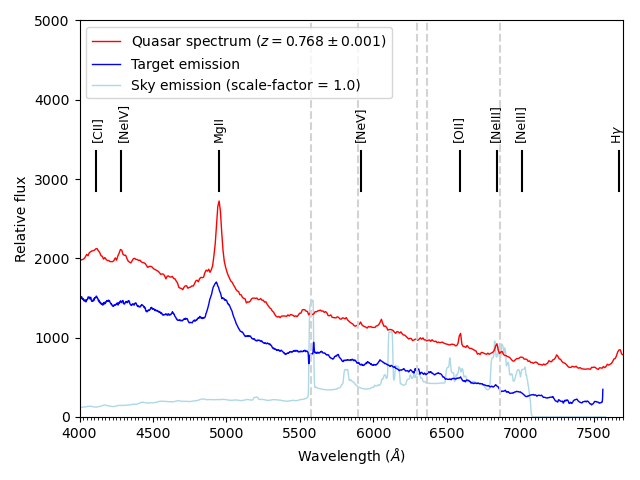}
	} 
\subfigure[G4Jy 665 at a redshift of $z = 0.091$]{
	\includegraphics[width=0.47\linewidth]{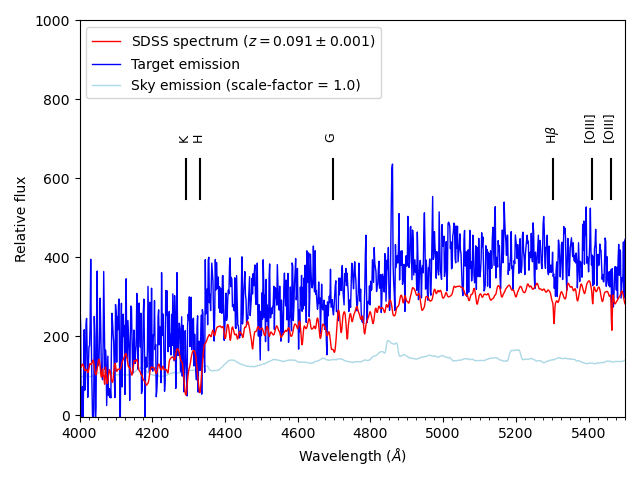}
	} \\[-0.05cm]
\subfigure[G4Jy 1708 at the 6dFGS-catalogue redshift, $z = 0.200$]{
	\includegraphics[width=0.49\linewidth]{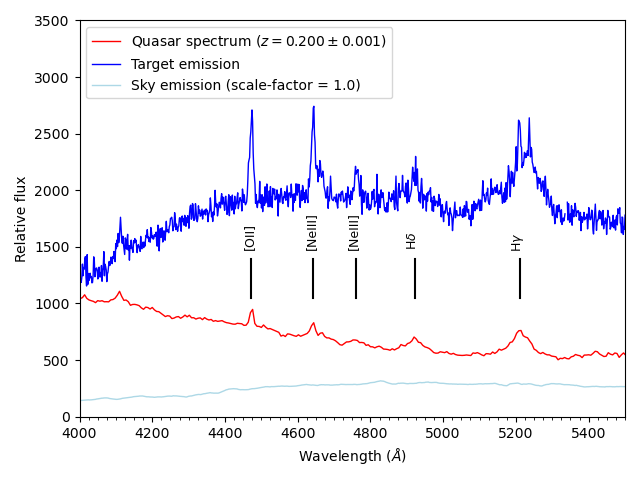} 
	}  
\subfigure[G4Jy 1753 at an alternative redshift, $z = 0.001$]{
	\includegraphics[width=0.49\linewidth]{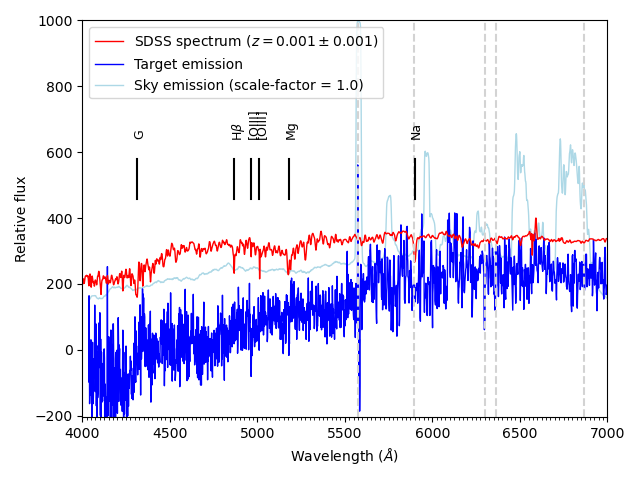} 
	} 

\caption{Re-fitting of the 6dFGS spectrum for G4Jy 28 (where we find the 6dFGS redshift to be robust), G4Jy 665 (with a noisy Ca H+K doublet that is discernible after zooming in), G4Jy 1708 (where the quoted redshift is correct but the spectrum had been mislabelled online), and G4Jy 1753. The suggested redshift for the latter leads to better agreement (between the 6dFGS spectrum and the template) for the Ca G absorption-line, and we think that misalignment of subsequent lines is due to poor wavelength calibration (Appendix~\ref{app:6dfgs_refitting}).  \label{fig:6dfgs_leftover_refits}}
\end{figure*}

Our note for G4Jy 28 (Figure~\ref{fig:6dfgs_leftover_refits}), G4Jy 411, and G4Jy 1842 is that they have ``a single emission line (``Mg{\sc ii}'')''. Re-fitting the spectrum shows that this is the correct line identification for all three of them, as identifying the emission as a different broad-line leads to the expectation of other bright lines being seen in the spectrum. This is not the case, and so we assign these sources a `$z$-Quality flag', Q, of `1'. 

\begin{figure}
%\vspace{-0.7cm}
\centering

\subfigure[At the 6dFGS-catalogue redshift, $z = 0.057$]{
	\includegraphics[width=0.98\linewidth]{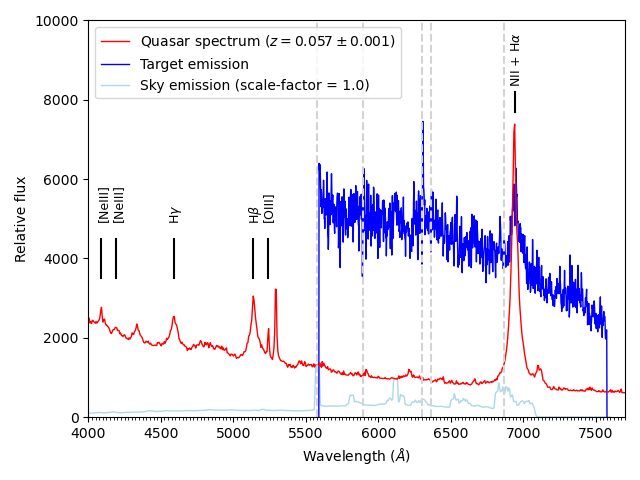}
	} \\[-0.05cm]
\subfigure[$z=1.483$ as a possible alternative redshift]{
	\includegraphics[width=0.98\linewidth]{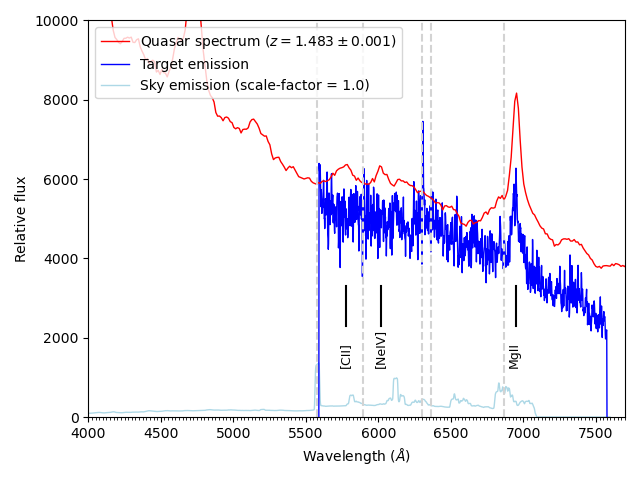} 
	}  \\[-0.05cm]
\subfigure[$z = 3.487$ as a less-likely alternative redshift]{
	\includegraphics[width=0.98\linewidth]{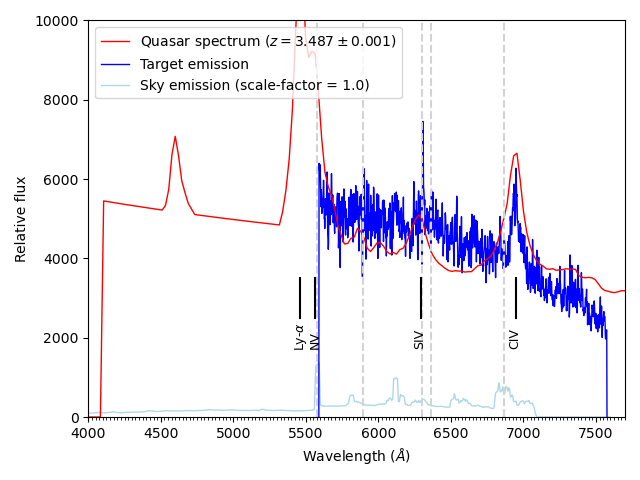} 
	} 

\caption{Re-fitting of the 6dFGS ($R$-band) spectrum for G4Jy 540, for different interpretations of the broad-line emission: (a) as a H$\alpha$ line, (b) as a Mg{\sc ii} line, and (c) as a CIV line (Appendix~\ref{app:6dfgs_refitting}). Nominal normalisations have been applied to the quasar template provided courtesy of P.C. Hewett \citep{Temple2021}. We retain a cautionary flag of Q = 4 for this source until we follow-up with optical spectroscopy at $\lesssim 5600$\,\AA. \label{fig:g4jy540}}
\end{figure}

For G4Jy 540, the catalogued redshift corresponds to identifying the broad emission-line as H$\alpha$, but as shown in Figure~\ref{fig:g4jy540}, there is the possibility that this is actually the Mg{\sc ii} line. Observations of the [O{\sc iii}] doublet in the $V$ band would have helped to differentiate between these two scenarios, but unfortunately the $V$-spectrum data appear to be corrupted. Another possibility is that the C{\sc iv} line has been detected instead (Figure~\ref{fig:g4jy540}c), with Lyman-$\alpha$ appearing in the $V$ band, but this is much less likely given the large redshift that this would involve ($z = 3.487$). We will follow-up this source with SALT spectroscopy to resolve the ambiguity (PI: White).

\setkeys{Gin}{draft}

In the case of G4Jy 1708, we find confirmation that the lines have been mislabelled, but {\it not} as a result of line misidentification. As viewed on the 6dFGS website, absorption-line labelling had been applied when emission-line labelling is required instead. The latter demonstrates that the redshift determination is robust (Figure~\ref{fig:6dfgs_leftover_refits}), and so we update the Q flag to `1'. 

\begin{figure*}
%\vspace{-0.7cm}
\centering
\subfigure[G4Jy 1127 at the 6dFGS-catalogue redshift, $z = 0.065$]{
	\includegraphics[width=0.49\linewidth]{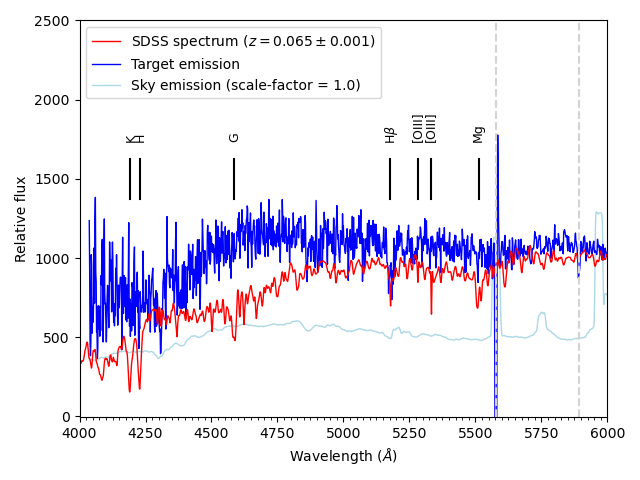}
	} 
\subfigure[G4Jy 1127 at an alternative redshift, $z = 0.001$]{
	\includegraphics[width=0.49\linewidth]{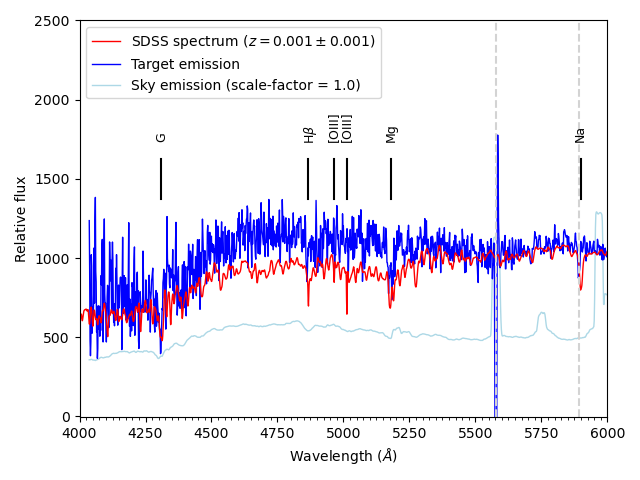}
	} \\[-0.05cm]
\subfigure[G4Jy 1533 at the 6dFGS-catalogue redshift, $z = 0.057$]{
	\includegraphics[width=0.49\linewidth]{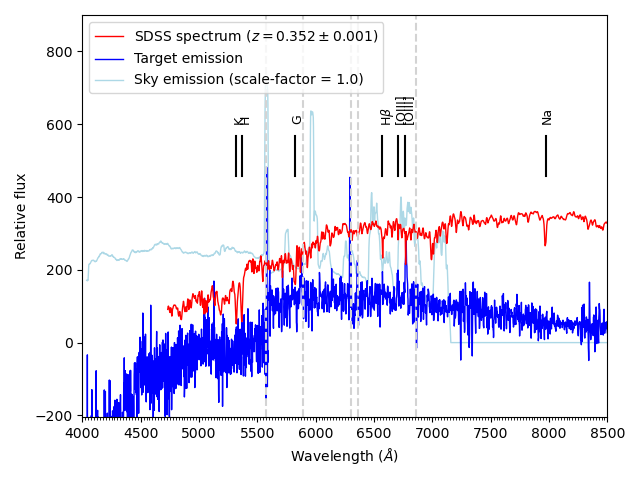} 
	}  
\subfigure[G4Jy 1533 at an alternative redshift, $z = 0.400$]{
	\includegraphics[width=0.49\linewidth]{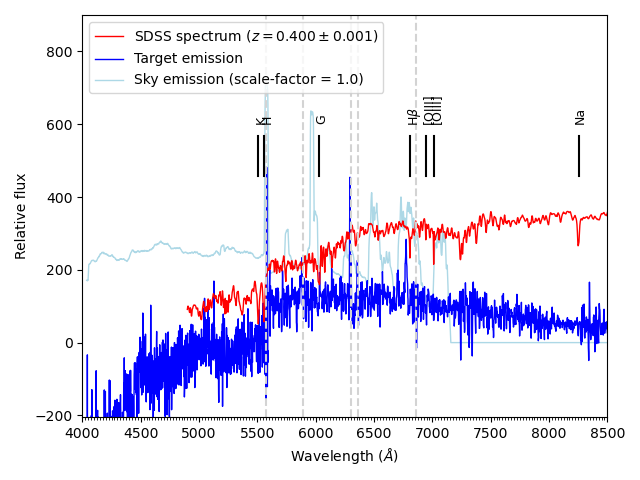} 
	} 

\caption{Re-fitting of the 6dFGS spectrum for (a)/(b) G4Jy 1127 and (c)/(d) G4Jy 1533 (Appendix~\ref{app:6dfgs_refitting}). This demonstrates the validity of alternative redshifts for the template-fitting, but the combination of noise and (possible) poor wavelength-calibration means that we cannot quote these as reliable alternatives in the G4Jy catalogue. Therefore, we retain the original (6dFGS-catalogue) redshifts for these sources, and accompany them with a cautionary flag of Q = 4. \label{fig:6dfgs_wavelength}}
\end{figure*}

The following sources are deemed to have ``too noisy a spectrum'' for a redshift to be determined robustly: G4Jy 994, G4Jy 1127, G4Jy 1533, and G4Jy 1554. As such, they are assigned `Q = 4'. Furthermore, Figure~\ref{fig:6dfgs_wavelength}a and Figure~\ref{fig:6dfgs_wavelength}b shows that the ``H$\beta$'' absorption-line may in fact be the Mg absorption-line, with the suggested redshift ($z = 0.001 \pm 0.001$)\footnote{As these spectra have been manually-fitted `by eye', an uncertainty in the redshift of $\Delta z = 0.001$ is the smallest that can be achieved.} showing better alignment with Ca G absorption. For G4Jy 1533 (Figure~\ref{fig:6dfgs_wavelength}cd), the 6dFGS redshift-fitting appears to have prioritised the putative presence of the [O{\sc iii}] emission-line doublet. If, instead, the spectral break is interpreted as the 4000-\AA\ break, we suggest that the redshift is higher, at $z = 0.400 \pm 0.001$. In either case, the inconsistent (mis-)alignment of several features suggests that the wavelength calibration is suboptimal. 

\setkeys{Gin}{draft=false}

\setkeys{Gin}{draft}

We cited ``no spectral features'' in Table~\ref{tab:6dfgs} for: G4Jy 229, G4Jy 543, G4Jy 665, G4Jy 893, G4Jy 1245, G4Jy 1664, and G4Jy 1753. This was based on visual inspection of the full, combined $V$--$R$ spectrum, spanning from 4000\AA\ to 8500\AA, but zooming in allows possible absorption lines to be discerned (in the cases of G4Jy 665 and G4Jy 1753; Figure~\ref{fig:6dfgs_leftover_refits}). Nonetheless, as all of the sources in this category need to have their redshifts confirmed, we assign a Q flag of `4' for each of them.

\section{Re-fitting SDSS DR16 spectra}
\label{app:sdss_refitting}

\setkeys{Gin}{draft=false}

\begin{figure}
%\vspace{-0.7cm}
\centering

\subfigure[G4Jy~179, which originally had a redshift of $z = 0.236$]{
	\includegraphics[width=0.98\linewidth]{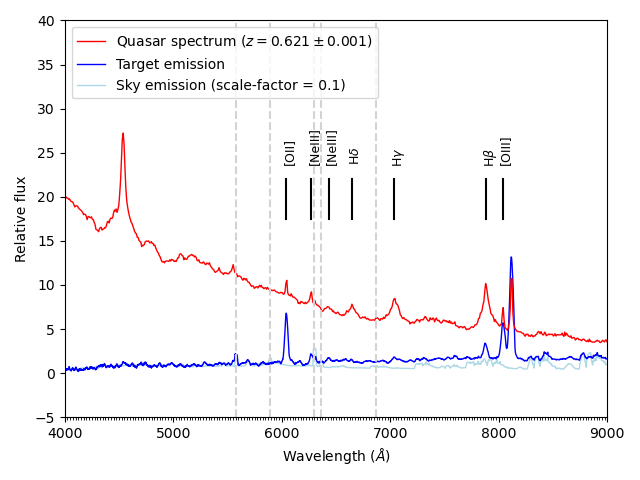}
	} \\[-0.05cm]
\subfigure[G4Jy~682, which originally had a redshift of $z = 1.968$]{
	\includegraphics[width=0.98\linewidth]{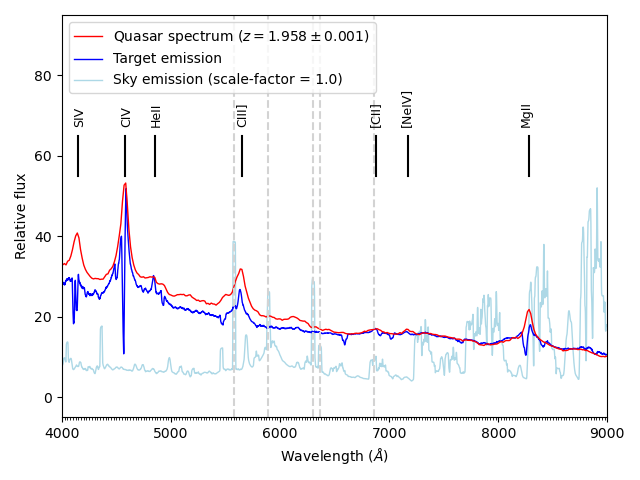} 
	}  \\[-0.05cm]
\subfigure[G4Jy~842, which originally had a redshift of $z = 1.736$]{
	\includegraphics[width=0.98\linewidth]{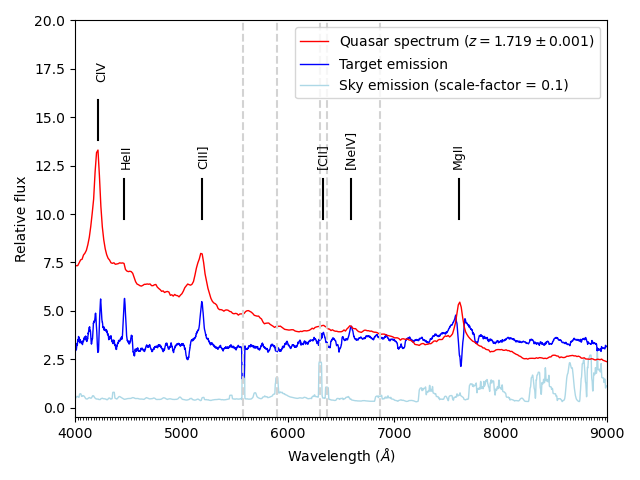} 
	} 

\caption{Re-fitting of the SDSS DR16 spectra for (a) G4Jy~179 (b) G4Jy~682, and (c) G4Jy~842 (Appendix~\ref{app:sdss_refitting}). Nominal normalisations have been applied to the quasar template provided courtesy of P.C. Hewett \citep{Temple2021}. \label{fig:refitted_sdss_spectra}}
\end{figure}

\setkeys{Gin}{draft}

We visually inspected the spectra crossmatched from SDSS DR16 online, via the Science Archive Server. As with the 6dFGS spectra, this allowed us to identify sources where the redshift could be improved via re-fitting. However, in one case (G4Jy 741, described below), two candidate redshifts remained viable, and so we leave the SDSS redshift `as is' but accompanied by z\_Quality\_flag = 4. The remainder have their flag updated to `1'.

For G4Jy 179, the brightest emission-line was misidentified as H$\alpha$. It is in fact the 5007-\AA\ half of the [O{\sc iii}] doublet, with Figure~\ref{fig:refitted_sdss_spectra}a showing good alignment (with the quasar template) for multiple emission-lines. We are therefore confident in assigning a new redshift for this source: $z = 0.621$. 

The redshift fitting for G4Jy 682 only required slight adjustment (Figure~\ref{fig:refitted_sdss_spectra}b), on account of the S{\sc iv}, C{\sc iv}, and Mg{\sc ii} lines displaying significant absorption. This causes the centroids of the emission lines to become shifted (equivalent to $\Delta z = 0.010$, in this particular case), and is flagged by SDSS pipelines as ``negative emission''.

\setkeys{Gin}{draft=false}

\begin{figure}
%\vspace{-0.7cm}
\centering

\subfigure[G4Jy~741, at the SDSS-DR16 redshift of $z = 0.230$]{
	\includegraphics[width=0.98\linewidth]{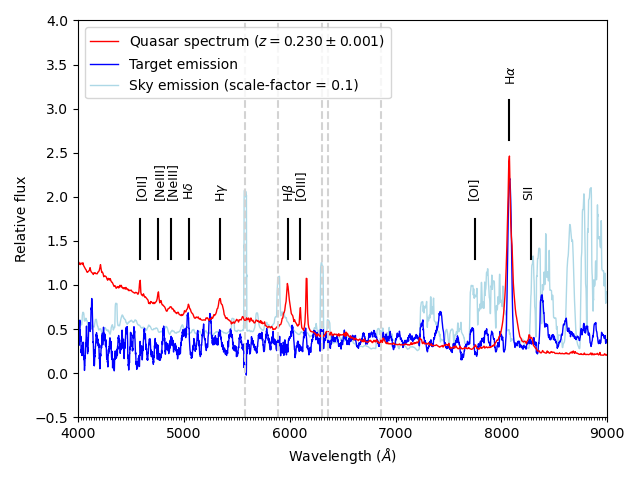}
	} 
\subfigure[G4Jy~741, at an alternative redshift of $z = 0.631$]{
	\includegraphics[width=0.98\linewidth]{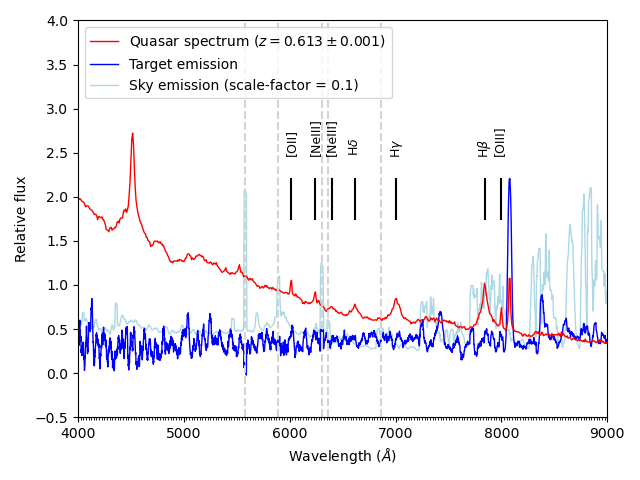} 
	}  

\caption{Re-fitting of the SDSS DR16 spectrum for G4Jy~741, showing that the brightest emission could be interpreted as being (a) H$\alpha$ or (b) [O{\sc iii}] (Appendix~\ref{app:sdss_refitting}).  Nominal normalisations have been applied to the quasar template provided courtesy of P.C. Hewett \citep{Temple2021}. \label{fig:g4jy741}}
\end{figure}

\setkeys{Gin}{draft}

The original redshift for G4Jy 741, $z=0.230$, is based on interpreting the strongest emission-line as H$\alpha$ (Figure~\ref{fig:g4jy741}a). However, the absence of other prominent emission-lines leads us to query this redshift. If we instead interpret the bright emission as part of the [O{\sc iii}] doublet at $z=0.613$ (Figure~\ref{fig:g4jy741}b), we again do not see other expected lines significantly above the continuum noise-level. Therefore, we proceed with the SDSS-DR16 redshift-value until additional spectroscopy shows otherwise.  

Absorption of the C{\sc iv} and Mg{\sc ii} lines is also seen for G4Jy~842 (Figure~\ref{fig:refitted_sdss_spectra}c). This time we adjust the redshift by $\Delta z = 0.017$, to $z = 1.719$.

\section{DESI DR1 spectra}
\label{app:desi_spectra}

\begin{figure*}
%\vspace{-0.7cm}
\centering
\subfigure[G4Jy 923 at the re-fitted redshift, $z = 0.6337$]{
	\includegraphics[width=0.98\linewidth]{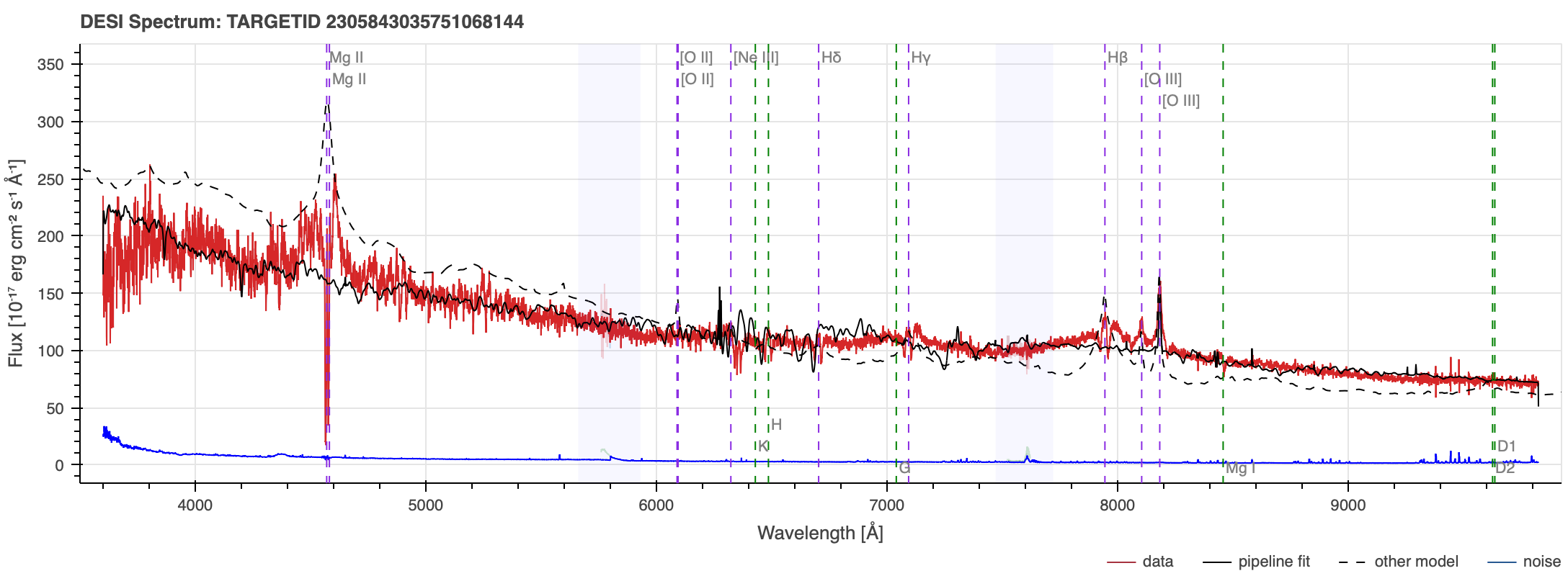}
	} 
\subfigure[G4Jy 1228 at the re-fitted redshift, $z = 1.7764$]{
	\includegraphics[width=0.98\linewidth]{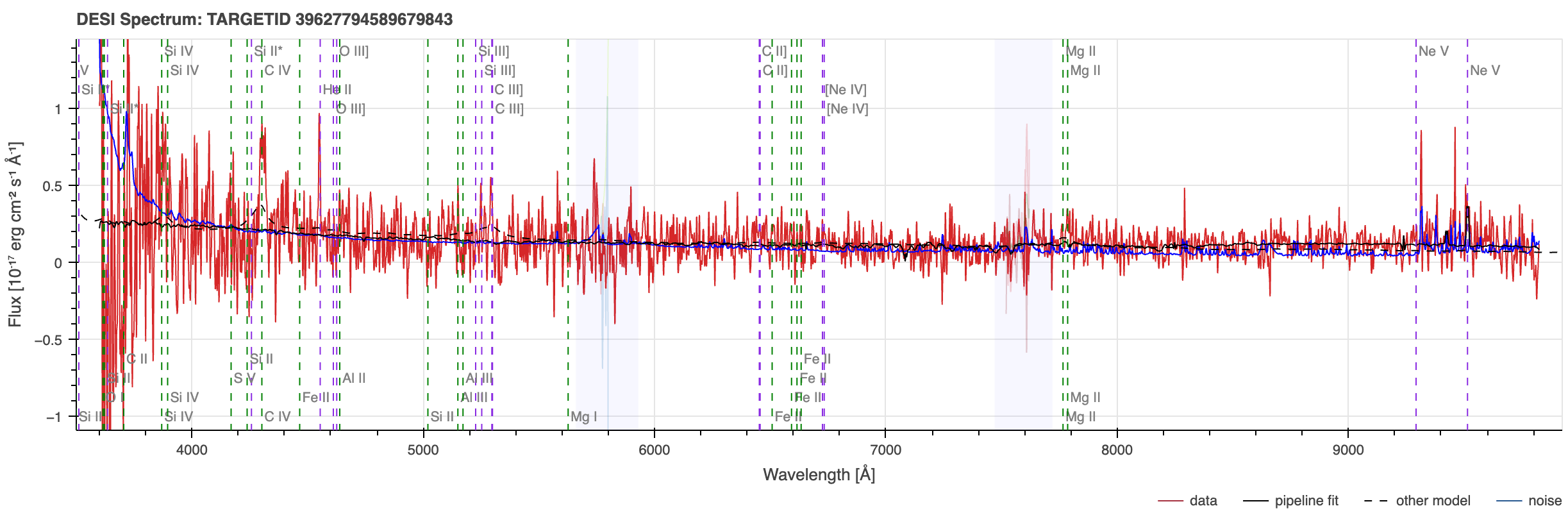}
	} 
\subfigure[G4Jy 1419 at the re-fitted redshift, $z = 1.3040$]{
	\includegraphics[width=0.98\linewidth]{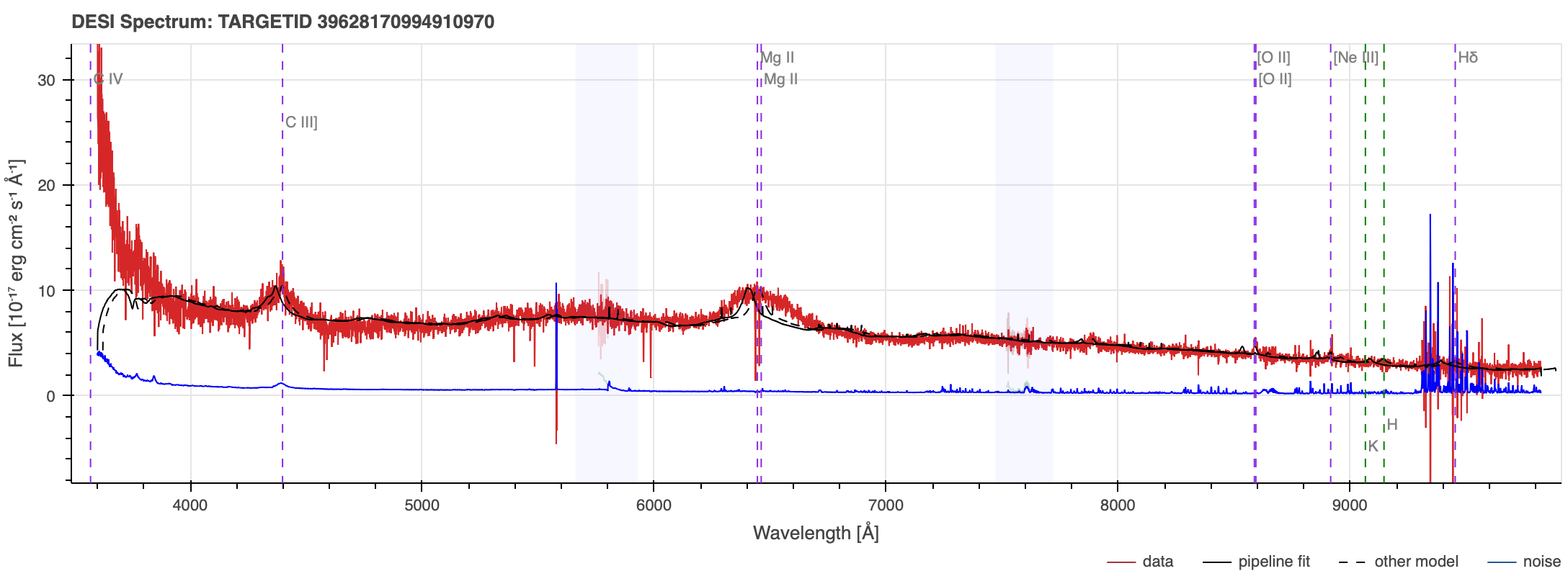} 
	}  

\caption{Re-fitting of DESI DR1 spectra, via the DESI Spectral Viewer, for (a) G4Jy~923, (b) G4Jy~1228, and (c) G4Jy 1419 (Appendix~\ref{app:desi_spectra}).  The target emission is shown in red, and the noise spectrum is shown in blue. Major spectral features are shown for panels (a) and (c), whilst minor features have been toggled `on' for panel (b). The pipeline fit to the spectrum is represented by a solid black line, whilst a dashed black line represents our alternative template fit. \label{fig:refitted_desi_spectra}}
\end{figure*}

Before adding new DESI DR1 spectra to our dataset, we again conduct visual inspection to identify any anomalies. These spectra were accessed through the Legacy Survey Sky Viewer\footnote{\url{https://www.legacysurvey.org/viewer}}, which links to a sleek graphical user interface (GUI) that allows the user to refine the redshift fitting, if they so wish. We completed such re-fitting for three sources (Figure~\ref{fig:refitted_desi_spectra}), as discussed below.

For G4Jy~923 we note that the spectrum is better-fit by a `QSO' (quasar) template than a `GAL' (galaxy) template. When we apply the former, we can see that the broad emission around 4600\,\AA\ can be explained by the Mg{\sc ii} emission line. In conjunction with aligning the [O{\sc iii}] emission at $\sim$8150\,\AA,\ we determine a redshift-fit of $z = 0.6337$ (rather than $z=0.6827$ via the DESI DR1 pipeline).

The pipeline fit for G4Jy~1228 is at $z=0.8999$, with a `GAL' template being chosen for this faint quasar. The detection of C{\sc iv} emission is supported by the narrow emission-line from Fe{\sc ii}, leading to a revised redshift of $z=1.7764$. However, to note the tentativeness of this interpretation, we accompany this spectroscopic redshift with a z\_Quality\_flag of `4' (Section~\ref{sec:xmatch_specz}).

The broad emission-lines for G4Jy 1419 are correctly fitted by the pipeline with a `QSO' template (at $z=1.2893$). However, we can make an improved fit to the data through manual adjustment of this template, resulting in a re-fitted redshift of $z = 1.3040$. 

\begin{figure*}
%\vspace{-0.7cm}
\centering
\subfigure[G4Jy 784 at the pipeline-fitted redshift, $z = 0.5918$]{
	\includegraphics[width=0.98\linewidth]{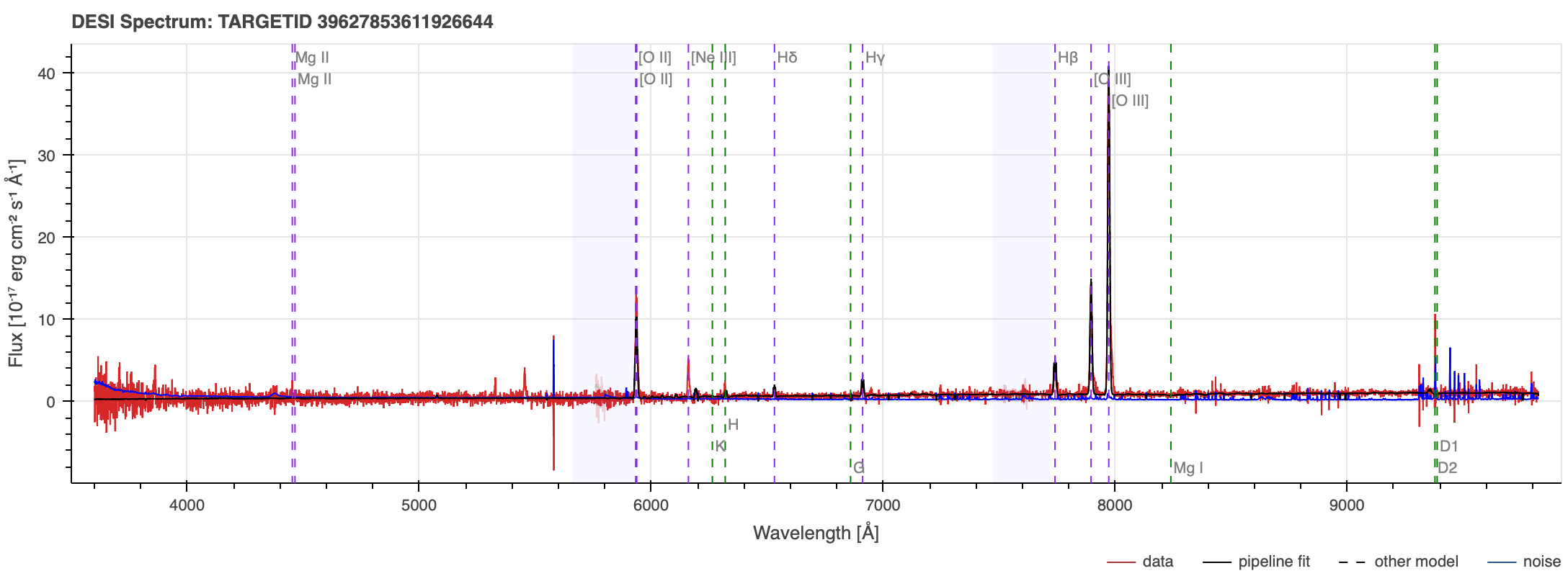}
	} 
\subfigure[G4Jy 1307 at the pipeline-fitted redshift, $z = 1.9785$]{
	\includegraphics[width=0.98\linewidth]{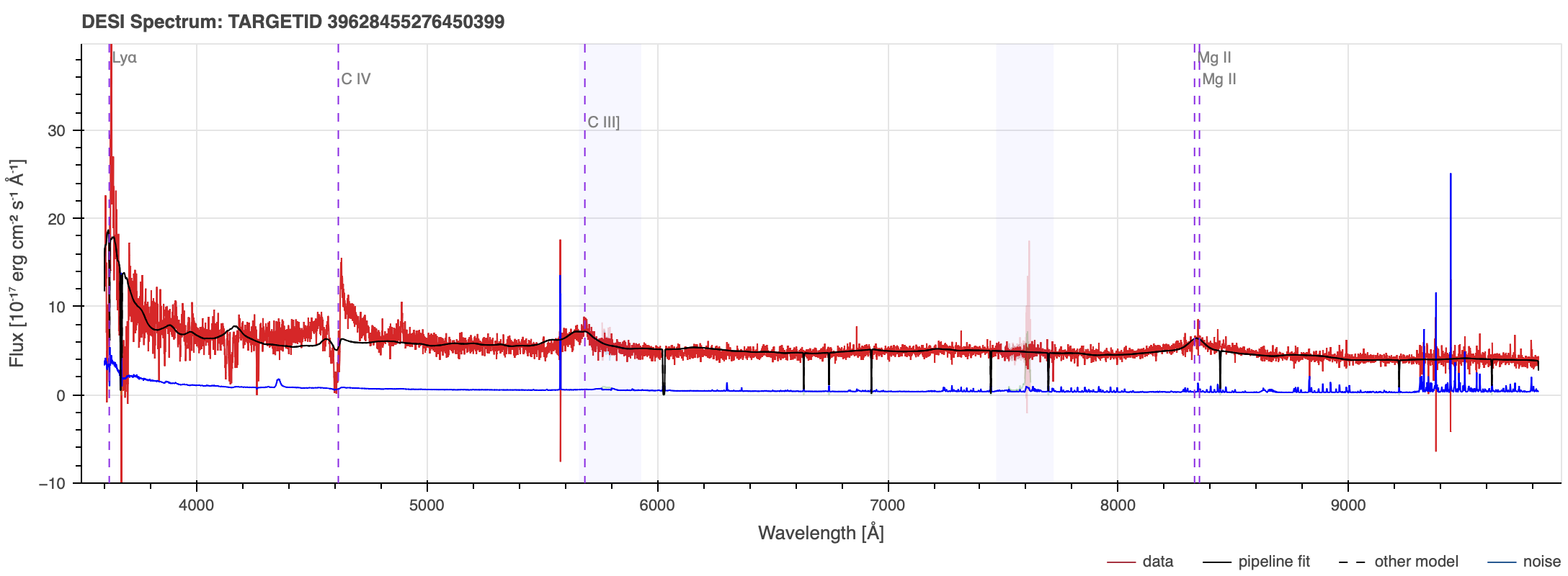}
	} 
\subfigure[G4Jy 1323 at the pipeline-fitted redshift, $z = 0.5559$]{
	\includegraphics[width=0.98\linewidth]{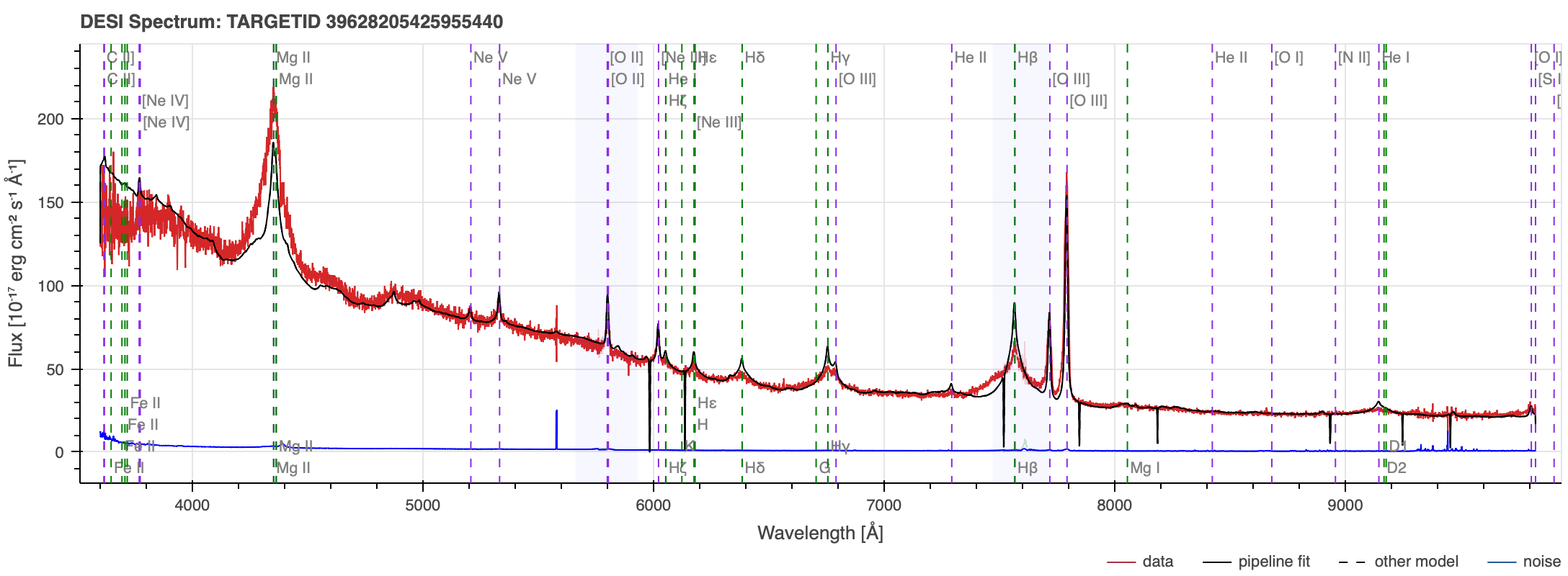} 
	}  

\caption{Examples of DESI DR1 spectra, showing (a) very little noise in the continuum emission, (b) well-accounted-for absorption of the C{\sc iv} line, and (c) numerous beautiful emission-lines (Appendix~\ref{app:desi_spectra}).  The target emission is shown in red, and the noise spectrum is shown in blue. Major spectral features are shown for panels (a) and (b), whilst minor features have been toggled `on' for panel (c). The DR1 pipeline fit to the spectrum is represented by a solid black line. \label{fig:desi_spectra_examples}}
\end{figure*}

Finally, we draw attention to beautiful examples of DESI DR1 spectra (Figure~\ref{fig:desi_spectra_examples}), noting the high signal-to-noise that helps to make high-precision redshift measurements for millions of sources across more than 9,000\,deg$^2$ \citep{AbdulKarim2025}. In addition, the first author thanks Dustin Lang for promptly updating the DESI Spectral-Viewer GUI so that it is more accessible for colour-blind users.  

\setkeys{Gin}{draft=false}

\setkeys{Gin}{draft}

%%%%%%%%%%%%%%%%%%%%%%%%%%%%%%%%%%%%%%%%%%%%%%%%%%

% Don't change these lines
\bsp	% typesetting comment
\label{lastpage}
\end{document}